\documentclass[preprint,12pt]{elsarticle}

\usepackage{graphicx}% Include figure files
\usepackage{dcolumn}% Align table columns on decimal point
\usepackage{bm}% bold math
\usepackage{subfigure}
\usepackage{longtable}
\usepackage{hyperref}

\usepackage{algorithm}
\usepackage{algpseudocode}
\usepackage{amsmath}
\usepackage[utf8]{inputenc}
\usepackage[T1]{fontenc}
\usepackage{mathptmx}
\usepackage{etoolbox}
\usepackage{float}
\usepackage{amssymb}
 \usepackage{booktabs}
\makeatletter
\def\@email#1#2{%
 \endgroup
 \patchcmd{\titleblock@produce}
  {\frontmatter@RRAPformat}
  {\frontmatter@RRAPformat{\produce@RRAP{*#1\href{mailto:#2}{#2}}}\frontmatter@RRAPformat}
  {}{}
}%
\makeatother
\begin{document}

%\preprint{AIP/123-QED}

\begin{frontmatter}

\title{Structural Dynamics of G5 Stock Markets During Exogenous Shocks: A Random Matrix Theory-Based Complexity Gap Approach}

\author[label1]{Kundan Mukhia}
\ead{kundanmukhia07@gmail.com}

\author[label2]{Imran Ansari\corref{cor1}}
\ead{imranansari@iisc.ac.in}

\author[label1]{Md.\ Nurujjaman}
\ead{md.nurujjaman@nitsikkim.ac.in}

\cortext[cor1]{Corresponding author}

\affiliation[label1]{organization={Department of Physics, National Institute of Technology Sikkim},
            addressline={},
            city={Sikkim},
            postcode={737139},
            state={},
            country={India}}

\affiliation[label2]{organization={Department of Management Studies, Indian Institute of Science},
            addressline={},
            city={Bengaluru},
            postcode={},
            state={Karnataka},
            country={India}}

%% Abstract
\begin{abstract}
We identify a robust structural signature of stock markets during exogenous shock events by analyzing collective return dynamics across G5 countries. Using Random Matrix Theory, we introduce the complexity gap, defined as the difference between the normalized largest eigenvalue and the average pairwise correlation, to quantify changes in market structure. This measure reveals a consistent three-phase pattern across multiple shocks, including the 2025 U.S. tariff event, the COVID-19 crisis, and country-specific shocks in Japan and China during 2024. Before a shock, markets show a positive complexity gap, reflecting a rich structure with multiple interacting factors. During shocks, the gap collapses to near zero, signaling strong synchronization under a single dominant mode. Post-shock recovery follows a nonmonotonic path: an initial widening (a false recovery), a temporary recollapse, and final sustained restoration. This pattern holds at both market and sector levels and across global and local shocks. Ordinal entropy analysis confirms the same sequence of collapse and false recovery in directional diversity. We further demonstrate that lower complexity gap values predict higher future portfolio volatility, especially after shocks, establishing its value as a state-dependent risk indicator. For investors, initial gap widening may mislead, while sustained widening signals genuine structural stabilization. These findings reveal a robust structural signature governing financial market dynamics during crisis and recovery periods.
\end{abstract}

\end{frontmatter}

\section{Introduction}

Financial markets are complex adaptive systems characterized by interactions among thousands of assets, where collective dynamics emerge from the aggregation of individual behaviors~\cite{mantegna1999introduction, bouchaud2003theory}. These dynamics evolve continuously in response to changing economic conditions, policy decisions, and external shocks~\cite{mantegna1999hierarchical, sornette2017stock,mukhia2025early}. Under normal conditions, asset returns exhibit heterogeneous correlation structures reflecting sectoral organization, economic linkages, and multiple sources of risk~\cite{onnela2003dynamics, tumminello2005tool, ansari2025novel}. However, during periods of financial shock, markets often undergo abrupt transitions toward strongly synchronized behavior, in which assets move collectively, and structural distinctions between sectors become less pronounced or even disappear~\cite{billio2012econometric, pawanesh2025exploring}. Such episodes, observed during events such as the COVID-19 crisis and more recent global shocks, including the 2025 U.S. tariff announcement, highlight the fragility of diversification and the emergence of dominant market-wide forces.

Correlation-based approaches provide a natural framework for studying the collective organization of financial markets \cite{mantegna1999hierarchical, onnela2003dynamics, bonanno2003topology, ansari2025novel, sharma2015study}. Empirical studies consistently report a significant increase in correlations during crisis periods, reflecting the growing influence of common factors \cite{junior2012correlation, song2011evolution, mukhia2024complex}. However, average correlation alone does not fully characterize changes in market structure \cite{plerou2002random, malevergne2004collective}. Two markets with similar average correlations may exhibit fundamentally different internal organizations, ranging from well-defined sectoral clustering to nearly uniform collective motion \cite{tumminello2005tool, kenett2010dominating}. Capturing this distinction requires measures that go beyond aggregate statistics and directly quantify structural organization. A characteristic feature of systemic events is the homogenization of correlations across assets, where pairwise correlations become increasingly similar, and a common mode dominates the dynamics of individual assets \cite{song2011evolution, preis2011switching, bouchaud2021radical}. In this regime, sectoral organization weakens, and diversification benefits diminish, indicating a loss of structural heterogeneity in the market \cite{onnela2003dynamics, pharasi2019complex, pan2007collective}. While this phenomenon has been widely documented, quantifying the degree of structural heterogeneity and its collapse remains challenging. Existing approaches, including network filtering techniques and spectral methods based on Random Matrix Theory (RMT), do not provide a direct, simple, and easily comparable scalar measure to capture this behavior consistently across different markets and time periods \cite{laloux1999noise, plerou2002random, rosenow2000application,mukhia2026core, tang2023looking}.

The application of the RMT to financial correlation matrices has emerged as a well-established framework for separating meaningful economic signals from random noise~\cite{potters2005financial, mehta2004random}. The pioneering studies of Laloux et al.~\cite{laloux1999noise} and Plerou et al.~\cite{plerou1999universal} showed that the bulk of the eigenvalue spectrum of empirical correlation matrices closely follows the Marčenko–Pastur distribution expected for purely random matrices, but the largest eigenvalue deviates significantly. This dominant eigenvalue, commonly referred to as the market mode, captures the collective co-movement shared by all assets in the market. Subsequent research has shown that these spectral properties undergo systematic changes during periods of financial stress. Empirical evidence from major crises, including the 2008 global financial crisis, the European Sovereign Debt Crisis, and the COVID-19 pandemic, consistently reveals a sharp increase in the largest eigenvalue, accompanied by a compression of the remaining spectrum~\cite{zheng2012changes, kenett2010dynamics, sandoval2014structure, junior2012correlation, dominguez2025correlation}. Together, these effects indicate a transition toward stronger market-wide synchronization and a reduction in effective dimensionality, implying that fewer independent factors are sufficient to explain asset returns. Related research has examined the average pairwise correlation coefficient as a more direct measure of market integration. Studies such as Sandoval and Franca~\cite{junior2012correlation} and Kenett et al.~\cite{kenett2012quantifying} show that average correlations increase markedly during crisis periods, reflecting the breakdown of diversification benefits. While both approaches capture heightened collective behavior, they do so from distinct perspectives: the largest eigenvalue reflects the dominance of a global factor, whereas average correlation aggregates both systemic and idiosyncratic interactions. Despite their widespread use, these measures have largely been analyzed in isolation. As a result, the structural relationship between the dominant market mode and aggregate pairwise co-movement remains underexplored, limiting our ability to directly quantify market heterogeneity~\cite{plerou2002random, malevergne2004collective, mukhia2026core}. Recent work by González et al.~\cite{molero2025random} shows that the largest eigenvalue serves as a proxy for market spillovers and corresponds to the Sharpe market factor, while the second-largest eigenvalue exhibits counter-cyclical behavior, suggesting a potential defensive role during periods of high volatility. However, the existing literature does not develop a unified framework linking the dominant market mode to the broader correlation structure and its evolution across market shocks.

Despite the extensive literature in this area, several important questions remain unresolved. Most studies have compared discrete market states, pre-crisis, crisis, and post-crisis, rather than tracking the continuous temporal evolution of market structure following a shock. Whether recovery is smooth and monotonic, or whether it passes through intermediate phases, remains largely unexplored. In addition, the majority of studies examine one shock event or one market in isolation, making it difficult to distinguish between responses that are generic to all crises and those that are specific to the nature of the shock. Sectoral dynamics within the RMT framework have also received limited systematic attention; market-level analyses may mask heterogeneous sector-level responses that are critical for understanding how shocks propagate through the economic system. Finally, the practical value of spectral complexity measures for portfolio risk management has not been evaluated in a systematic and comparative way across multiple markets and shocks.

To address these gaps, we introduce the complexity gap, defined as the difference between the normalized largest eigenvalue and the average pairwise correlation of the return correlation matrix. This scalar measure captures the deviation between the dominant market mode and the overall level of co-movement, and we argue that it provides a natural and interpretable indicator of structural market heterogeneity. When the complexity gap is large, the market is organized around a strong dominant factor while retaining meaningful secondary structure; when it collapses, the market transitions toward uniform collective motion. We apply this measure to multiple markets and sectors across several distinct shock events, including the COVID-19 pandemic, the 2025 U.S. tariff announcement, and country-specific shocks in Japan and China in 2024. This allows us to examine whether the complexity gap exhibits a consistent three-phase temporal pattern around exogenous shocks, characterized by collapse, false recovery, and sustained structural restoration. We find that markets exhibit a robust three-phase response pattern consisting of an initial collapse of the gap, a transient 'false recovery' widening, and a final sustained restoration of structural diversity. We further show that while the sign of the complexity gap inverts at the sectoral level due to higher intra-industry homogeneity, the three-phase dynamic remains a scale-invariant feature of the system. Finally, we validate these findings using a distinct ordinal entropy framework and demonstrate that the complexity gap contains significant predictive information for portfolio risk, particularly as a state-dependent indicator of post-shock volatility. We further show that while domestic shocks produce localized disruptions, the 2025 U.S. tariff announcement triggers a globally synchronized response across all G5 markets and sectors.

The remainder of the paper is organized as follows. Section~\ref{methodlogy} describes the data and outlines the methodological framework employed in this study, including the construction of correlation matrices, the RMT-based spectral decomposition, and the formulation of the Complexity Gap as a measure of structural market dynamics. Section~\ref{Results} presents the empirical findings, covering both market-wide and sectoral analyses, the identification of recovery dynamics across different shock events, and the implications of these results for portfolio risk assessment. Finally, Section~\ref{Conclusion} summarizes the findings of our study.

\section{Method of Analysis}
\label{methodlogy}

To identify the structural signature of stock markets during exogenous shock events, we applied several analytical methods. Initially, we use Random Matrix Theory (RMT) to decompose the rolling correlation matrix of stock returns and extract the normalized largest eigenvalue $\lambda^{\rm norm}_{\rm max}(t)$ and the average pairwise correlation $\rho(t)$. This is followed by the definition of the complexity gap $\Delta(t) = \lambda^{\rm norm}_{\rm max}(t) - \rho(t)$, which serves as our central measure for tracking structural changes in market dynamics across pre-shock, shock, and post-shock regimes. We support this with an ordinal entropy analysis of cross-sectional return patterns to validate the observed structural shifts. Finally, we employ portfolio optimization to evaluate whether the complexity gap carries predictive information about future portfolio risk. Further details on each of these steps are discussed below.

\subsection{Random Matrix Theory}
\label{RMT}

We analyze daily closing prices $P_i(t)$ for stocks across the G5 markets. To mitigate the non-stationarity inherent in price levels, the time series are converted into logarithmic returns defined as

\begin{equation}
    r_i(t) = \ln \frac{P_i(t)}{P_i(t-1)},
\end{equation}
where $P_i(t)$ denotes the closing price of asset $i$ at time $t$. Within each rolling window, the return series are standardized using $z$-score normalization~\cite{nayak2014impact,bhanja2018impact}:
\begin{equation}
    \tilde{r}_i(t) = \frac{r_i(t) - \langle r_i \rangle}{\sigma_i},
\end{equation}
where $\langle r_i \rangle$ and $\sigma_i$ denote the sample mean and standard deviation of stock $i$ computed over the corresponding time window. This normalization ensures that the standardized returns satisfy $\langle \tilde{r}_i \rangle = 0$ and $\mathrm{Var}(\tilde{r}_i) = 1$.

A rolling time window of fixed length $T = 60$ trading days is applied to capture the temporal evolution of market correlations. For each time index $\tau$, we construct the return submatrix
\begin{equation}
    \mathbf{R}_\tau = \bigl\{r_i(t) \mid i = 1,\ldots,N_c,\; 
    t = \tau - T + 1,\ldots,\tau\bigr\},
\end{equation}
which contains the returns of all $N_c$ assets over the preceding $T$ trading days. The matrix $\mathbf{R}_\tau \in \mathbb{R}^{N_c \times T}$ represents the local return dynamics used to compute the correlation matrix at time $\tau$.

The equal-time Pearson correlation matrix is constructed within each rolling window as
\begin{equation}
    C_{ij}(t) = \frac{1}{T} \sum_{t=\tau-T+1}^{\tau} 
    \tilde{r}_i(t)\,\tilde{r}_j(t),
\end{equation}
where $\tilde{r}_i(t)$ denotes the standardized return of asset $i$ at time $t$ and $T$ is the window length. This yields a symmetric positive semi-definite correlation matrix $\mathbf{C}(t) \in \mathbb{R}^{N_c \times N_c}$, with diagonal elements equal to unity and off-diagonal elements $C_{ij}(t) \in [-1, 1]$ capturing the pairwise linear co-movement between assets $i$ and $j$.

The correlation matrix $\mathbf{C}(t)$ is diagonalized through the eigenvalue equation
\begin{equation}
    \mathbf{C}(t)\,\mathbf{v}_k(t) = \lambda_k(t)\,\mathbf{v}_k(t),
\end{equation}
yielding $N_c$ real eigenvalues ordered as
\begin{equation}
    \lambda_0(t) \geq \lambda_1(t) \geq \cdots \geq \lambda_{N_c - 1}(t),
\end{equation}
and their corresponding orthonormal eigenvectors $\mathbf{v}_k(t)$. In the RMT framework, for a purely random correlation matrix, all eigenvalues are expected to fall within the theoretical bounds $[\lambda_-, \lambda_+]$, where
\begin{equation}
    \lambda_{\pm} = \left(1 \pm \sqrt{\frac{1}{Q}}\right)^2, 
    \qquad Q = \frac{T}{N_c},
\end{equation}
with $T$ the window length and $N_c$ the number of stocks~\cite{laloux1999noise,plerou2002random,potters2005financial}. Eigenvalues lying within this interval are consistent with random noise and carry no significant economic information, whereas eigenvalues 
exceeding $\lambda_+$ reflect genuine collective behavior in the market~\cite{plerou1999universal}. The largest eigenvalue $\lambda_0(t)$ consistently exceeds $\lambda_+$ across all markets and time periods studied here, confirming that it carries genuine collective market information. This dominant eigenvalue, commonly referred to as the market mode, captures the degree of global synchronization shared by all assets and is widely interpreted as an indicator of systemic risk~\cite{plerou2002random, zheng2012changes}.

To facilitate comparison across markets with different numbers of assets $N_c$, the largest eigenvalue is normalized as
\begin{equation}
    \lambda^{\rm norm}(t) = \frac{\lambda_0(t) - 1}{N_c - 1}, 
    \qquad 0 \leq \lambda^{\rm norm}(t) \leq 1,
\end{equation}
where $\lambda^{\rm norm} = 0$ corresponds to a fully uncorrelated market with $\lambda_0 \approx 1$, and $\lambda^{\rm norm} = 1$ indicates the theoretical limit of perfect synchronization in which all assets move identically.

\subsection{The Complexity Gap}
\label{comp gap}

To characterize the structural organization of the market beyond the dominant mode alone, we compute the average pairwise correlation as the mean of all off-diagonal elements of $\mathbf{C}(t)$:
\begin{equation}
    \rho(t) = \langle C_{ij}(t) \rangle_{i \neq j}.
\end{equation}
While $\lambda^{\rm norm}(t)$ reflects the strength of the single dominant market factor, $\rho(t)$ aggregates all pairwise interactions, including sectoral and idiosyncratic components. These two quantities therefore, characterize collective market behavior from complementary perspectives. We define the complexity gap as the difference:
\begin{equation}
    \Delta(t) = \lambda^{\rm norm}(t) - \rho(t).
\end{equation}
This scalar measure captures the deviation between the dominant market mode and the overall level of pairwise co-movement. A large positive gap, $\Delta(t) > 0$, indicates a structurally rich market in which multiple interacting components, such as industry sectors, contribute to the correlation structure alongside the common market trend. When the gap collapses toward zero, $\Delta(t) \approx 0$, the correlation matrix approaches an effective rank-one structure, indicating that stock dynamics are governed almost entirely by a single global factor and structural diversity is suppressed. The complexity gap, therefore, provides a direct, interpretable, and easily comparable scalar indicator of structural market heterogeneity that can be tracked continuously over time across different markets and shock events.

\subsection{Ordinal Entropy }
\label{Ord Entropy }

The ordinal pattern approach transforms a time series into a sequence of rank‑based symbols, providing a robust description of its temporal structure that is invariant under monotonic transformations and resistant to outliers \cite{PhysRevLett.88.174102}. Given a time series $\{x_t\}$ of length $T$, one constructs $d$-dimensional state vectors 
\begin{equation}
X_t = (x_t, x_{t+\tau}, \dots, x_{t+(d-1)\tau}),
\end{equation}
where $d$ is the embedding dimension and $\tau$ is the time delay. Each vector $X_t$ is mapped to a permutation $\pi$ of $\{0,1,\dots,d-1\}$ that sorts its components in ascending order. The relative frequency $p(\pi)$ of each of the $d!$ possible ordinal patterns is then computed, and the permutation entropy is defined as the Shannon entropy of this distribution:
\begin{equation}
H = -\sum_{\pi} p(\pi) \log p(\pi).
\end{equation}
When all patterns appear equally often, $H$ attains its maximum $\log(d!)$, indicating maximal complexity; a low entropy value signals that a small number of patterns dominate the dynamics.

In this work, we apply the ordinal pattern framework to the cross‑section of stock returns rather than a single time series. For each 60‑day rolling window, we consider the final three daily returns of every stock ($d=3$, $\tau=1$). For each stock, we determine its 3‑day ordinal pattern, yielding one of the $3!=6$ possible permutations. Aggregating over all $N$ stocks in the window, we obtain the empirical distribution $p_k(t)$ of patterns across the market. The Ordinal Entropy of the market at time $t$ is then
\begin{equation}
H_{\mathrm{ord}}(t) = -\sum_{k=1}^{6} p_k(t) \log p_k(t),
\end{equation}
which quantifies the diversity of short‑term directional behavior: a value near $\log 6 \approx 1.79$~nats reflects heterogeneous trading sequences, whereas a drop in $H_{\mathrm{ord}}(t)$ indicates that a large fraction of stocks synchronize onto the same ordinal pattern. By construction, $H_{\mathrm{ord}}(t)$ is consistent with the RMT Complexity Gap $\Delta(t)$, enabling a direct comparison between macro-structural and micro-behavioral aspects of market dynamics.

\subsection{Portfolio Optimization}
\label{Port Opt}

To examine whether the complexity gap contains information about future portfolio risk, we consider a financial market of $N$ stocks with return vector $\mathbf{r}_t = (r_{1,t}, \dots, r_{N,t})^\top$. A portfolio with weight vector $\mathbf{q}$ has return
\begin{equation}
    r_{p,t} = \mathbf{q}^\top \mathbf{r}_t.
\end{equation}
Let $\mathbf{V}$ denote the covariance matrix of returns estimated within the formation window. We construct the minimum-variance portfolio (MVP)~\cite{markowitz1952modern}, which minimizes portfolio risk for a given level of investment, by solving
\begin{equation}
    \min_{\mathbf{q}}\; \mathbf{q}^\top \mathbf{V}\,\mathbf{q}
    \quad \text{s.t.} \quad \mathbf{q}^\top \mathbf{1} = 1,
\end{equation}
with the closed-form solution
\begin{equation}
    \mathbf{q}_{\rm MVP} = 
    \frac{\mathbf{V}^{-1}\mathbf{1}}
    {\mathbf{1}^\top \mathbf{V}^{-1}\mathbf{1}},
\end{equation}
where the Moore--Penrose pseudo-inverse is used for numerical stability. As a naive benchmark, we also consider the equally weighted portfolio (EWP)~\cite{maillard2010properties} with weights $q_i^{\rm EW} = 1/N$, which requires no parameter estimation.

To link portfolio risk to market structure, we compute the correlation matrix $\mathbf{C}$ of standardized returns within each formation window. The normalized largest eigenvalue is
\begin{equation}
    \lambda_{\rm norm} = \frac{\lambda_{\max} - 1}{N - 1},
\end{equation}
and the average pairwise correlation is
\begin{equation}
    \bar{\rho} = \langle C_{ij} \rangle_{i \neq j}.
\end{equation}
The complexity gap is then defined as
\begin{equation}
    \Delta = \lambda_{\rm norm} - \bar{\rho},
\end{equation}
which captures the deviation between the dominant collective market mode and the average level of pairwise correlations. A large $\Delta > 0$ reflects a structurally diverse market, while $\Delta \approx 0$ signals a loss of structural heterogeneity under strong market-wide synchronization.

We employ a rolling window framework consisting of a $T$-day formation window followed by an $h$-day test window. In each formation window, we randomly select $N = 10$ stocks and generate $P = 500$ portfolios via Monte Carlo sampling~\cite{detemple2003monte, michaud2007estimation}. Portfolio weights are estimated in the formation window and applied out-of-sample to the test window. The realized portfolio volatility is computed as
\begin{equation}
    \sigma^{\rm test} = \sqrt{\mathrm{Var}(r_{p,t}^{\rm test})},
\end{equation}
and annualized for comparability across windows. For each portfolio, the complexity gap $\Delta$ is recorded from the formation window. The analysis spans all five G5 markets from January to December 2025. With $W_c$ rolling windows and $P = 500$ portfolios per window, the total number of portfolio observations per country is $\mathcal{N}_c = W_c P$, giving 416,500 observations in total.

To evaluate the predictive content of $\Delta$ for future portfolio risk, we use three complementary approaches: Spearman rank correlations between $\Delta$ and realized MVP volatility, incremental $R^2$ measures that quantify the additional explanatory power of $\Delta$ beyond standard benchmark predictors, and a quintile analysis in which portfolios are sorted into five groups based on $\Delta$ to examine whether higher structural complexity is systematically associated with lower realized volatility. The full computational procedure is summarized in Algorithm~\ref{alg:portfolio}.

\begin{algorithm}[htbp]
\caption{Portfolio Risk Evaluation}
\label{alg:portfolio}
\begin{algorithmic}[1]
\For{each rolling window $w = 1, \dots, W$}
    \State Define formation window of length $T$
    \State Define subsequent test window of length $h$
    \State Restrict to stocks with complete data in both windows
    \For{$p = 1, \dots, P$}
        \State Randomly select $N = 10$ stocks
        \State Standardize returns in formation window
        \State Compute correlation matrix $\mathbf{C}$
        \State Estimate covariance matrix $\mathbf{V}$
        \State Compute MVP weights
        \[
            \mathbf{q} = 
            \frac{\mathbf{V}^{-1}\mathbf{1}}
            {\mathbf{1}^\top \mathbf{V}^{-1}\mathbf{1}}
        \]
        \State Compute complexity gap
        \[
            \Delta = \frac{\lambda_{\max} - 1}{N - 1} - \bar{\rho}
        \]
        \State Compute test-window portfolio returns 
               $r_{p,t}^{\rm test}$
        \State Compute realized annualized volatility 
               $\sigma^{\rm test}$
        \State Store $(\Delta,\, \sigma^{\rm test})$
    \EndFor
\EndFor
\State Sort all portfolios by $\Delta$
\State Partition into quintiles $Q_k$, $k = 0, 1, 2, 3, 4$
\State Compute average volatility per quintile
\[
    \bar{\sigma}^{(k)} = \frac{1}{|Q_k|} 
    \sum_{i \in Q_k} \sigma_i^{\rm test}
\]
\end{algorithmic}

\end{algorithm}

\subsection{Data Analyzed}

To identify structural patterns in stock markets during shock periods, we conduct the analysis across the G5 countries, namely, the United States, China, Japan, Germany, and India. The analysis period covers two global shocks, the COVID-19 shock and the 2025 U.S. tariff shock, along with local shocks observed in Japan and China in 2024. We selected 120 stocks from each G5 country, comprising five sectors: Information Technology (IT), Industrials, Healthcare, Consumer, and Finance. The stock data were obtained from Yahoo Finance~\cite{yahoofinance} using the Python package yfinance, and the corresponding list of stock tickers is provided in Appendix~\ref{List of Stock Tickers}. We use daily closing prices from January 1, 2019, to December 31, 2022, to analyze the COVID-19 shock, and from January 1, 2023, to December 31, 2025, to examine the 2025 U.S. tariff shock. These time windows are chosen to capture both the pre- and post-shock periods. The results of the analysis are presented in the following sections.

\section{Results and Discussion}
\label{Results}

This section presents the results of identifying structural patterns in stock markets during the COVID-19 shock, the 2025 U.S. tariff shock, and local shocks observed in Japan and China in 2024. In subsections~\ref{RMT-Based Market Dynamics and Complexity Gap} and~\ref{Sectoral Market}, we identify a three-phase structural pattern in both the overall market and sector-wise dynamics using the RMT-based complexity gap. Subsection~\ref{Ordinal Entropy} provides complementary evidence through an ordinal entropy analysis of cross-sectional return patterns, confirming the observed sequence of collapse and false recovery. Furthermore, using heatmap analysis in subsection~\ref{Heatmap}, we show that the 2025 U.S. tariff shock acts as a global shock impacting the G5 stock markets. Finally, in subsection~\ref{Portfolio Risk and the Complexity Gap}, we discuss the implications of the complexity gap for portfolio risk.

\subsection{RMT-Based Market Dynamics and Complexity Gap}
\label{RMT-Based Market Dynamics and Complexity Gap}

To understand how the financial markets collectively absorb and recover from exogenous shocks, we apply a dynamic RMT analysis. We calculate the cross-correlation $C(t)$ using a 60-day rolling window across stocks in each market of the G5 countries. Using $C(t)$, as discussed in the subsection~\ref{RMT}, we calculated the normalized largest eigenvalue $\lambda_{\max}^{\mathrm{norm}}(t) = (\lambda_{\max}(t)-1)/(N-1)$, which measures the strength of the market mode. The raw average correlation $\rho(t) = \langle C_{ij}(t) \rangle_{i \neq j}$, defined as the unfiltered mean of all off-diagonal elements of $C(t)$, captures both positive and negative co-movements among stocks. We define the difference between these two quantities, $\Delta(t) = \lambda_{\max}^{\mathrm{norm}}(t) - \rho(t)$, as the complexity gap. A large gap $\Delta(t) > 0$ indicates a structurally complex market influenced by multiple factors beyond a single common trend, whereas a small or near-zero gap $\Delta(t) \approx 0$ suggests a loss of structural diversity and market behavior dominated by one global factor. Fig.~\ref{fig:complexity_mode} shows the time evolution of $\lambda_{\max}^{\mathrm{norm}}(t)$, $\rho(t)$, and $\Delta(t)$ for the stock markets of the G5 countries namely, the United States, China, Japan, India, and Germany. The quantities $\lambda_{\max}^{\mathrm{norm}}(t)$ and $\rho(t)$ are represented by the red and green lines, respectively, while $\Delta(t)$ is shown as a shaded blue region. The red dashed vertical line indicates the date on which the administration of U.S.\ President Donald Trump announced a global tariff policy for 2025. This announcement represents an exogenous shock and provides a natural experiment to examine market behavior. Despite geographic and economic diversity, all five markets exhibit a consistent three-phase pattern in the time evolution of these quantities.

\subsubsection{Pre-Shock Market Structure}

\begin{figure}[htbp]
    \centering

    \subfigure[ China\label{fig:raw_complexity_china}]{
        \includegraphics[width=0.45\linewidth]{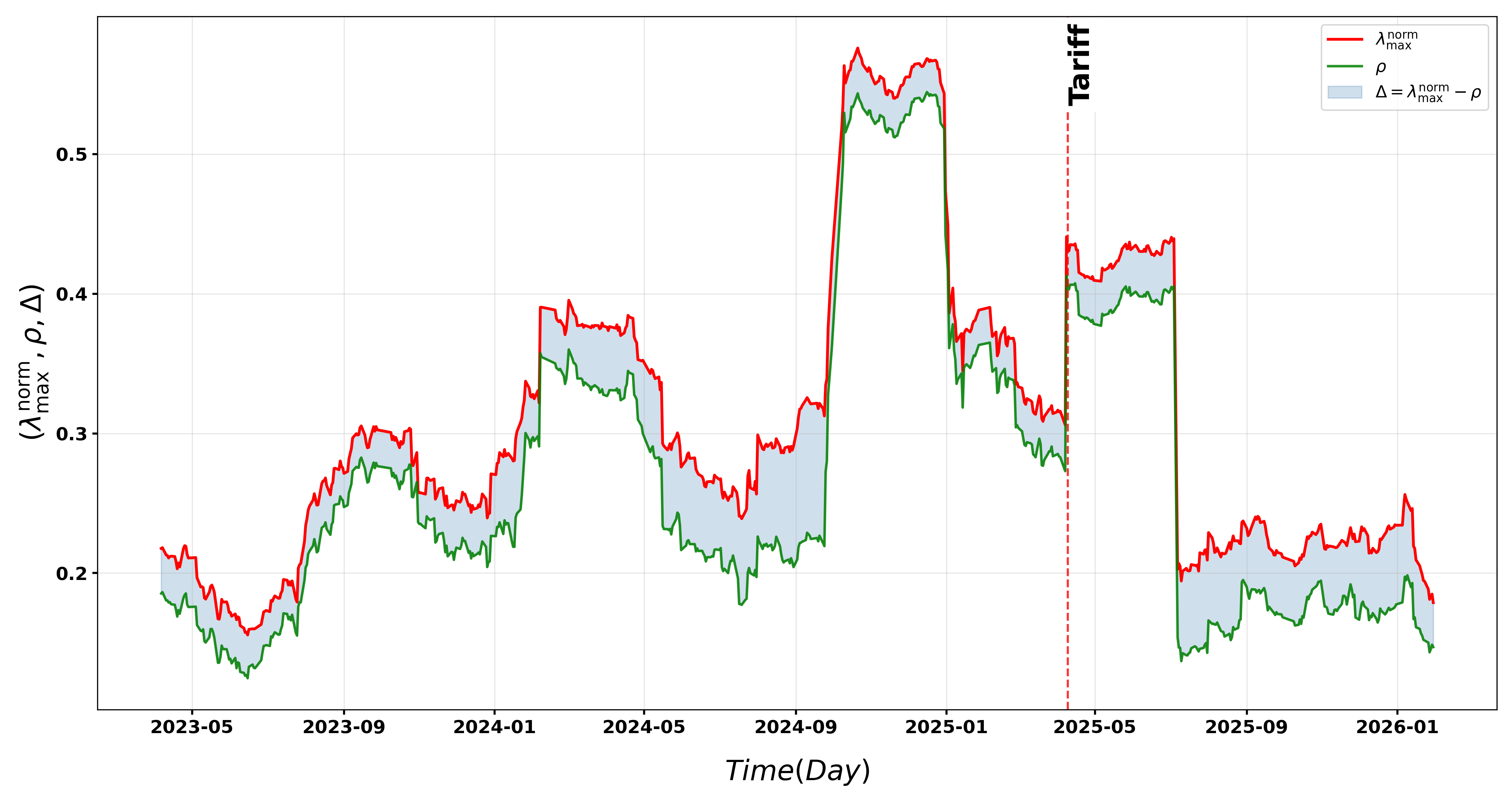}
    }
    \hspace{0.01\linewidth}
    \subfigure[ Japan\label{fig:raw_complexity_japan}]{
        \includegraphics[width=0.45\linewidth]{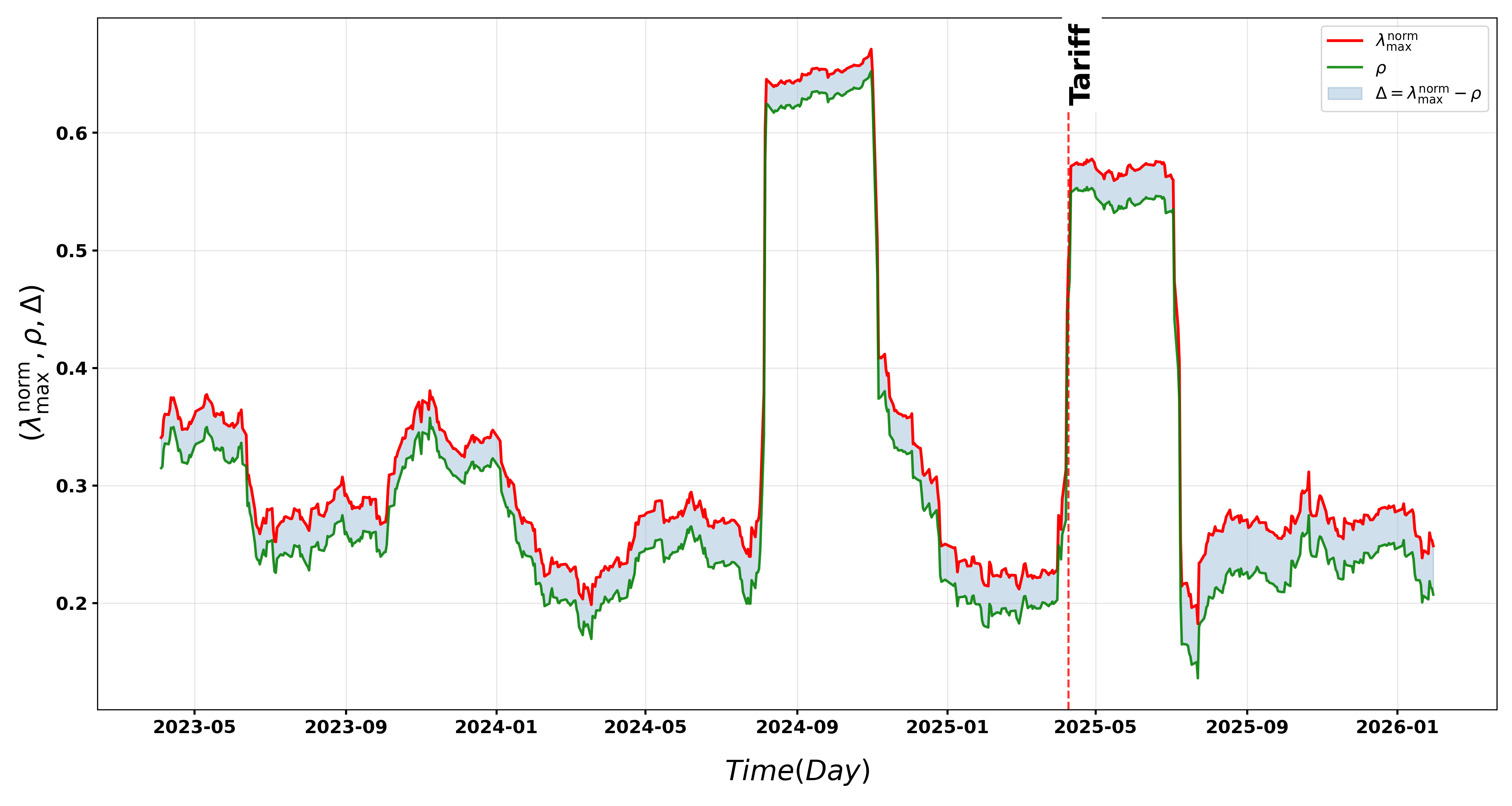}
    }

    \vspace{0.1 cm}

    \subfigure[India\label{fig:raw_complexity_india}]{
        \includegraphics[width=0.45\linewidth]{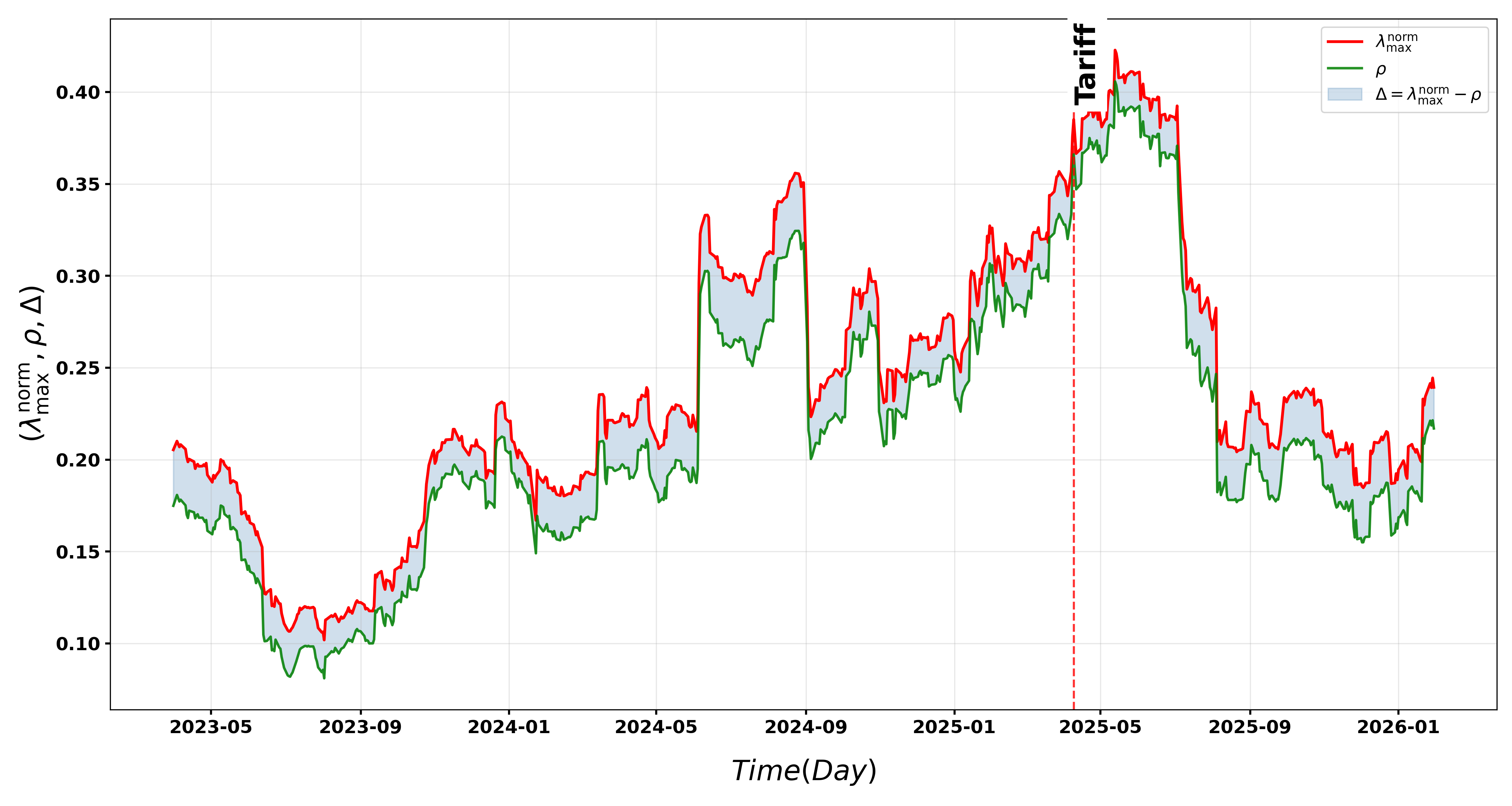}
    }
    \hspace{0.01\linewidth}
    \subfigure[ Germany\label{fig:raw_complexity_germany}]{
        \includegraphics[width=0.45\linewidth]{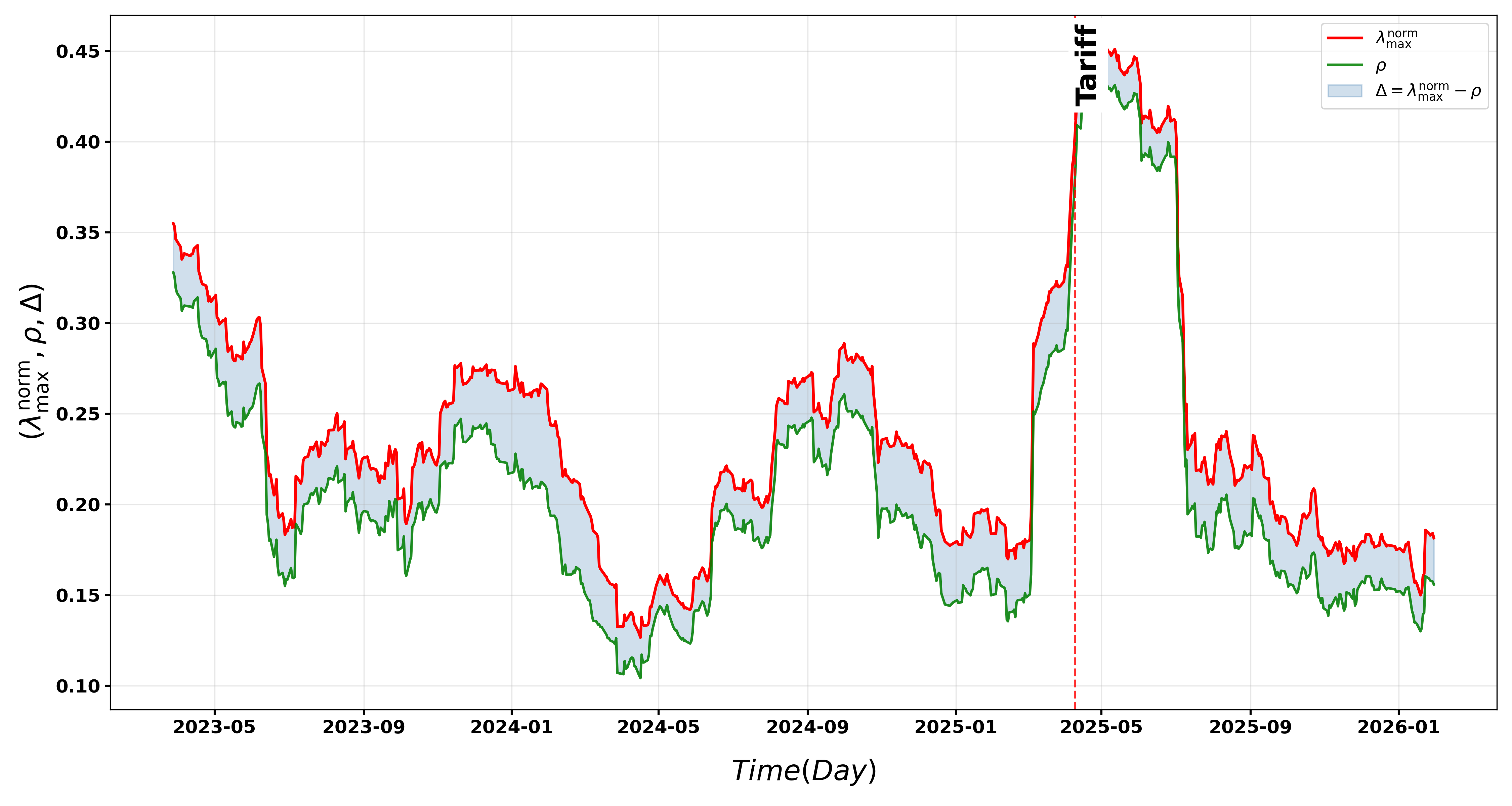}
    }

    \vspace{0.1cm}

    \subfigure[United States\label{fig:raw_complexity_us}]{
        \includegraphics[width=0.80\linewidth]{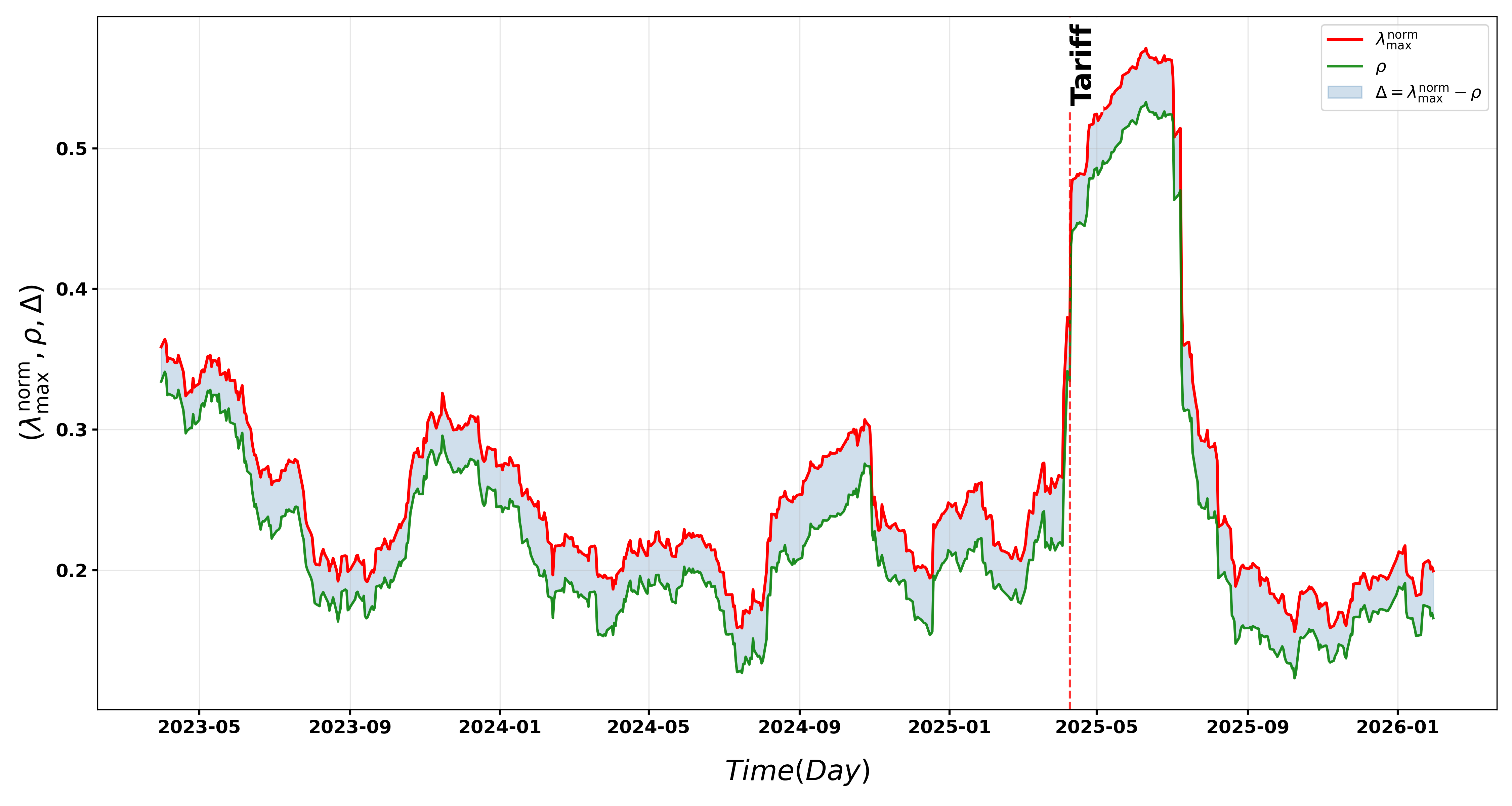}
    }

    \caption{Time evolution of RMT-based complexity metrics for G5 stock markets. The red line represents the normalized largest eigenvalue $\lambda_{\max}^{\text{norm}}(t)$, the green line shows the raw average correlation $\rho(t)$, and the shaded region between them denotes the complexity gap $\Delta(t)$. The vertical red dashed line marks the U.S. tariff announcement in April 2025. Across all markets, we observe a consistent three-phase response: a pre-event state with a persistent gap $\Delta(t) > 0$, a sharp convergence with $\Delta(t) \approx 0$ immediately following the shock, and a post-event recovery pattern characterized by gap re-widening, a secondary convergence, and finally sustained structural recovery.}

    \label{fig:complexity_mode}
\end{figure}

In this study, we define the period before the tariff announcement by the administration of U.S.\ President Donald Trump as the pre-event period. As shown in Figs.~\ref{fig:raw_complexity_china}–\ref{fig:raw_complexity_us}, during this period, the stock markets of the G5 countries exhibit distinct country-specific dynamics and patterns reflecting their respective domestic economic conditions and market structures. In this pre-event period, both $\lambda_{\max}^{\mathrm{norm}}(t)$ and $\rho(t)$ fluctuate according to the individual dynamics of each market. Throughout this period, $\lambda_{\max}^{\mathrm{norm}}(t)$ generally exceeds $\rho(t)$, resulting in a positive complexity gap $\Delta(t) > 0$. Although temporary deviations are observed in some markets during short time intervals, the overall separation between the two quantities remains positive. These variations in $\Delta(t)$ arise from country-specific market dynamics and local shocks affecting individual markets. The persistence of $\Delta(t) > 0$ during most of the pre-event period indicates the presence of additional collective modes beyond the dominant market factor. Such excess eigenvalue contributions arise from internal market structures, including industry sectors and other economically related stock groupings that exhibit correlated price movements independent of the global trend. The quantity $\rho(t)$ captures the net effect of all pairwise interactions, incorporating both positive and negative co-movements. Its systematic separation from $\lambda_{\max}^{\mathrm{norm}}(t)$, quantified by the complexity gap $\Delta(t)$, therefore serves as a measure of internal market structure. A wide gap implies that market dynamics are not fully governed by a single common factor but instead reflect the coexistence of multiple interacting components within the correlation matrix. Therefore, the pre-event period is characterized by a high-complexity market state in which sector-specific information and intra-market interactions play a significant role. Although brief intervals of gap contraction are observed in some markets due to local shocks, the overall dominance of $\Delta(t) > 0$ suggests that this structural complexity represents the baseline condition of the financial system before perturbation by the exogenous shock.

\subsubsection{Shock-Induced Synchronization}

After the tariff announcement by the U.S administration in April, 2025, the market structure undergoes a clear structural transition across all G5 countries, as shown in Figs.~\ref{fig:raw_complexity_china}–\ref{fig:raw_complexity_us}. The vertical red dashed line in Figs.~\ref{fig:raw_complexity_china}–\ref{fig:raw_complexity_us} marks the date of the tariff announcement. Following this point, both $\lambda_{\max}^{\text{norm}}(t)$ and $\rho(t)$ increase sharply across all G5 markets, while the complexity gap $\Delta(t)$ rapidly contracts toward zero. This behavior reflects the rapid onset of market-wide synchronization, where stock prices across sectors begin to move simultaneously. During this period, both quantities $\lambda_{\max}^{\text{norm}}(t)$ and $\rho(t)$ increase sharply and move toward each other, leading to a near-merging of the two lines across all G5 markets, resulting in the $\Delta(t) \approx 0$, indicating a collapse of market complexity and structural information. In this period, the dynamics of individual stocks become increasingly dominated by the global market mode rather than sector-specific interactions. This near-merging of $\lambda_{\max}^{\text{norm}}$ and $\rho(t)$ implies that the correlation matrix approaches an effective rank-one structure. In this period, fluctuations are dominated by a single collective market mode, while residual structural components, such as sectoral organization, are temporarily suppressed. This behavior represents a collapse of market dimensionality and corresponds to a regime where diversification benefits largely disappear, as individual stock dynamics become driven by a common global factor. The simultaneous synchronization of $\lambda_{\max}^{\text{norm}}(t)$ and $\rho(t)$, with $\Delta(t) \approx 0$, across all G5 markets suggests that the U.S. tariff policy triggered a global regime shift in the stock market. This consistent cross-market response to the tariff shock confirms that the tariff announcement functioned as an exogenous shock to the global financial system.

Before this tariff shock in the stock market, we also observed similar $\Delta(t) \approx 0$ patterns in individual markets corresponding to country-specific domestic shocks in 2024. The Japan stock market exhibited $\Delta(t) \approx 0$ during  August 2024, following the Bank of Japan's unexpected policy adjustment, which disrupted carry trades and triggered a sharp sell-off in Japanese equities. China experienced a significant volatility in October 2024, driven by a stimulus-fueled rally that abruptly reversed due to disappointment over the lack of fiscal detail from the National Development and Reform Commission. In each case, the $\lambda_{\max}^{\text{norm}}$, $\rho(t)$, and $\Delta(t)$ metrics successfully captured these domestic shock events. Domestic shocks produce $\Delta(t) \approx 0$ patterns that are limited to individual countries, whereas the U.S. tariff announcement leads to a simultaneous convergence across all G5 stock markets. This coordinated cross-market response indicates that the tariff policy acted as a global exogenous shock, in contrast to country-specific events that affect only local markets. The ability of the proposed RMT-based metrics to capture both local and global disturbances highlights their usefulness for monitoring market stability and identifying regime shifts at different scales.

\begin{figure}[htbp]
    \centering

    \subfigure[ China\label{fig:raw_complexity_china_covid}]{
        \includegraphics[width=0.42\linewidth]{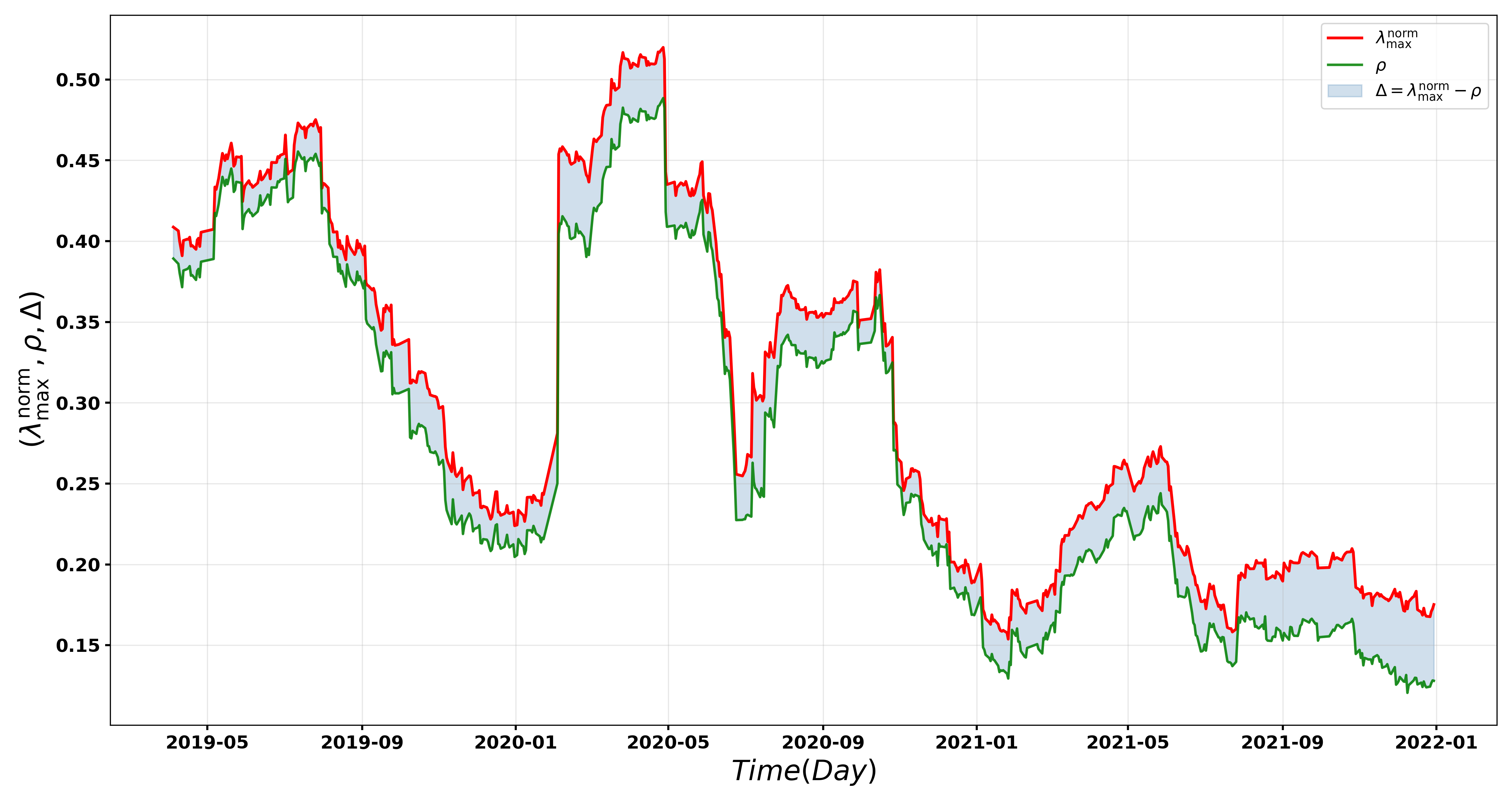}
    }
    \hspace{0.01\linewidth}
    \subfigure[ Japan\label{fig:raw_complexity_japan_covid}]{
        \includegraphics[width=0.42\linewidth]{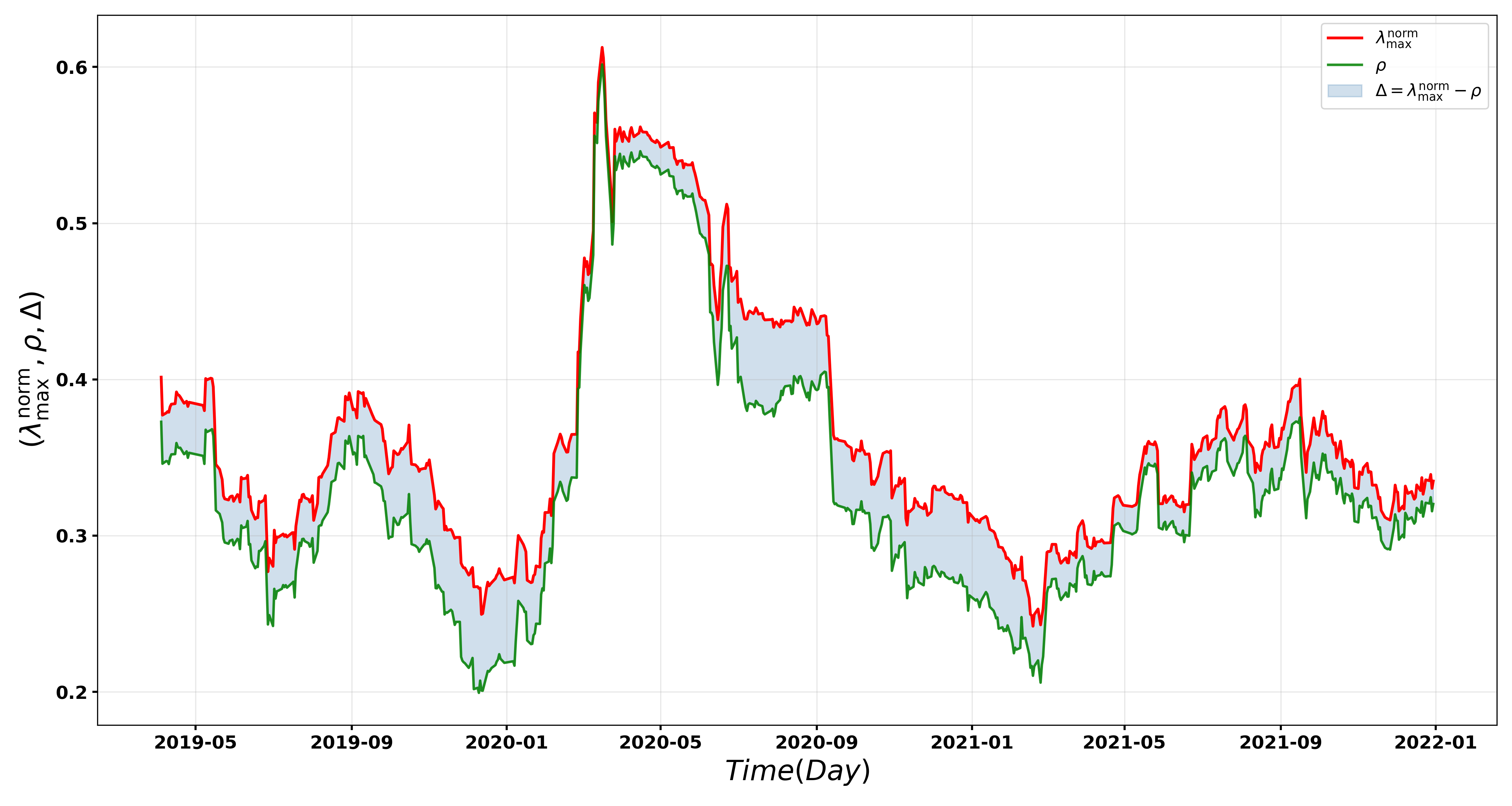}
    }

    \vspace{0.05 cm}

    \subfigure[India\label{fig:raw_complexity_india_covid}]{
        \includegraphics[width=0.42\linewidth]{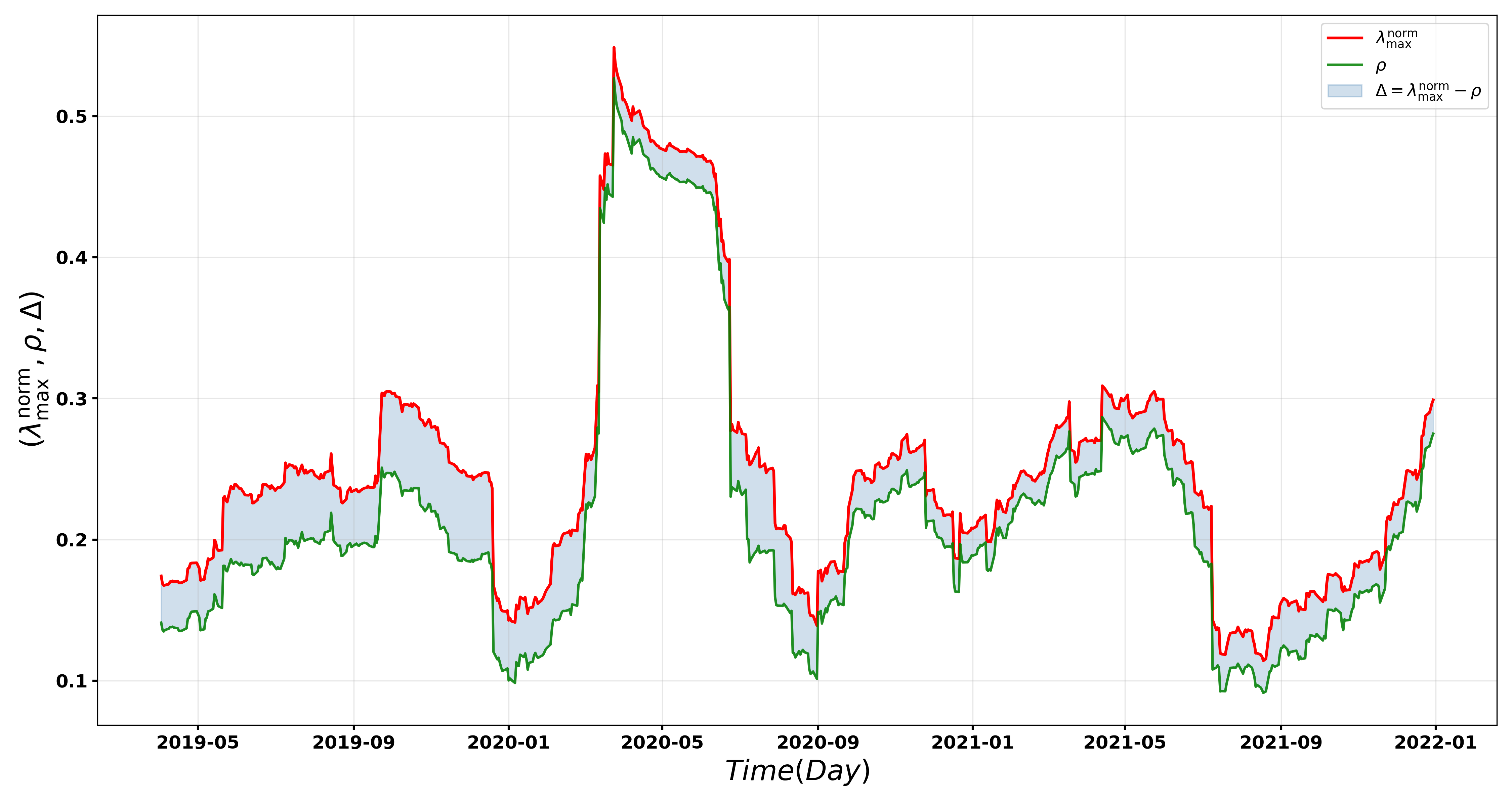}
    }
    \hspace{0.01\linewidth}
    \subfigure[ Germany\label{fig:raw_complexity_germany_covid}]{
        \includegraphics[width=0.42\linewidth]{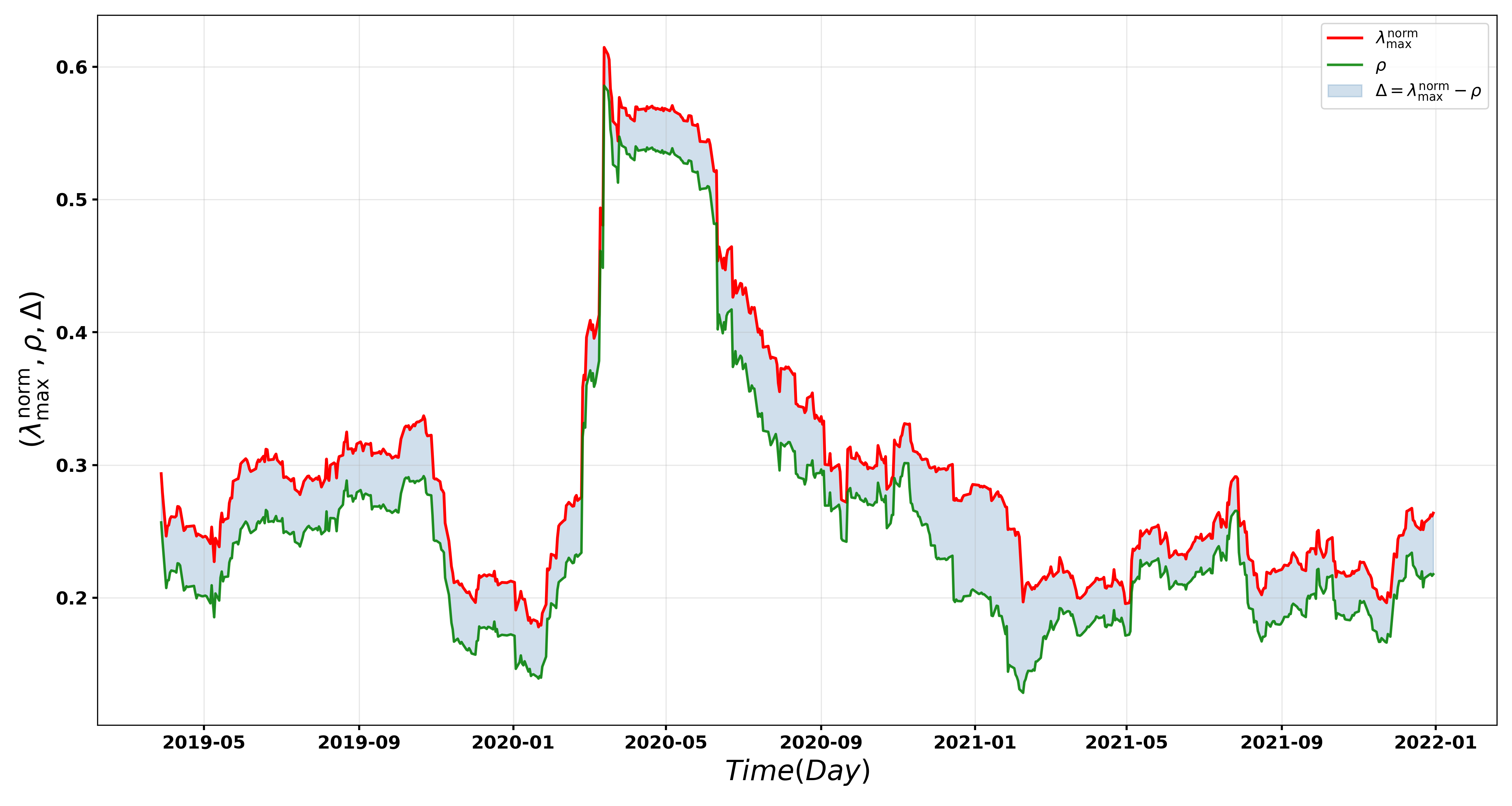}
    }

    \vspace{0.05cm}

    \subfigure[United States\label{fig:raw_complexity_us_covid}]{
        \includegraphics[width=0.70\linewidth]{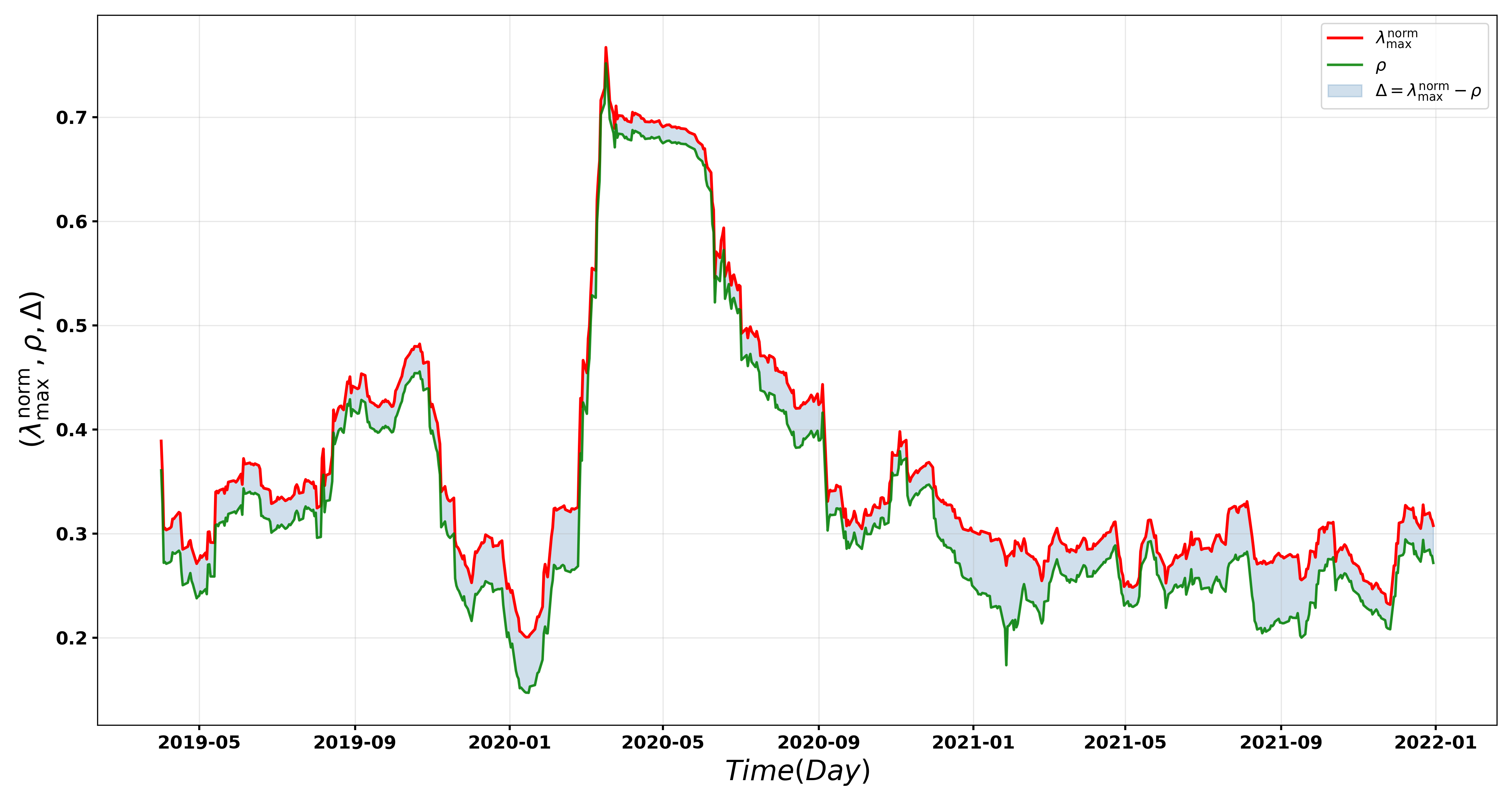}
    }
\caption{
Time evolution of RMT-based complexity metrics for G5 stock markets during the COVID-19 market shock. The red line represents the normalized largest eigenvalue $\lambda_{\max}^{\text{norm}}(t)$, the green line shows the average pairwise correlation $\rho(t)$, and the shaded region between them denotes the complexity gap $\Delta(t)$. Across all markets, a consistent three-phase pattern is observed: a pre-event regime characterized by a persistent gap $\Delta(t) > 0$, a sharp convergence with $\Delta(t) \approx 0$ during the shock, indicating strong market-wide synchronization, and a post-event recovery phase marked by gap re-widening, intermittent secondary convergence, and gradual restoration of structural complexity.}

    \label{fig:complexity_mode_covid}
\end{figure}

\subsubsection{Post-Shock Recovery Dynamics}

After the emergence of a similar event-driven pattern in $\lambda_{\max}^{\text{norm}}(t)$, $\rho(t)$, and $\Delta(t) \approx 0$ across all G5 markets in the event-driven period, the system enters a more complex post-event period. As shown in Figs.~\ref{fig:raw_complexity_china}–\ref{fig:raw_complexity_us}, the near-merging of $\lambda_{\max}^{\text{norm}}(t)$ and $\rho(t)$ with $\Delta(t) \approx 0$ is followed by a reopening of the complexity gap, resulting in $\Delta(t) > 0$ after the shock. This behavior arises from the different decay rates of $\lambda_{\max}^{\text{norm}}(t)$ and $\rho(t)$. In this regime, $\rho(t)$ decreases more rapidly than $\lambda_{\max}^{\text{norm}}(t)$, causing the complexity gap to widen, i.e., $\Delta(t) > 0$. This re-widening after the merger of the two metrics $\lambda_{\max}^{\text{norm}}(t)$ and $\rho(t)$ suggests that the market structure begins to reconstruct following the shock. As the dominant influence of the common shock weakens, sector-specific dynamics and intra-market interactions gradually re-emerge. After this initial widening of the gap, a second convergence occurs in which $\lambda_{\max}^{\text{norm}}(t)$ and $\rho(t)$ move closer together, causing $\Delta(t) \approx 0$ once more. Unlike the first event-driven convergence, which occurred at elevated levels, this secondary merger is followed by a sharp decline in both $\lambda_{\max}^{\text{norm}}(t)$ and $\rho(t)$. This pattern suggests a temporary re-synchronization of stocks as the market collectively adjusts while absorbing new information and resolving uncertainty. Both quantities continue to decline with merging to lower levels, and the gap $\Delta(t)$ gradually widens again over time. This final widening of $\Delta(t)$ indicates that market complexity has been fully restored, with stocks no longer moving uniformly and heterogeneous responses across industries, economic sectors, and individual G5 markets becoming observable again. This similar pattern is also observed during the COVID-19 market shock, as shown in fig.~\ref{fig:complexity_mode_covid} and in the domestic shocks of 2024 for the Japanese market crash in August 2024 and China’s market volatility in October 2024 across all G5 markets. Despite differences in the nature and timing of the shock, the convergence of $\lambda_{\max}^{\text{norm}}(t)$ and $\rho(t)$ and the subsequent non-monotonic recovery of $\Delta(t)$ remain qualitatively consistent across all G5 markets. This further supports the interpretation of the three-stage dynamics as a robust structural signature of crisis-driven market behavior.

To ensure the statistical robustness of these patterns, we performed a sensitivity analysis using different rolling window sizes with $L = 30$ and $L = 90$ days. In addition, we verified the results using the mean absolute correlation, defined as the arithmetic mean of the absolute off-diagonal elements of the correlation matrix $C(t)$, i.e., $\bar{\rho}(t) = \langle |C_{ij}| \rangle$. In all cases, the observed three-stage recovery pattern remained consistent across markets, confirming that the identified dynamics reflect the intrinsic market response rather than artifacts arising from parameter choices or window length.

\subsection{Sectoral Market}
\label{Sectoral Market}

The market-level analysis across the G5 countries reveals a consistent three-phase complexity response to the shock. However, this overall behavior may hide heterogeneous dynamics present at the sectoral level. To examine this, we apply the same RMT-based metrics $\lambda_{\max}^{\text{norm}}(t)$, $\rho(t)$, and $\Delta(t)$ within five sectors, namely Information Technology (IT), Industrial, Healthcare, Consumer, and Finance of the G5 stock market. This sector-wise analysis allows us to investigate whether the three-phase complexity collapse and recovery pattern observed at the market level also emerges within individual sectors, and whether different sectors exhibit varying sensitivities to the shock. Since sectoral interactions are the underlying source of the market-wide complexity gap, this decomposition provides a better understanding of how sector-specific dynamics shape the aggregate market response.

\begin{figure}[htbp]
    \centering

    \subfigure[ China\label{fig:sector_IT_china}]{
        \includegraphics[width=0.45\linewidth]{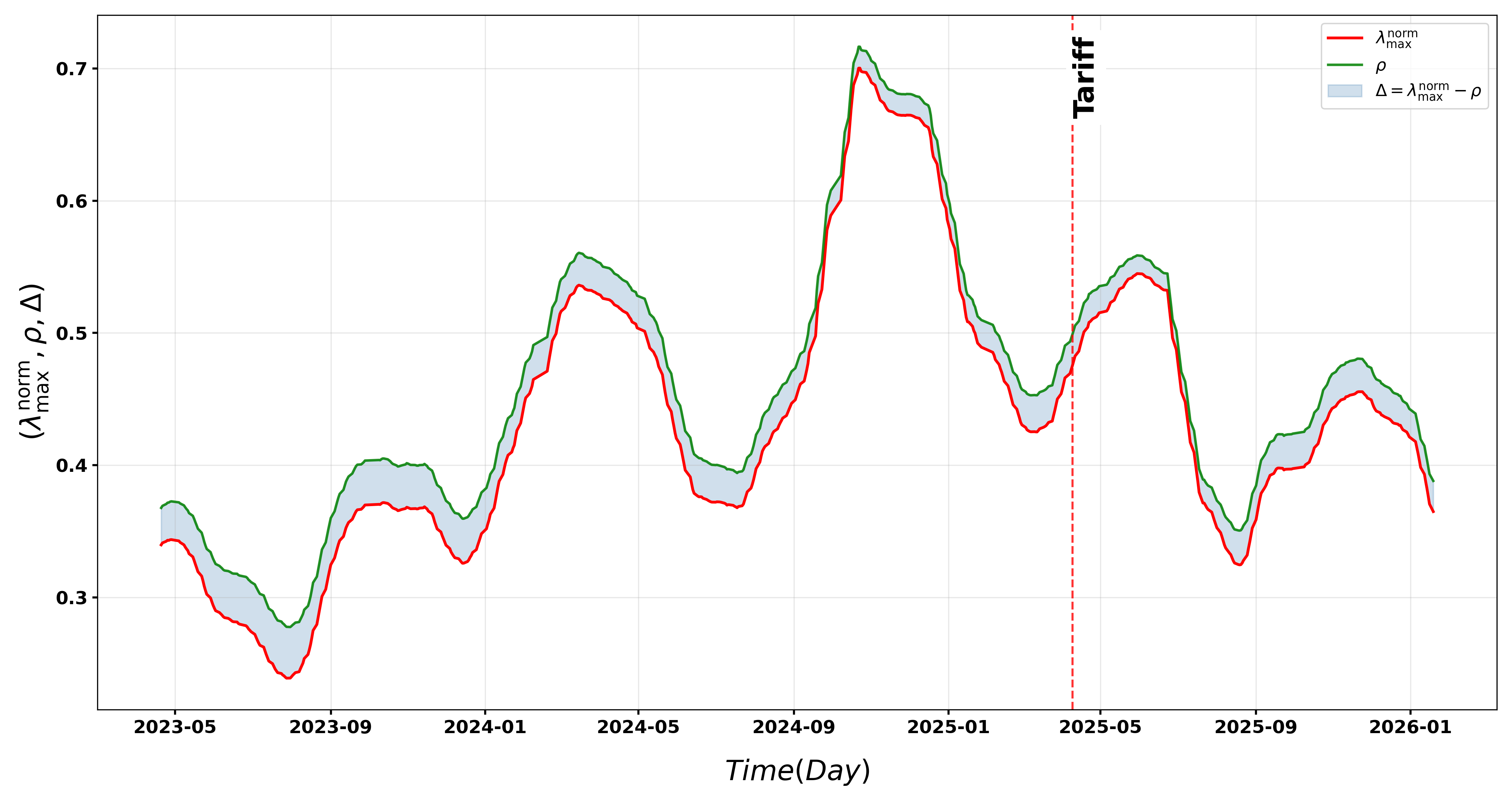}
    }
    \hspace{0.01\linewidth}
    \subfigure[ Japan\label{fig:sector_IT_japan}]{
        \includegraphics[width=0.45\linewidth]{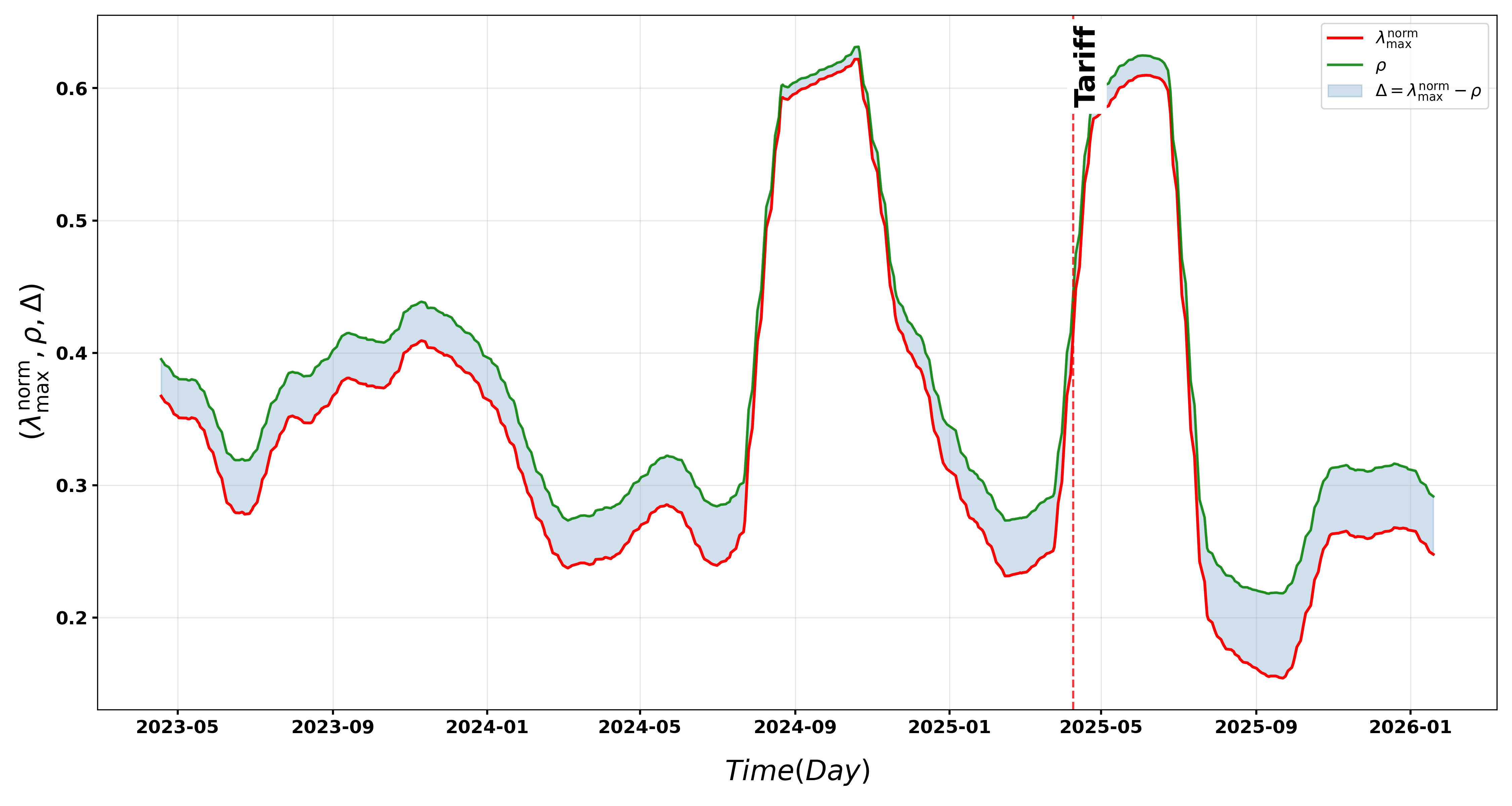}
    }

    \vspace{0.1 cm}

    \subfigure[India\label{fig:sector_IT_india}]{
        \includegraphics[width=0.45\linewidth]{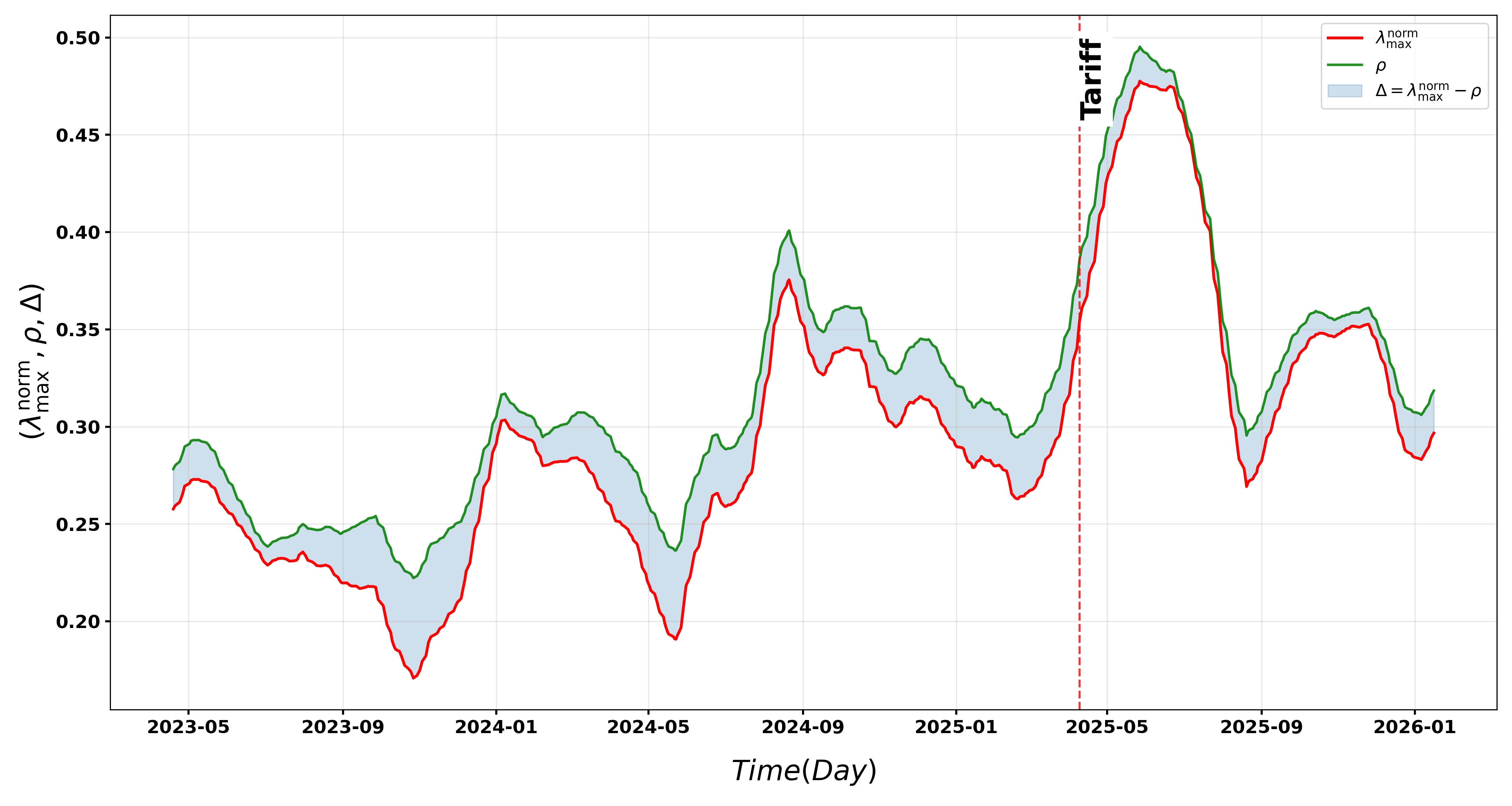}
    }
    \hspace{0.01\linewidth}
    \subfigure[ Germany\label{fig:sector_IT_germany}]{
        \includegraphics[width=0.45\linewidth]{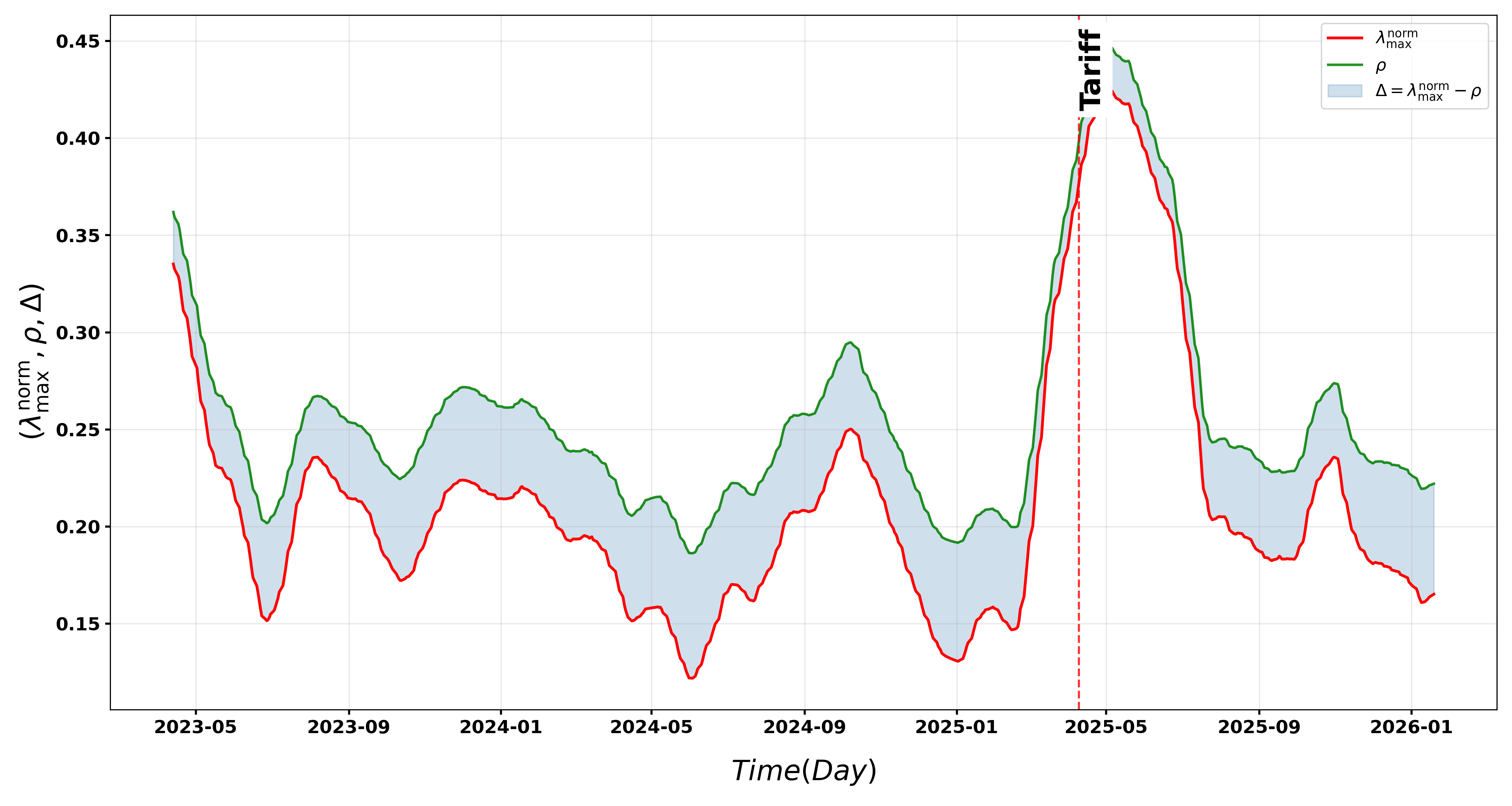}
    }

    \vspace{0.1cm}

    \subfigure[United States\label{fig:sector_IT_us}]{
        \includegraphics[width=0.80\linewidth]{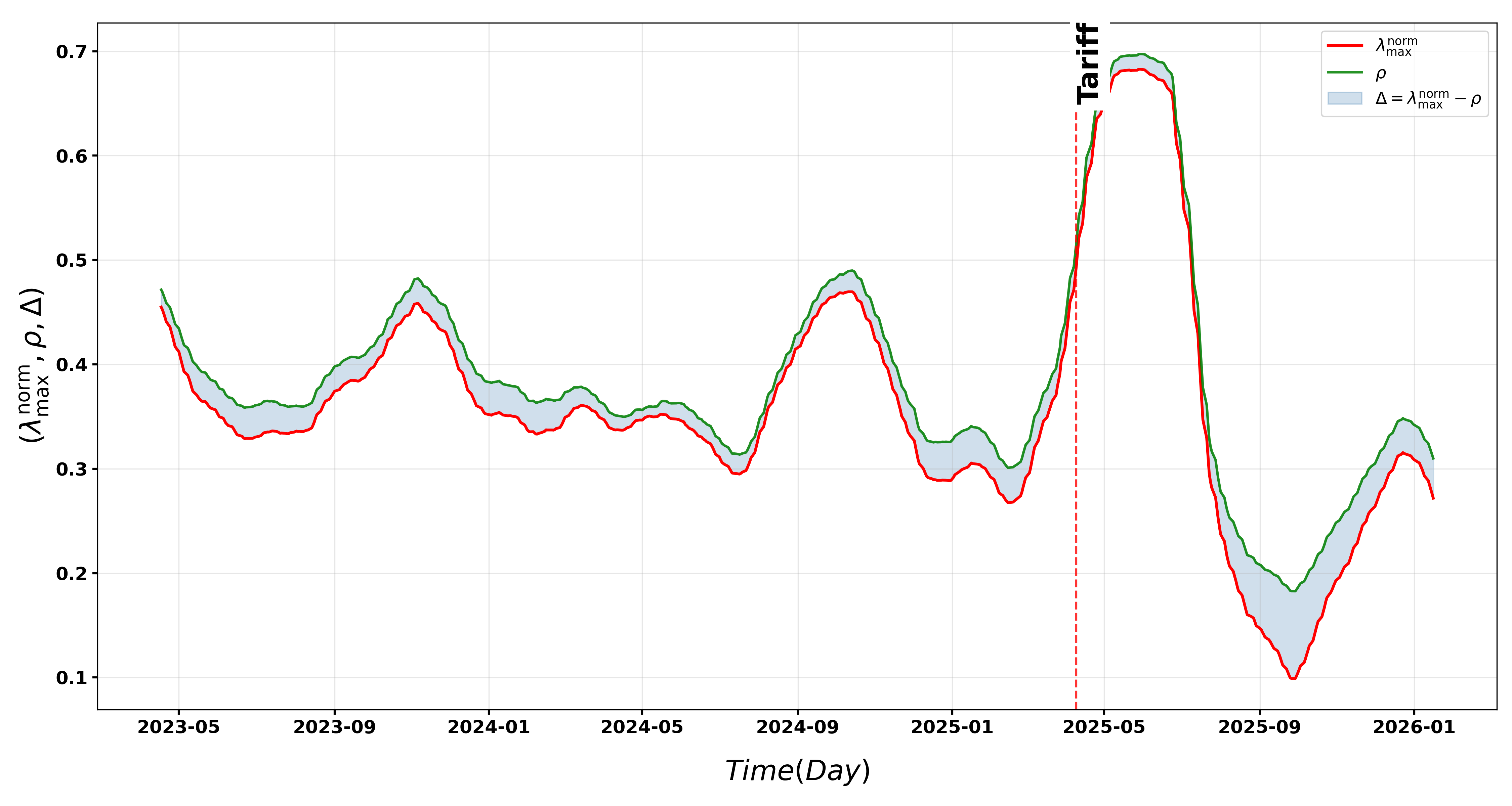}
    }

    \caption{Time evolution of RMT-based complexity metrics for G5 stock markets for the IT sector. The blue line represents the normalized largest eigenvalue $\lambda_{\max}^{\text{norm}}(t)$, the orange line shows the raw average correlation $\rho(t)$, and the shaded region between them denotes the complexity gap $\Delta(t)$. The vertical red dashed line marks the U.S. tariff announcement in April 2025. Across all markets, we observe a consistent three-phase pattern: a pre-event state with a persistent gap $\Delta(t) < 0$, a sharp convergence with $\Delta(t) \approx 0$ immediately following the shock, and a post-event recovery pattern characterized by gap re-widening, a secondary convergence, and finally sustained structural recovery.}
    \label{fig:sector_IT}
\end{figure}

\subsubsection{Sector-stock Complexity Gap}
\label{Sector-stock Complexity Gap}

Fig.~\ref{fig:sector_IT} shows the time evolution of $\lambda_{\max}^{\text{norm}}(t)$, $\rho(t)$, and $\Delta(t)$ for the Information Technology(IT) across the G5 countries. The sectoral analysis reveals a clear structural difference compared to the market-level results. At the market level, the complexity gap $\Delta(t) > 0$, indicating that $\lambda_{\max}^{\text{norm}}(t)$ consistently exceeds $\rho(t)$ throughout the period. However, at the sectoral level, this relationship is reversed with $\rho(t)$ exceeding $\lambda_{\max}^{\text{norm}}(t)$, resulting in a negative complexity gap  $\Delta(t) < 0$. This inversion in $\Delta(t)$, reflects the intrinsic homogeneity of stocks within the same sector. Within a single sector, stocks belong to the same industry and are subject to similar economic forces, making them inherently more correlated with one another than stocks drawn from the full market. As a result, the $\rho(t)$ within a sector is higher, reflecting the underlying homogeneity of intra-sector stock movements. The negative gap ($\Delta(t) < 0$), therefore, indicates that, within a sector, stock dynamics are already largely governed by a common sectoral factor even under normal market conditions.

Despite this structural difference, the three-phase response to the shock remains clearly visible at the sectoral level. Before the shock, the negative gap $\Delta(t) < 0$ persists in a relatively stable manner, reflecting the baseline intra-sector homogeneity. After the tariff shock, both $\lambda_{\max}^{\text{norm}}(t)$ and $\rho(t)$ rise sharply and converge, with $\Delta(t) \approx 0$, indicating a further loss of internal sectoral structure as all stocks within each sector synchronize under the shock-driven market mode. In the post-event period, the same non-monotonic recovery pattern is observed: an initial re-widening of the gap, followed by a secondary convergence, and finally a sustained recovery toward the pre-event baseline. This confirms that the three-phase complexity response is not limited to the aggregate market level but is also a characteristic feature of individual sector dynamics. Similar three-phase patterns are also observed for the Industrial, Finance, Health, and Consumer sectors, as shown in Figs. ~\ref{fig:sector_Industries}, ~\ref{fig:sector_Fin},  ~\ref{fig:sector_Health}, and ~\ref{fig:sector_Consumer} in Appendix~\ref{App_Secotr}. While the three-phase pattern is broadly preserved at the sectoral level, its strength and clarity vary across sectors. This variation arises because sectors are intrinsically more homogeneous systems, with higher baseline correlations and fewer independent modes, which compresses the observable structural transition and leads to weaker or temporally shifted convergence patterns. To ensure robustness, we repeated the analysis using different rolling window sizes ($L=30$ and $L=90$ days) and mean absolute correlation, $\rho(t)=\langle |C_{ij}| \rangle$. In all cases, the three-phase recovery pattern remained consistent, confirming that the observed dynamics are not sensitive to parameter choices. These findings suggest that although the sign of the complexity gap, $\Delta(t)$, differs between the market and sectoral levels, the three-phase complexity response pattern remains consistent across both scales. The sectoral analysis, therefore, supports the market-level results by showing that the tariff announcement not only disrupted the broad market structure but also induced the same three-phase reorganization within individual sectors.

For investors, these structural patterns provide a clear advantage in understanding post-shock market conditions. Our analysis identifies a distinct false-recovery pattern: immediately after a shock, the complexity gap $\Delta(t)$ may temporarily widen as the initial panic subsides, giving the appearance of a return to normal conditions. However, this regime is often misleading. Our results show that this early stabilization is followed by a secondary down-merge, in which $\lambda_{\max}^{\text{norm}}(t)$ and $\rho(t)$ re-synchronize and the gap $\Delta(t) \approx 0$ merges again. Therefore, investors should view the first re-emergence of $\Delta(t) > 0$ with caution. A premature re-entry during this phase may expose portfolios to the secondary synchronization risk. The robust signal for re-entry may appear only when the gap widens in a sustained manner, accompanied by both metrics stabilizing at lower levels. By recognizing this three-stage process, investors can distinguish false signals from genuine stability, avoiding the trap of premature allocation and waiting for confirmed structural restoration before rebuilding risk positions.

\subsection{Sectoral Collective Mode Strength: Heatmap Analysis}
\label{Heatmap}

\begin{figure}[htbp]
    \centering

    \subfigure[ China\label{fig:Heat_china}]{
        \includegraphics[width=0.45\linewidth]{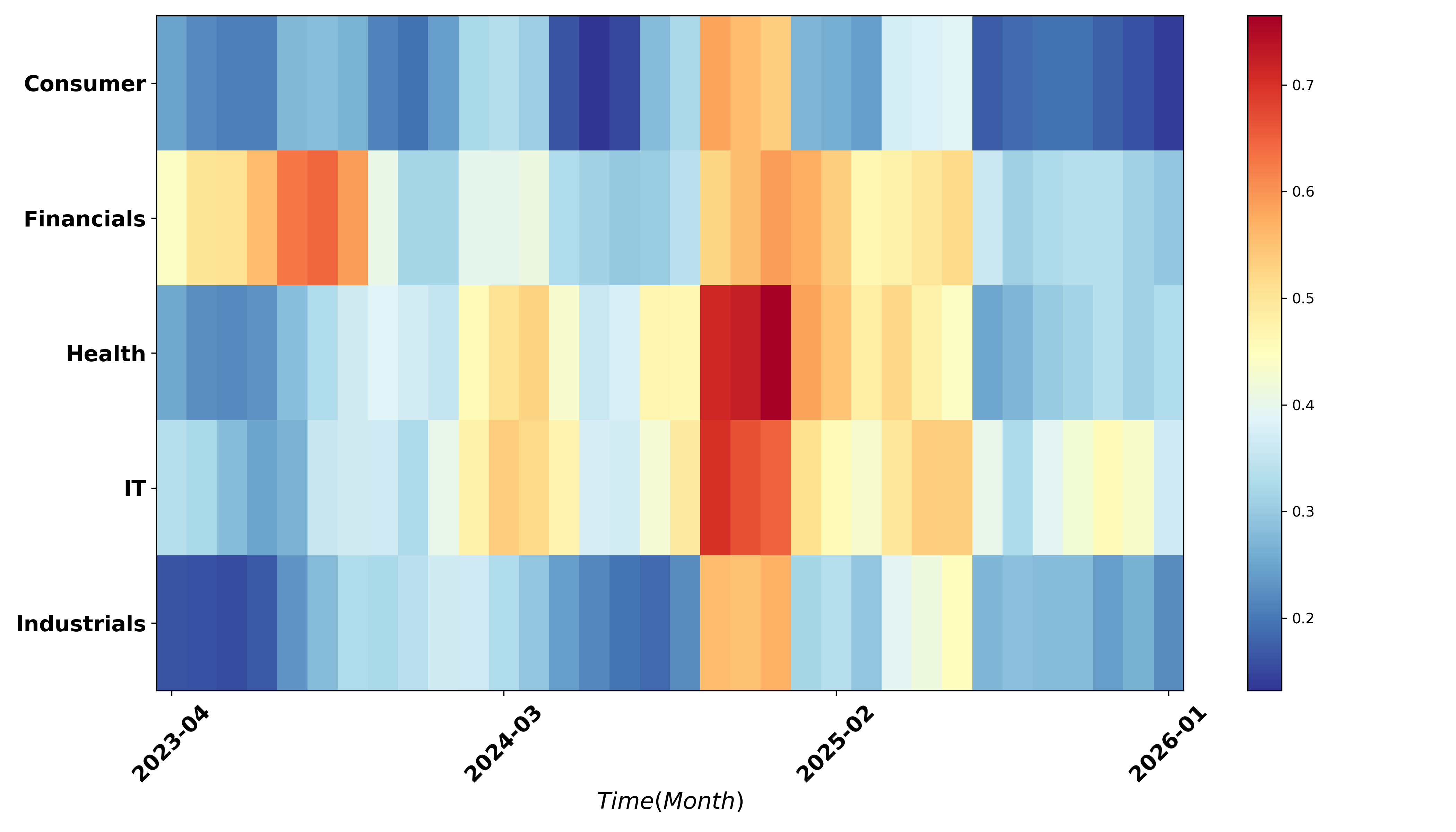}
    }
    \hspace{0.01\linewidth}
    \subfigure[ Japan\label{fig:Heat_japan}]{
        \includegraphics[width=0.45\linewidth]{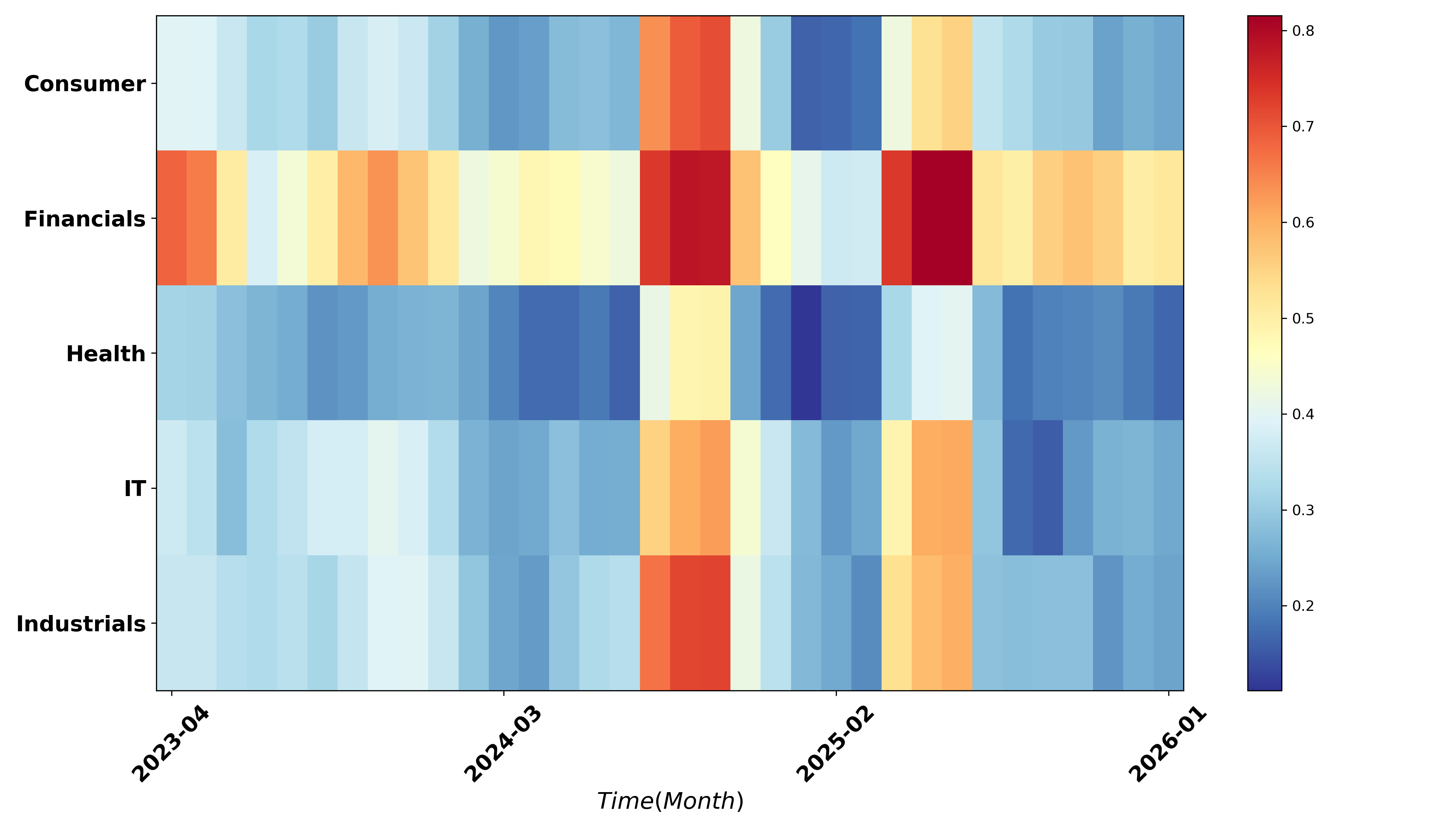}
    }

    \vspace{0.1 cm}

    \subfigure[India\label{fig:Heat_india}]{
        \includegraphics[width=0.45\linewidth]{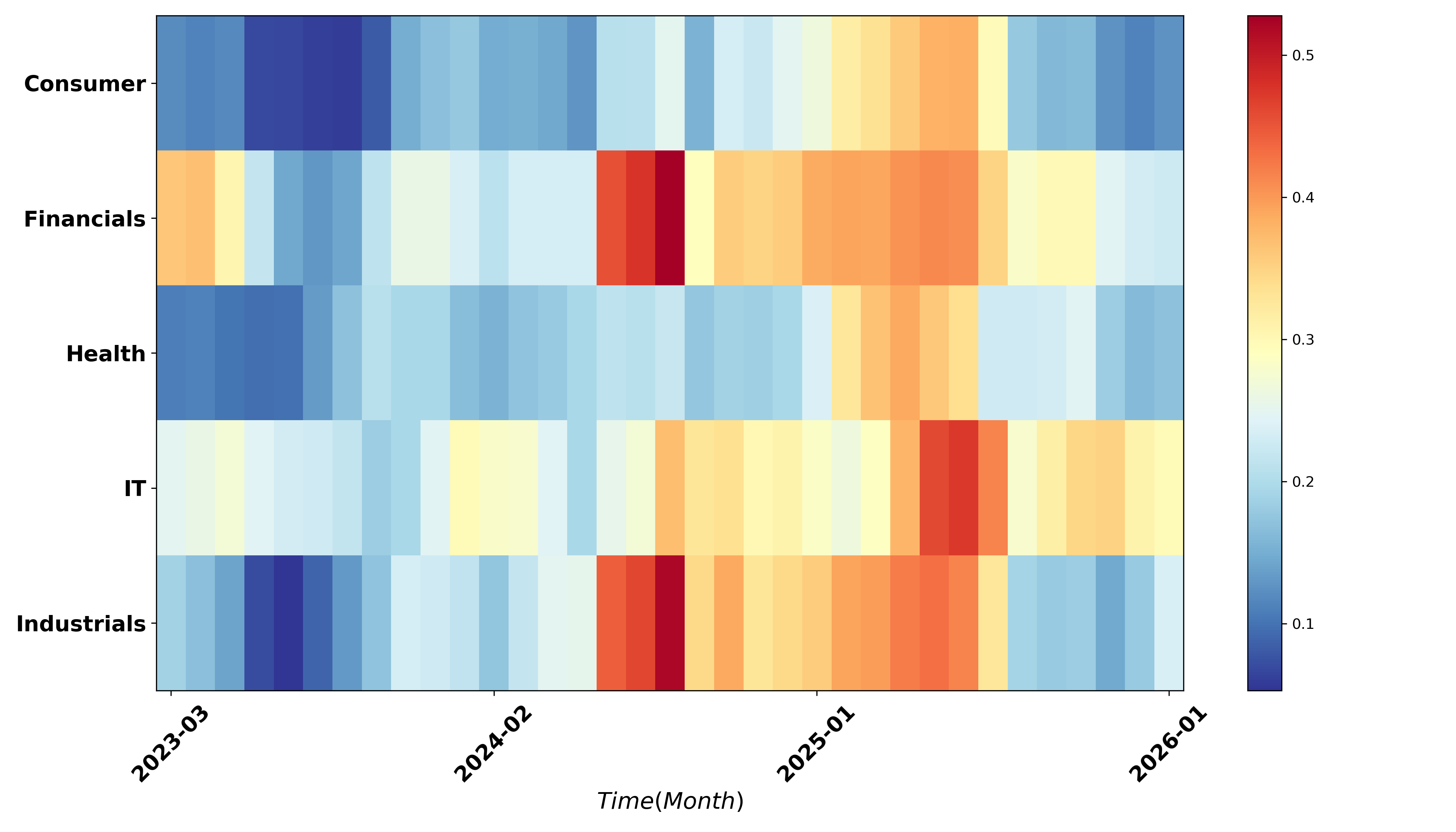}
    }
    \hspace{0.01\linewidth}
    \subfigure[ Germany\label{fig:Heat_germany}]{
        \includegraphics[width=0.45\linewidth]{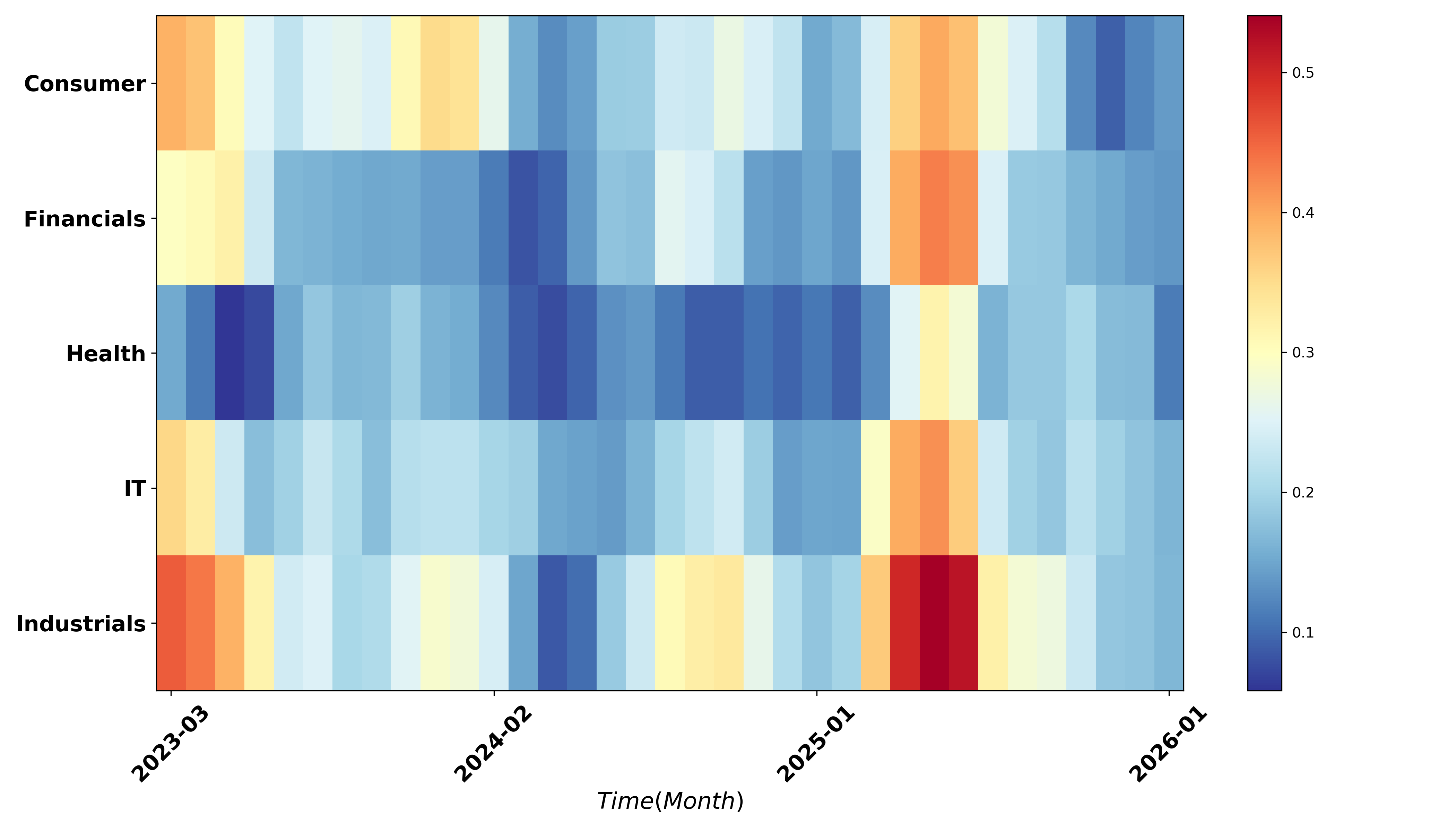}
    }

    \vspace{0.1cm}

    \subfigure[United States\label{fig:Heat__us}]{
        \includegraphics[width=0.80\linewidth]{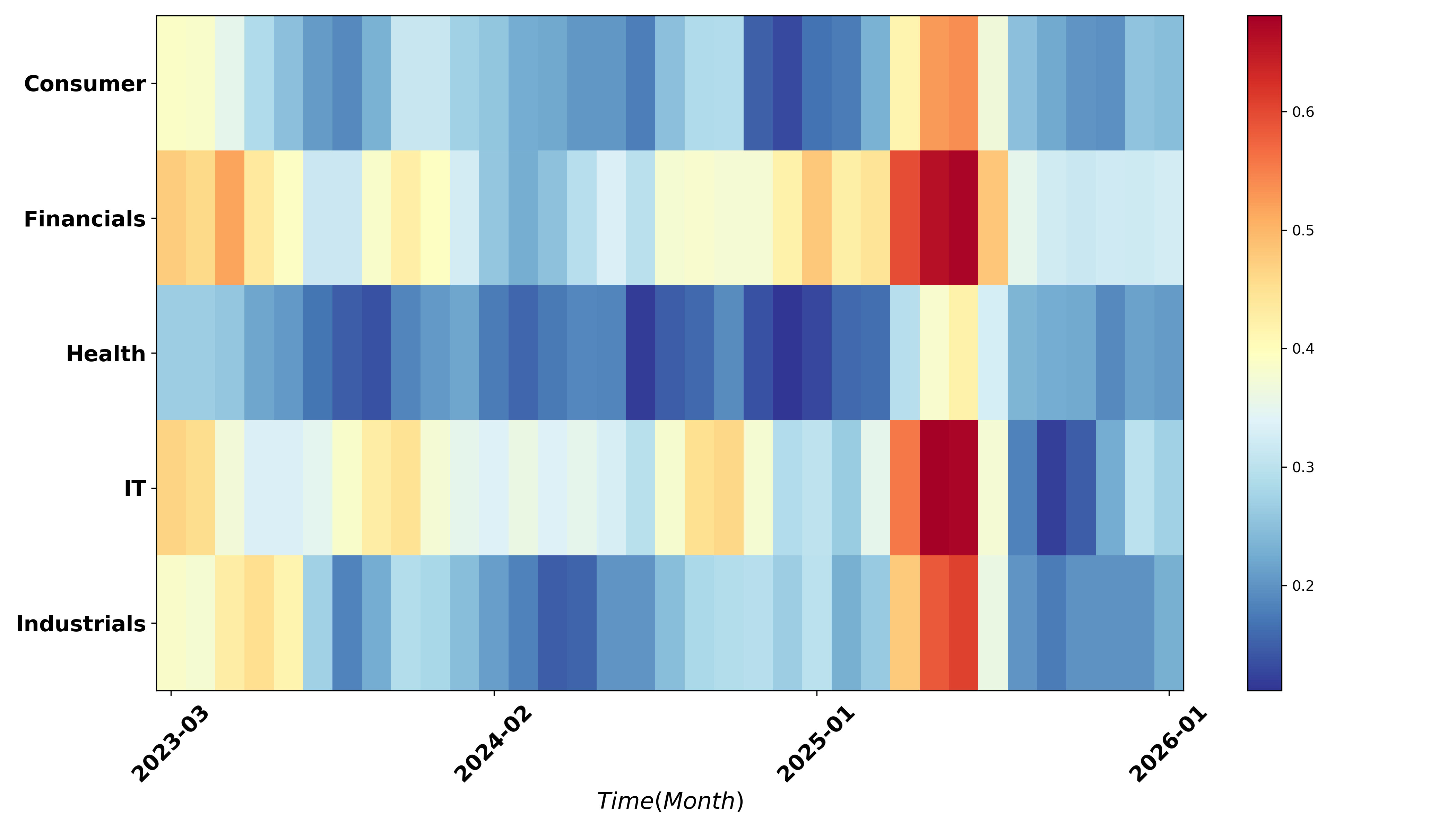}
    }

    \caption{Monthly evolution of sectoral collective mode strength measured by the normalized largest eigenvalue $\lambda_{\max}^{\mathrm{norm}}(t)$ across five sectors and five G5 markets. Warmer colors indicate stronger synchronization. The U.S. tariff announcement in April 2025 triggers a simultaneous, cross-sectoral surge in collective dynamics across all markets, confirming its nature as a global exogenous shock, in contrast to the localized responses observed during the 2024 domestic events in Japan and China.}
\label{fig:Heatmap}
\end{figure}

To further examine how shocks propagate across sectors and countries, we visualize the temporal evolution of sectoral collective dynamics using heatmaps for the G5 markets. Figure~\ref{fig:Heatmap} shows the monthly evolution of sectoral collective mode strength across the stock sectors, namely Consumer, Financial, Healthcare, Information Technology(IT), and Industrial sectors for China, Japan, India, Germany, and the United States. The heatmaps are constructed using the monthly average of the normalized largest eigenvalue, $\lambda_{\max}^{\mathrm{norm}}(t)$, obtained from rolling correlation matrices within each sector. This quantity measures the strength of the dominant collective market mode. Higher values of $\lambda_{\max}^{\mathrm{norm}}(t)$ indicate stronger synchronization among stocks within the sector and therefore stronger collective market behavior. In the heatmaps, rows correspond to each sector and columns represent monthly time intervals, while the color scale reflects the magnitude of collective synchronization. Warmer colors (yellow–red) indicate stronger collective market behavior and higher synchronization among stocks, whereas cooler colors (blue) represent weaker synchronization and more heterogeneous stock movements.

During normal market conditions, the heatmaps exhibit relatively dispersed patterns, with moderate variations across sectors. This indicates that stock movements are influenced by heterogeneous sector-specific dynamics rather than a single dominant market mode. From Fig.~\ref{fig:Heatmap}, we observe localized red patches corresponding to country-specific domestic shocks during 2024. The Japanese market shows a pronounced increase in collective mode strength during August 2024, coinciding with the market disruption triggered by the Bank of Japan’s unexpected policy adjustment. Similarly, the Chinese market exhibits a clear spike in October 2024 associated with the volatility following the stimulus-driven rally and subsequent disappointment regarding fiscal policy announcements. These events appear in the heatmaps as temporally localized regions of strong collective synchronization confined to the affected country, while other markets remain relatively stable during the same period. During the period of the U.S. tariff announcement in April 2025, a different pattern emerges. Around this period, the heatmaps for all G5 stock markets simultaneously display large regions of elevated collective mode strength across multiple sectors. Unlike the localized domestic shocks observed earlier, this tariff event produces broad and synchronized increases in collective dynamics across all countries and sectors. The simultaneous emergence of high-intensity regions across geographically and economically diverse markets indicates a global synchronization of sectoral dynamics. The cross-sectoral nature of these high-intensity regions suggests that the tariff impact was not limited to a single industry but propagated across the full market structure. Financials, IT, Industrials, Consumer, and Healthcare sectors across all countries simultaneously exhibit elevated collective mode strength. This synchronized sectoral response aligns with the RMT results presented earlier, where the complexity gap $\Delta(t)$ collapsed across all markets, indicating a temporary loss of structural differentiation and the dominance of a single global market mode. Together, the heatmap analysis reinforces the conclusions derived from the complexity-gap dynamics. While domestic shocks generate localized and country-specific disturbances, the U.S. tariff announcement produces simultaneous synchronization across multiple sectors and markets, confirming its role as a global exogenous event. The consistency between the RMT-based metrics and the sectoral heatmap visualization provides robust evidence that the tariff shock triggered a coordinated global restructuring of financial market dynamics.

\subsection{Ordinal Entropy During the Tariff Shock}
\label{Ordinal Entropy}

The Complexity Gap $\Delta(t)$ captures structural changes at the spectral level. To assess whether the observed collapse and recovery are also reflected in the directional dynamics of individual stocks, we examine the Ordinal Entropy, $H_{\mathrm{ord}}(t)$, which measures the diversity of three-day ordinal patterns across all stocks. A value near $\log 6 \approx 1.79$ nats indicates a rich mixture of directional return patterns, whereas a sharp decline suggests that most stocks follow a common trajectory. We define four analysis windows to isolate the market response. The pre-shock reference period spans January~2 to March~31, 2025, representing three months of relatively stable conditions prior to the tariff announcement. The shock window is defined as April~7--11, 2025, corresponding to $\pm2$ trading days around the announcement. The false recovery phase extends from April~14 until the last day before $H_{\mathrm{ord}}(t)$ remains continuously above $1.0$ nats for at least 20 consecutive trading days. The stabilized period is defined as June~16--July~15, 2025. Table~\ref{tab:ordinal_results} presents the mean entropy for the pre-shock and stabilized intervals, the mean over the shock window, and the 95th percentile observed during the recovery phase.

\begin{table}[htbp]
\centering
\caption{Ordinal entropy statistics across G5 markets around the 2025 U.S. tariff shock. Pre‑shock and post‑shock values are means $\pm$ standard deviation over the indicated intervals. The shock window is April~7--11, 2025 ($\pm2$ trading days). The false recovery 95th percentile (95th \%ile)  is taken over the false recovery phase. Entropy values are in nats. The sharp drop during the shock window and the subsequent false recovery confirm the three‑phase pattern observed in the Complexity Gap.}
\label{tab:ordinal_results}
\begin{tabular}{l c c c c}
\hline
Country & Pre‑Shock Mean & Shock Window Mean & False Recovery & Post‑Shock Mean \\
 &  &  & 95th \%ile &  \\
\hline
United States & $1.50 \pm 0.12$ & $0.89 \pm 0.21$ & $1.72$ & $1.60 \pm 0.08$ \\
China & $1.53 \pm 0.14$ & $1.10 \pm 0.48$ & $1.67$ & $1.62 \pm 0.09$ \\
Japan & $1.48 \pm 0.11$ & $0.64 \pm 0.29$ & $1.68$ & $1.55 \pm 0.10$ \\
India & $1.49 \pm 0.13$ & $0.95 \pm 0.19$ & $1.63$ & $1.58 \pm 0.08$ \\
Germany & $1.47 \pm 0.10$ & $0.86 \pm 0.17$ & $1.74$ & $1.55 \pm 0.09$ \\
\hline
\end{tabular}
\end{table}

Table~\ref{tab:ordinal_results} represents the mean $H_{\mathrm{ord}}(t)$ across the four analysis windows. Before the shock, $H_{\mathrm{ord}}$ averages between $1.47$ and $1.53$ nats across all five markets, establishing a baseline of healthy directional diversity. During the April~7-11 shock window, this diversity collapses sharply, to $0.64$ nats in Japan and $0.86$ nats in Germany, as stocks synchronize onto a narrow set of ordinal sequences. China's shock‑window average is higher ($1.10 \pm 0.48$ nats), as the initial drop was followed by a rapid rebound within the same week. This may reflect the fact that Chinese equities were already under pressure from ongoing U.S.–China trade tensions and domestic policy adjustments, potentially reducing the incremental impact of the tariff announcement. In the false recovery phase, entropy rebounds to a 95th percentile of $1.63$-$1.74$ nats, approaching pre-shock levels. However, this recovery is short-lived: in the United States, entropy subsequently dips below $1.0$ nats, while Japan, India, and Germany exhibit notable secondary declines of $0.2$ to $0.4$ nats before finally stabilizing. China is the exception; after its rapid spike, entropy remains elevated without a secondary collapse. By the stabilized period, $H_{\mathrm{ord}}$ settles between $1.55$ and $1.65$ nats across all markets, with standard deviations notably smaller than before the shock, indicating a new, more tightly bounded equilibrium of directional behavior. The three-phase pattern in $H_{\mathrm{ord}}(t)$ mirrors the dynamics of the Complexity Gap $\Delta(t)$, confirming that the structural collapse, false recovery, and stabilization observed at the spectral level are accompanied by a corresponding loss and temporary restoration of directional diversity. This convergent evidence reinforces the interpretation of the tariff announcement as a genuine, multi-faceted shock to global equity markets.

\subsection{Portfolio Risk and the Complexity Gap}
\label{Portfolio Risk and the Complexity Gap}

To examine whether the complexity gap $\Delta$ contains information about future portfolio risk, we perform a large-scale portfolio simulation across the G5 markets over the period January to December 2025. This sample period covers three distinct periods: a pre-event period, the tariff shock in April 2025, and the subsequent post-event recovery period. Our approach uses a rolling window design, with 60-day estimation periods followed by 20-day out-of-sample test windows. At each step, we randomly selected 500 portfolios, each containing 10 stocks. For every portfolio, we calculated the complexity gap $\Delta (t) = \lambda_{\max}^{\text{norm}}(t) - \rho(t)$ using data from the estimation window. During each rolling step, we constructed two types of portfolios using the estimation window: minimum-variance portfolios (MVP), based on the estimated covariance matrix, and simple equal-weight portfolios (EW), which serve as a naive benchmark. These portfolio weights were then applied to the subsequent test window, where portfolio risk was measured as the annualized volatility of returns. For comparison, we also computed two traditional predictors from the estimation window: the average correlation $\rho$ and the historical equal-weight volatility $\sigma_{\text{hist}}$. In total, our analysis includes 416,500 portfolio observations. To evaluate the relationship between $\Delta$ and future portfolio risk, we employ three complementary approaches: Spearman rank correlations between $\Delta$ and realized MVP volatility, incremental $R^2$ measures that quantify the additional explanatory power of $\Delta$ beyond each benchmark predictor, and a quintile analysis in which portfolios are sorted into five groups based on $\Delta$ to examine whether higher complexity is systematically associated with lower realized volatility.

\begin{table}[htbp]
\centering
\footnotesize
\caption{This table reports Spearman rank correlations between the complexity gap ($\Delta$) and out-of-sample portfolio volatility for both minimum-variance portfolios (MVP) and equal-weight portfolios (EW). It also presents annualized MVP volatility across $\Delta$ quintiles. The long--short (L/S) spread is defined as the difference in volatility between the highest and lowest $\Delta$ quintiles (Q4 $-$ Q0). All volatilities are annualized (\%). In total, the analysis includes 416,500 portfolio observations.}
\label{tab:main_results}

\begin{tabular*}{0.75\textwidth}{@{\extracolsep{\fill}} l cc c ccccc }
\toprule
& \multicolumn{2}{c}{\textbf{Spearman Corr.}} 
& \textbf{L/S}
& \multicolumn{5}{c}{\textbf{MVP Volatility by $\Delta$ Quintile (\%)}} \\
\cmidrule(lr){2-3}\cmidrule(lr){4-4}\cmidrule(l){5-9}
\textbf{Country} 
& $\rho(\Delta, \sigma_{\text{MVP}})$ 
& $\rho(\Delta, \sigma_{\text{EW}})$
& Spread (\%)
& Q0 & Q1 & Q2 & Q3 & Q4 \\
\midrule
\textbf{US}      & $-0.176^{***}$ & $-0.046^{***}$ & $-2.033$ & 16.98 & 16.28 & 15.73 & 15.43 & 14.95 \\
\textbf{China}   & $-0.150^{***}$ & $+0.191^{***}$ & $-1.701$ & 13.33 & 12.78 & 12.32 & 11.81 & 11.63 \\
\textbf{India}   & $-0.160^{***}$ & $-0.074^{***}$ & $-2.210$ & 14.85 & 14.36 & 13.83 & 13.36 & 12.64 \\
\textbf{Japan}   & $-0.235^{***}$ & $+0.077^{***}$ & $-4.465$ & 20.18 & 17.86 & 16.96 & 16.26 & 15.71 \\
\textbf{Germany} & $-0.144^{***}$ & $-0.035^{***}$ & $-2.056$ & 17.54 & 16.55 & 15.94 & 15.85 & 15.49 \\
\bottomrule
\multicolumn{9}{l}{\footnotesize $^{***}p<0.001$. Q0: lowest $\Delta$ (complexity collapse); Q4: highest $\Delta$ (structural diversity).}
\end{tabular*}
\end{table}

Table~\ref{tab:main_results} presents the relationship between the complexity gap ($\Delta$) and future portfolio volatility. The Spearman correlations between $\Delta$ and MVP realized volatility are negative and statistically significant across all G5 stock markets, ranging from -0.144 to -0.235. This indicates that lower levels of complexity are associated with higher subsequent portfolio risk. The quintile analysis reinforces this finding in all G5 stock markets. Average portfolio volatility declines monotonically as we move from the lowest $\Delta$ group (Q0) to the highest $\Delta$ group (Q4). Here, Q0 represents the set of portfolios with the lowest values of the complexity gap, corresponding to periods when $\lambda_{\max}^{\text{norm}}(t)$ and $\rho(t)$ converge during market stress. In contrast, Q4 represents portfolios with the highest complexity gap, reflecting periods of normal market conditions with well-developed structural diversity. The annualized volatility difference between the extreme quintiles is reaching up to 4.47 percent in Japan. The correlations between $\Delta$ and equal-weight portfolio volatility vary in both sign and magnitude across markets, indicating that the predictive content of $\Delta$ arises primarily from the correlation structure rather than broad market movements. Table~\ref{tab:robustness} reports results for benchmark predictors and the additional contribution of the complexity gap $\Delta$. Average correlation and historical volatility show a positive association with future MVP volatility, reflecting persistence in market risk. In contrast, $\Delta$ exhibits a negative relationship, indicating that it captures a distinct structural feature of market dynamics. The incremental $R^2$ results show that $\Delta$ adds modest explanatory power beyond these benchmarks, suggesting that it does not operate as a purely linear predictor but varies across market conditions. During the pre-shock period, the relationship between $\Delta$ and future volatility is weak in most markets. Following the tariff shock, however, the relationship turns strongly negative across all markets, with correlations ranging from -0.141 to -0.239. This indicates that $\Delta$ becomes informative during periods of market stress, when the underlying market structure breaks down. During stable periods, when the complexity gap remains steady, its predictive power is limited.

\begin{table}[htbp]
\centering
\footnotesize
\caption{The table reports Spearman rank correlations between benchmark predictors, namely average correlation ($\rho$) and historical equal-weight volatility ($\sigma_h$), and out-of-sample MVP volatility. It also reports the incremental $R^2$ obtained by adding the complexity gap ($\Delta$) to each benchmark predictor. In addition, it presents the Spearman correlation between $\Delta$ and MVP volatility in the pre-shock and post-shock periods surrounding the April 2025 tariff event.}
\label{tab:robustness}

\begin{tabular*}{0.75\textwidth}{@{\extracolsep{\fill}} l cc cc cc}
\toprule
& \multicolumn{2}{c}{\textbf{Benchmark Spearman}}
& \multicolumn{2}{c}{\textbf{Incr.\ $R^2$ of $\Delta$}}
& \multicolumn{2}{c}{\textbf{Subperiod Spearman}} \\
\cmidrule(lr){2-3}\cmidrule(lr){4-5}\cmidrule(l){6-7}
\textbf{Country}
& $\rho(\rho,\,\sigma_{\text{MVP}})$
& $\rho(\sigma_h,\,\sigma_{\text{MVP}})$
& over $\rho$
& over $\sigma_h$
& Pre-shock
& Post-shock \\
\midrule
\textbf{US}      & $+0.522$ & $+0.578$ & $+0.0008$ & $+0.0000$ & $+0.045$ & $-0.206$ \\
\textbf{China}   & $+0.145$ & $+0.079$ & $+0.0043$ & $+0.0118$ & $+0.041$ & $-0.141$ \\
\textbf{India}   & $+0.443$ & $+0.555$ & $+0.0033$ & $+0.0054$ & $-0.072$ & $-0.150$ \\
\textbf{Japan}   & $+0.285$ & $+0.333$ & $+0.0037$ & $+0.0055$ & $+0.001$ & $-0.239$ \\
\textbf{Germany} & $+0.285$ & $+0.334$ & $+0.0024$ & $+0.0028$ & $+0.018$ & $-0.158$ \\
\bottomrule
\end{tabular*}
\end{table}

\section{Conclusion}
\label{Conclusion}

In this study, we introduced the complexity gap $\Delta(t)$, derived from RMT, to uncover a consistent and economically meaningful structural pattern in stock markets during exogenous shock events. We analyze both market-level and sector-level stock dynamics across the G5 economies, namely the United States, China, Germany, India, and Japan, covering shock events, the U.S. tariff announcement of 2025, and the COVID-19 crisis. Using a rolling-window RMT framework, we define the complexity gap as the difference between the normalized largest eigenvalue, $\lambda_{\max}^{\mathrm{norm}}(t)$, and the average correlation, $\rho(t)$. This measure captures whether market dynamics are dominated by a single common factor or by multiple interacting components.

Across all the G5 stock markets, we observe a consistent three-phase structural pattern at both the market and sector levels during these exogenous shock events. Before the shock events, markets exhibit a positive complexity gap $\Delta(t) > 0$, indicating a well-developed internal structure with sector-specific dynamics operating alongside the common market mode. However, during the shock periods, this structure collapses across all G5 markets, as both $\lambda_{\max}^{\mathrm{norm}}(t)$ and $\rho(t)$ increase sharply and converge, leading to a rapid contraction of the complexity gap toward zero, $\Delta(t) \approx 0$. This reflects a state of strong market-wide synchronization in which stock dynamics become dominated by a single global factor, resulting in a loss of structural diversity and an effective reduction in the informational content of the correlation matrix. In this regime, sector-specific and idiosyncratic interactions are suppressed, and market behavior is largely governed by a single collective mode. In the post-shock regime, we observe a non-monotonic recovery pattern characterized by an initial re-widening of the complexity gap, followed by a secondary convergence during which both $\lambda_{\max}^{\mathrm{norm}}(t)$ and $\rho(t)$ decline simultaneously and move closer together, leading to a renewed contraction of $\Delta(t) \approx 0$. Finally, the gap widens again in a sustained manner, restoring $\Delta(t) > 0$. This three-stage recovery reveals intermediate false recovery phases, where the complexity gap initially widens after the shock, giving the appearance of stabilization, but the market becomes synchronized again briefly before fully recovering. At the sectoral level, we observe a structurally distinct behavior in which the complexity gap is negative, $\Delta(t) < 0$, reflecting the inherent homogeneity and stronger correlations within individual sectors. Despite this inversion, the same three-phase structural pattern persists across sectors, demonstrating that the collapse and recovery of market structure is a robust phenomenon operating across both market and sector levels. The same three-phase pattern is observed across both global and country-specific shocks, including the U.S. tariff event, the COVID-19 crisis, and localized shocks in Japan and China, highlighting the robustness of the complexity gap in capturing structural changes across different types of shocks.

The heatmap analysis confirms that the U.S. tariff event acts as a global shock, producing simultaneous synchronization across all G5 markets and sectors, whereas domestic shocks remain localized within individual countries. This finding establishes that the 2025 tariff announcement fundamentally altered the global stock market structure, inducing a coordinated regime shift across geographically and economically diverse markets. Ordinal entropy analysis of cross-sectional return patterns further corroborates these results, revealing a parallel collapse and false restoration of directional diversity that aligns closely with the three-phase complexity gap dynamics. We also analyze portfolio risk using the complexity gap. We find that lower values of $\Delta(t)$ are associated with higher future portfolio volatility, while higher values correspond to more stable market conditions. This indicates that the complexity gap provides useful information about portfolio risk, particularly during periods of market stress. Overall, our results show that the complexity gap can be used to characterize structural dynamics in stock markets. By linking correlation structure, eigenvalue dynamics, and collective synchronization, this framework provides a unified view of how markets respond to both global and local shocks. The consistent emergence of the three-phase pattern across countries, sectors, and different shock events suggests that it represents a fundamental structural signature of financial systems under stress.

For investors, recognizing the false-recovery signal embedded in the initial post-shock gap widening may help avoid premature re-entry during periods when market structure remains fragile. The sustained re-widening of $\Delta$, accompanied by stabilization of both spectral measures at lower levels, appears to be a more reliable indicator of genuine structural recovery.

\clearpage

\appendix

\section{Complexity Gap During the Tariff 2025 Shock: Sector-level Analysis}
\label{App_Secotr}

\begin{figure}[htb!]
    \centering

    \subfigure[ China\label{fig:sector_Industries_china}]{
        \includegraphics[width=0.45\linewidth]{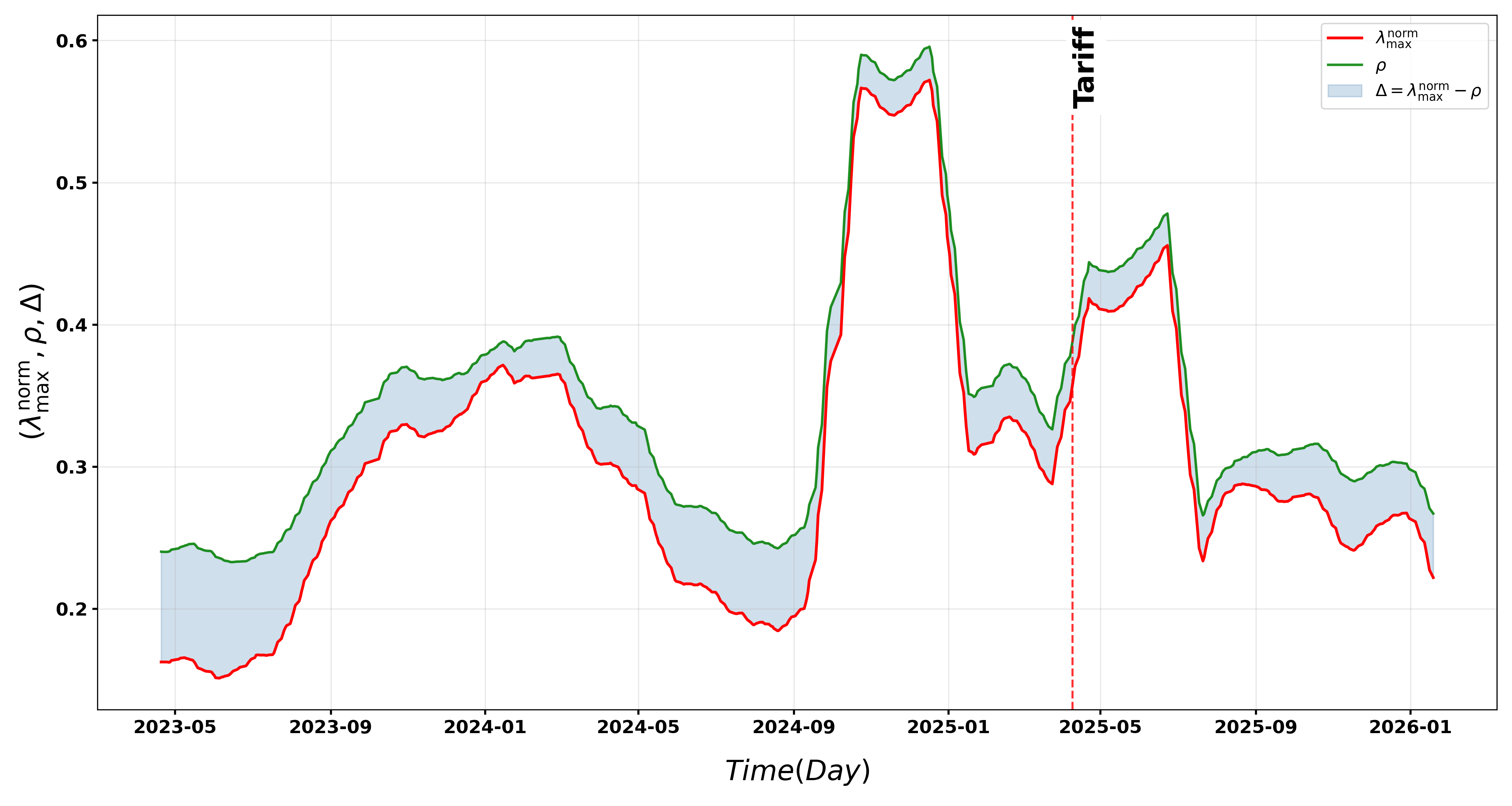}
    }
    \hspace{0.01\linewidth}
    \subfigure[ Japan\label{fig:sector_Industries_japan}]{
        \includegraphics[width=0.45\linewidth]{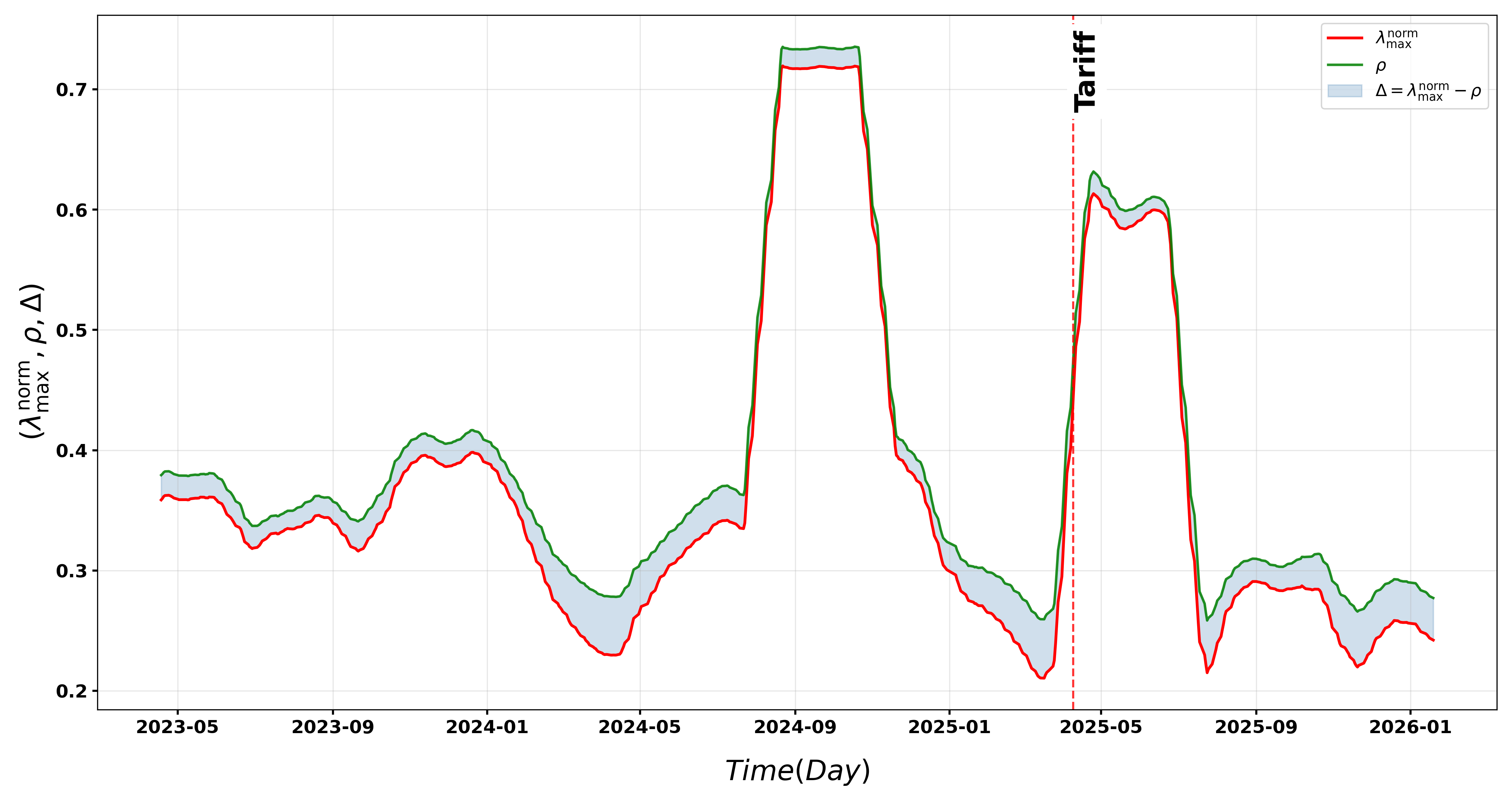}
    }

    \vspace{0.1 cm}

    \subfigure[India\label{fig:sector_Industries_india}]{
        \includegraphics[width=0.45\linewidth]{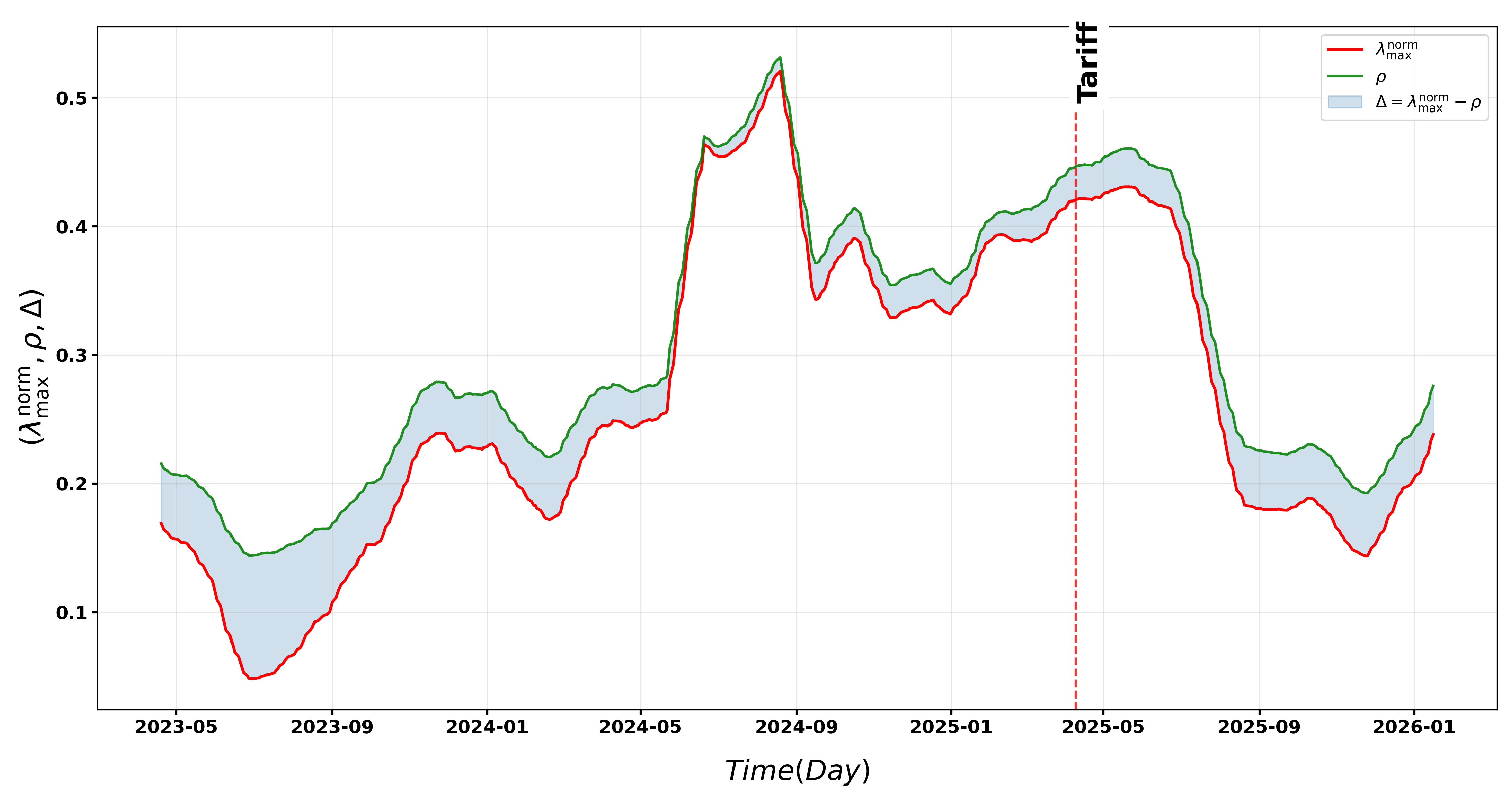}
    }
    \hspace{0.01\linewidth}
    \subfigure[ Germany\label{fig:sector_Industries_germany}]{
        \includegraphics[width=0.45\linewidth]{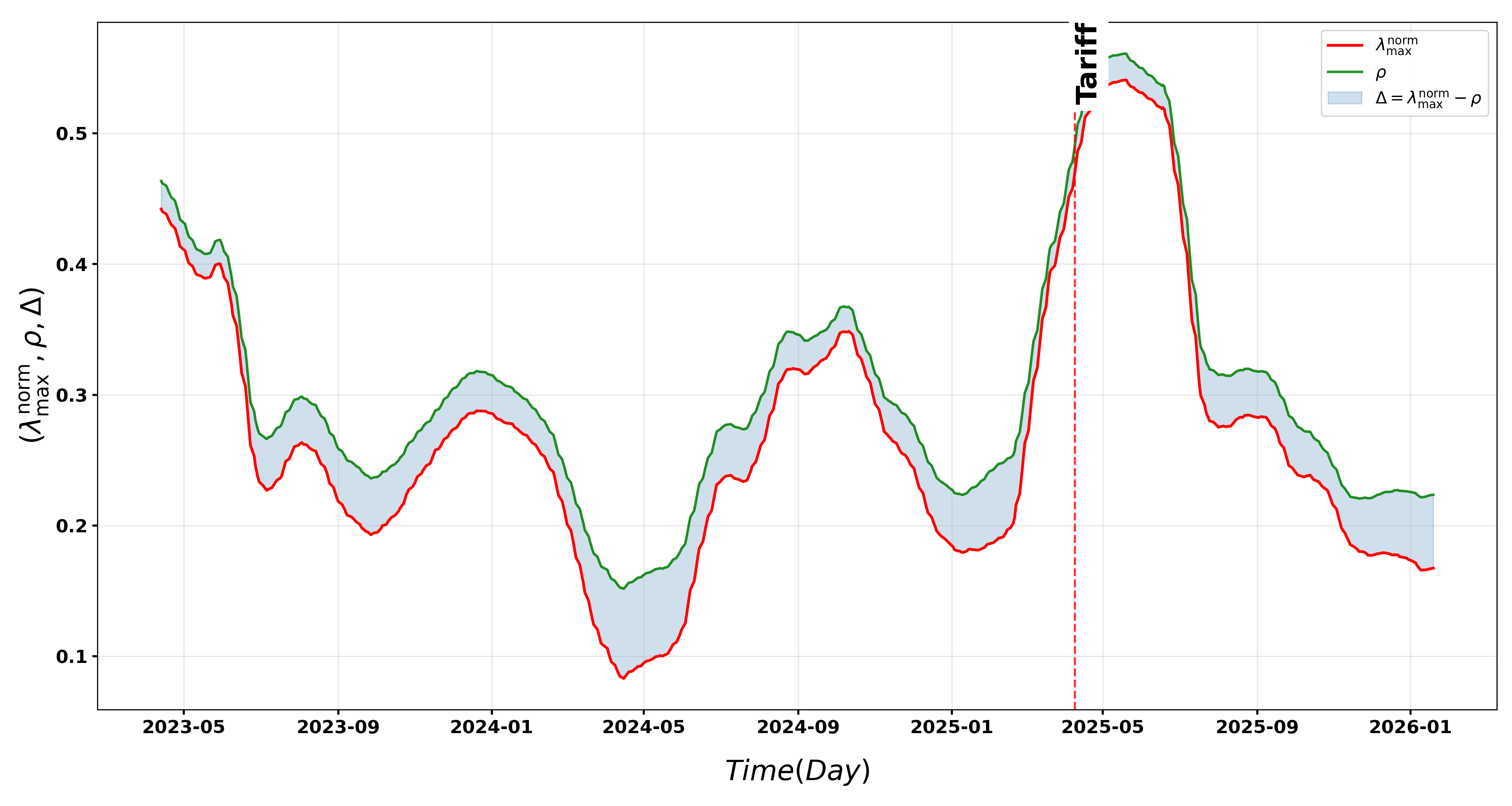}
    }

    \vspace{0.1cm}

    \subfigure[United States\label{fig:sector_Industries_us}]{
        \includegraphics[width=0.80\linewidth]{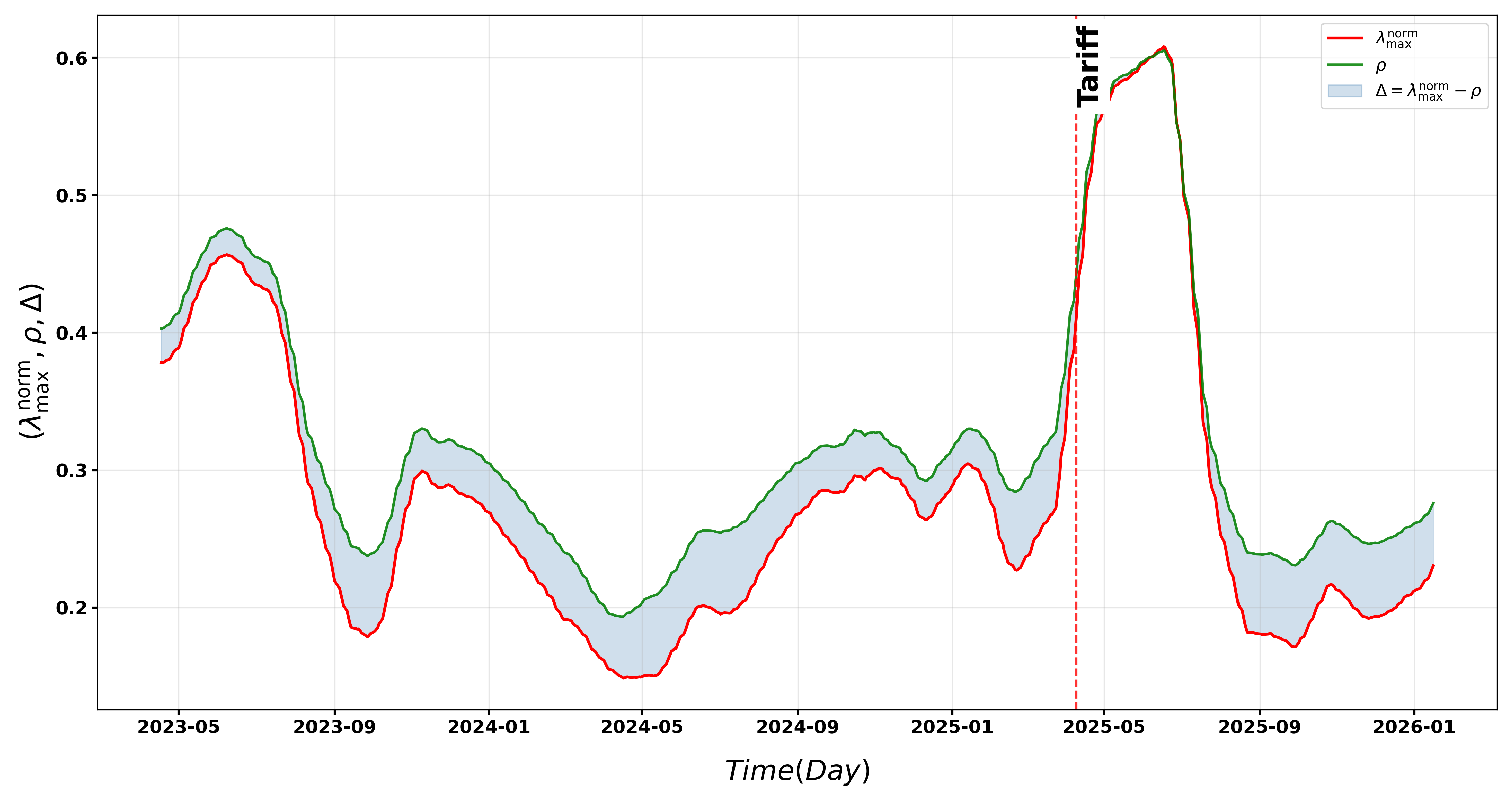}
    }

    \caption{Time evolution of RMT-based complexity metrics for G5 stock markets for the Industrials sector. The red line represents the normalized largest eigenvalue $\lambda_{\max}^{\text{norm}}(t)$, the green line shows the raw average correlation $\rho(t)$, and the shaded region between them denotes the complexity gap $\Delta(t)$. The vertical red dashed line marks the U.S. tariff announcement in April 2025. Across all markets, we observe a consistent three-phase pattern: a pre-event state with a persistent gap $\Delta(t) < 0$, a sharp convergence with $\Delta(t) \approx 0$ immediately following the shock, and a post-event recovery pattern characterized by gap re-widening, a secondary convergence, and finally sustained structural recovery.}
    \label{fig:sector_Industries}
\end{figure}

\clearpage

\begin{figure}[htbp]
    \centering

    \subfigure[ China\label{fig:sector_Fin_china}]{
        \includegraphics[width=0.45\linewidth]{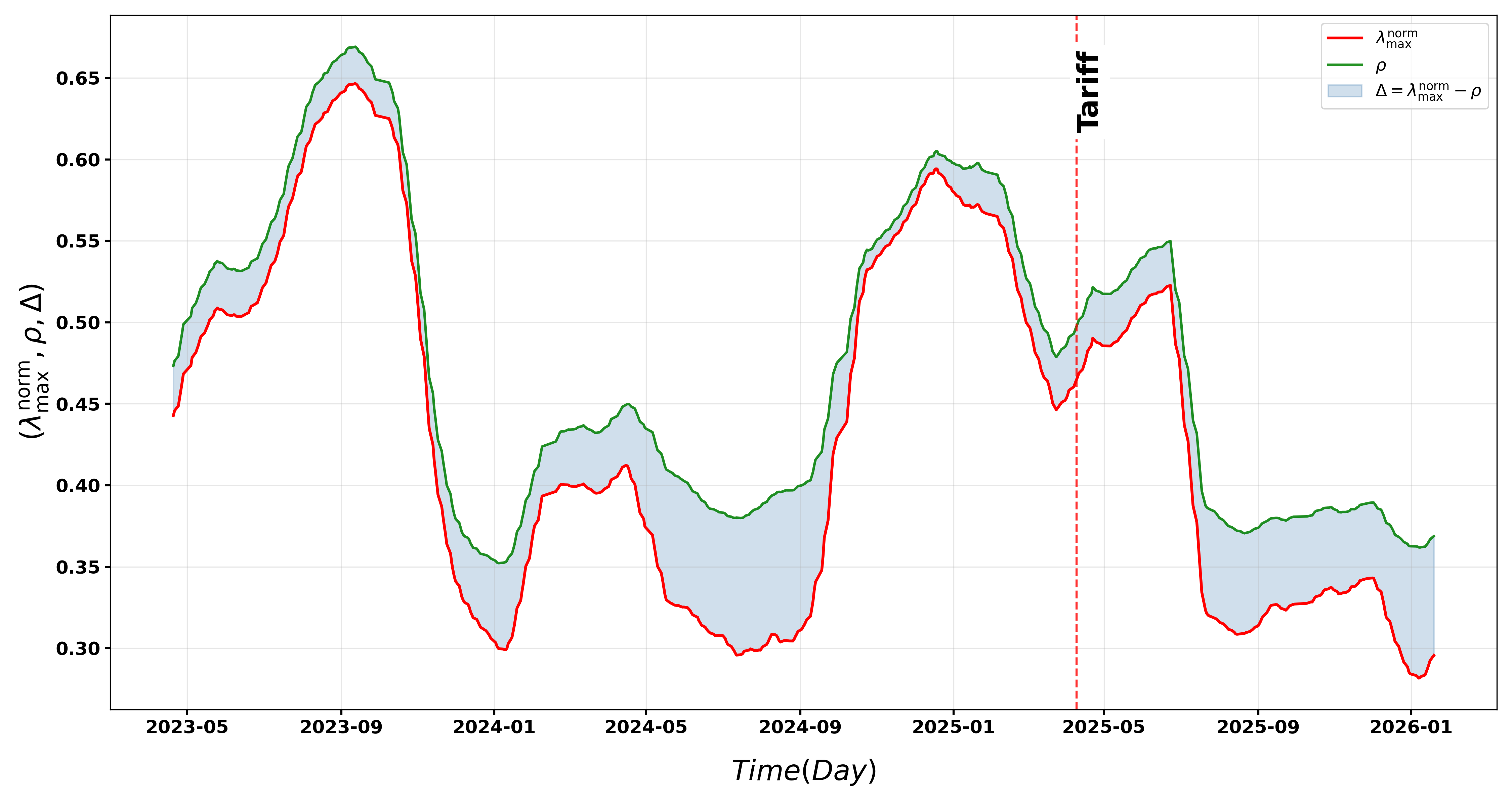}
    }
    \hspace{0.01\linewidth}
    \subfigure[ Japan\label{fig:sector_Fin_japan}]{
        \includegraphics[width=0.45\linewidth]{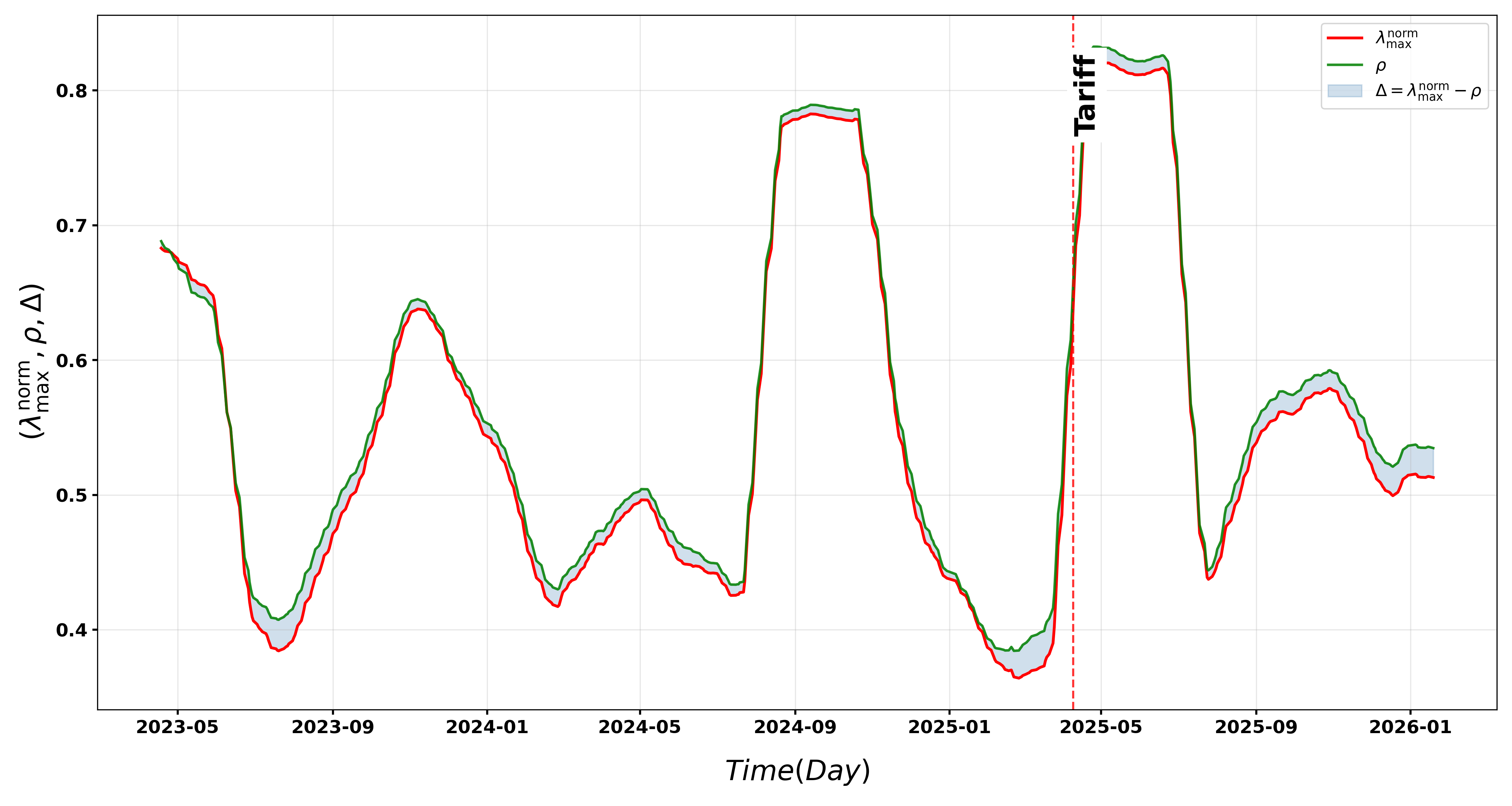}
    }

    \vspace{0.1 cm}

    \subfigure[India\label{fig:sector_Fin_india}]{
        \includegraphics[width=0.45\linewidth]{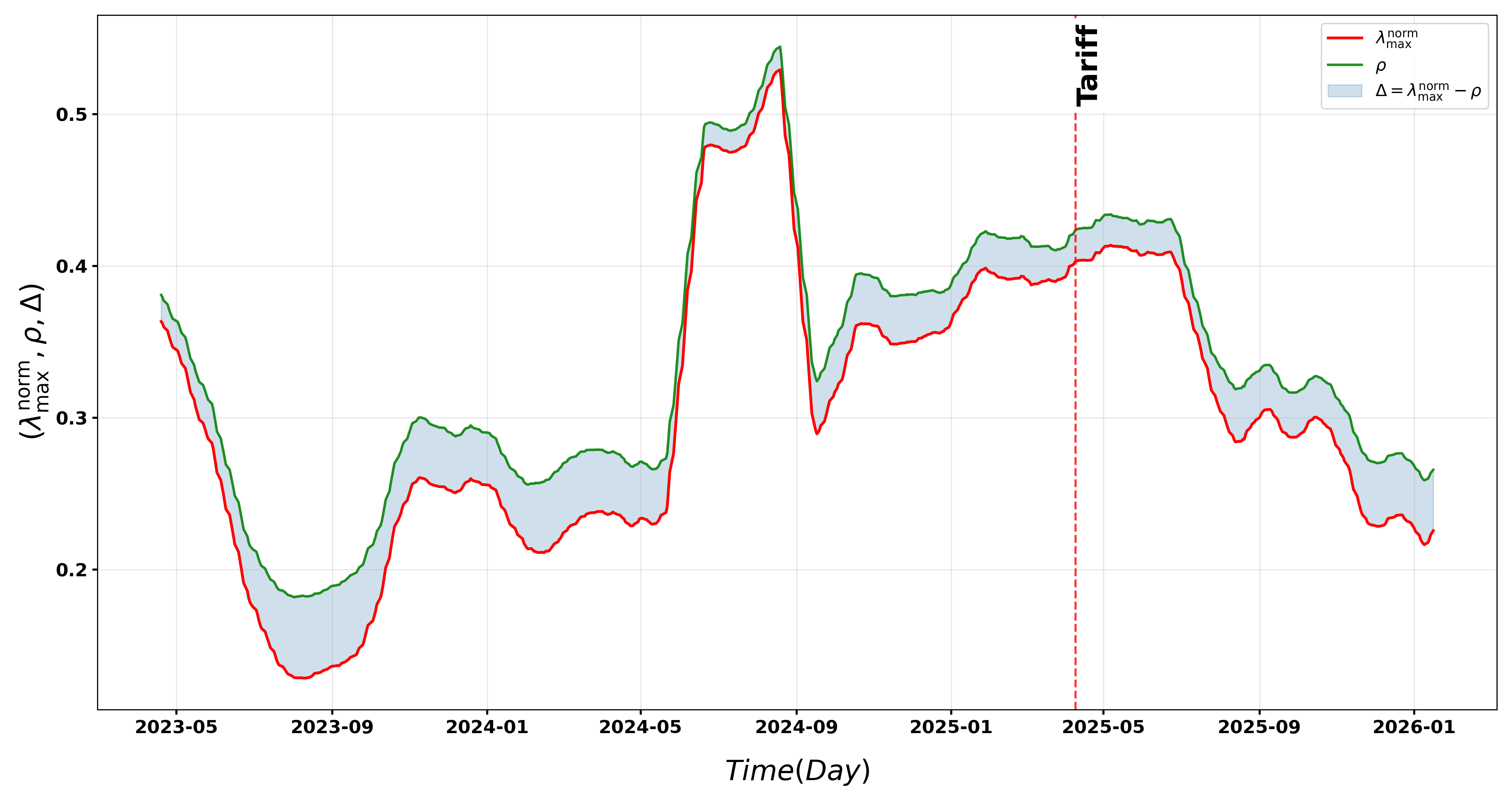}
    }
    \hspace{0.01\linewidth}
    \subfigure[ Germany\label{fig:sector_Fin_germany}]{
        \includegraphics[width=0.45\linewidth]{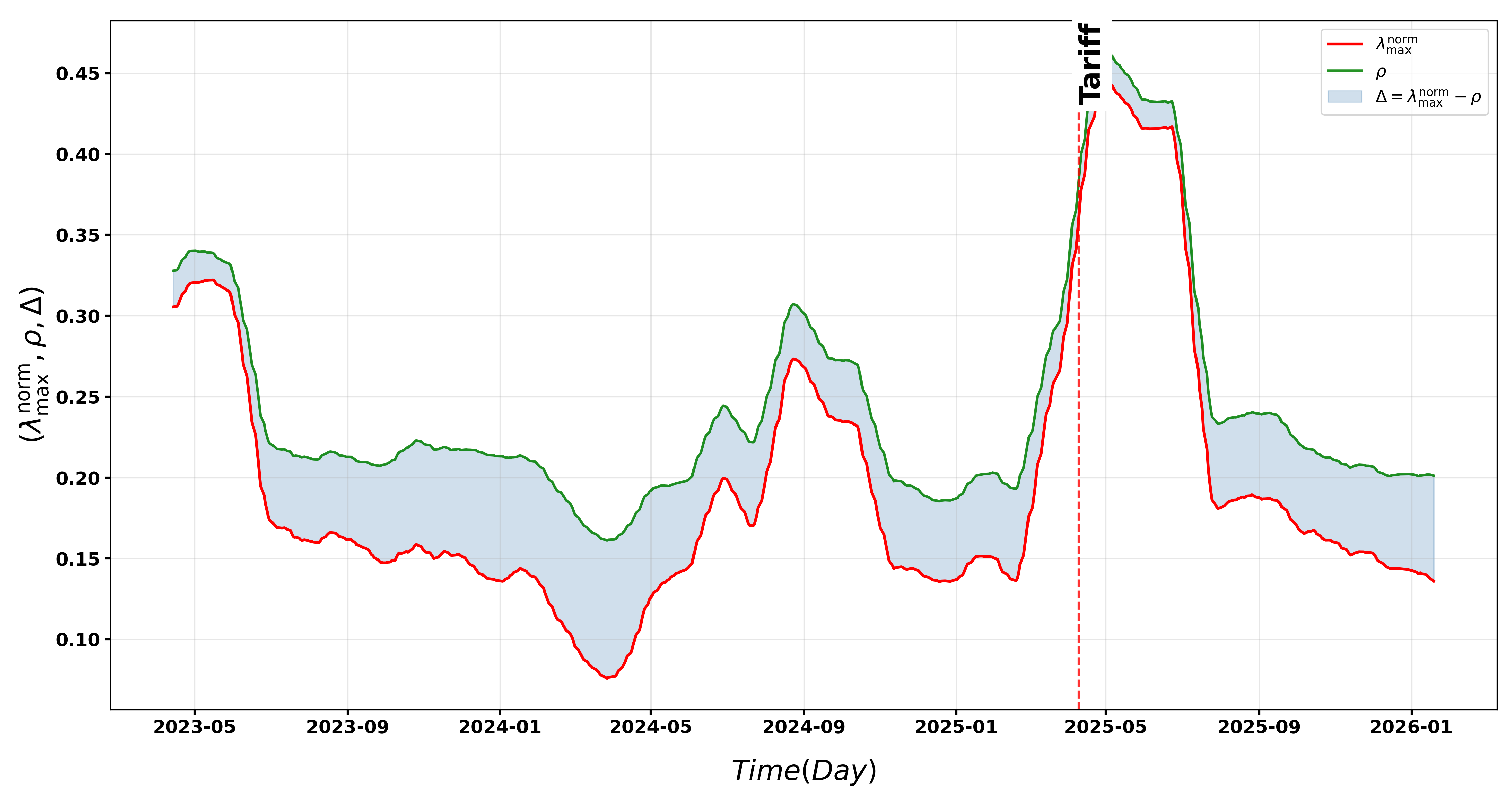}
    }

    \vspace{0.1cm}

    \subfigure[United States\label{fig:sector_Fin_us}]{
        \includegraphics[width=0.80\linewidth]{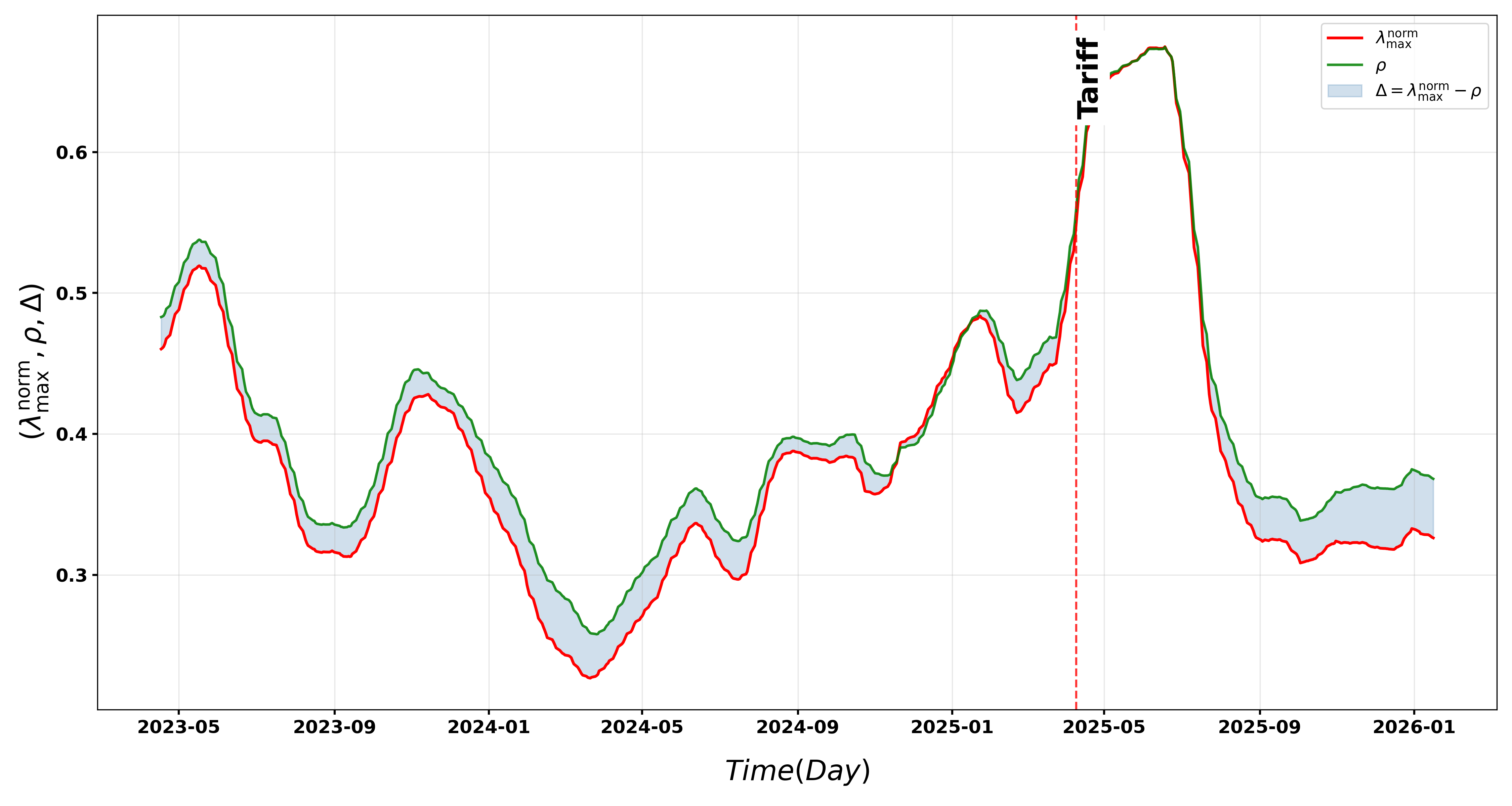}
    }

    \caption{Time evolution of RMT-based complexity metrics for G5 stock markets for the Financial sector. The red line represents the normalized largest eigenvalue $\lambda_{\max}^{\text{norm}}(t)$, the green line shows the raw average correlation $\rho(t)$, and the shaded region between them denotes the complexity gap $\Delta(t)$. The vertical red dashed line marks the U.S. tariff announcement in April 2025. Across all markets, we observe a consistent three-phase pattern: a pre-event state with a persistent gap $\Delta(t) < 0$, a sharp convergence with $\Delta(t) \approx 0$ immediately following the shock, and a post-event recovery pattern characterized by gap re-widening, a secondary convergence, and finally sustained structural recovery.}
    \label{fig:sector_Fin}
\end{figure}

\clearpage

\begin{figure}[htbp]
    \centering

    \subfigure[ China\label{fig:sector_Health_china}]{
        \includegraphics[width=0.45\linewidth]{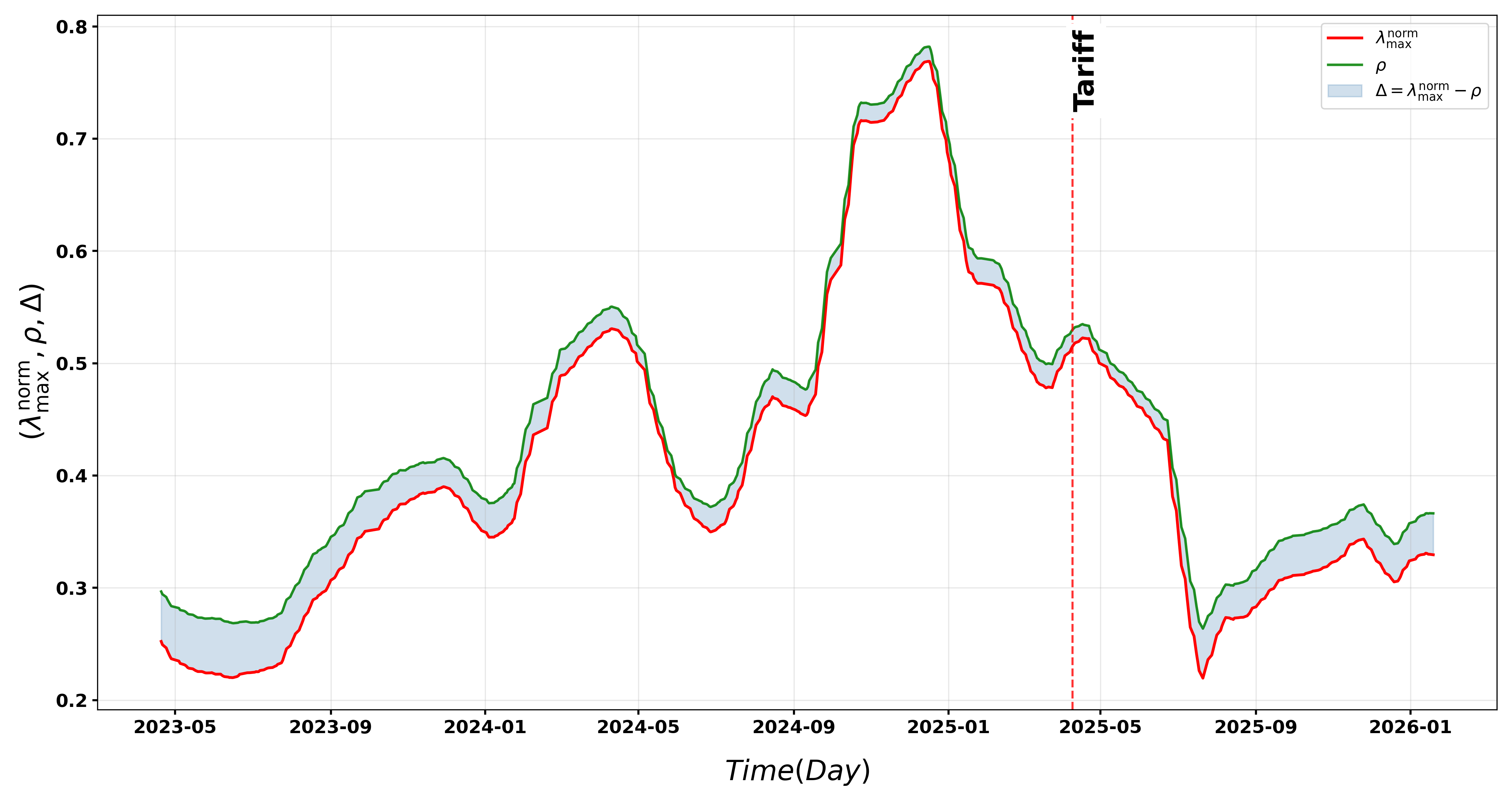}
    }
    \hspace{0.01\linewidth}
    \subfigure[ Japan\label{fig:sector_Health_japan}]{
        \includegraphics[width=0.45\linewidth]{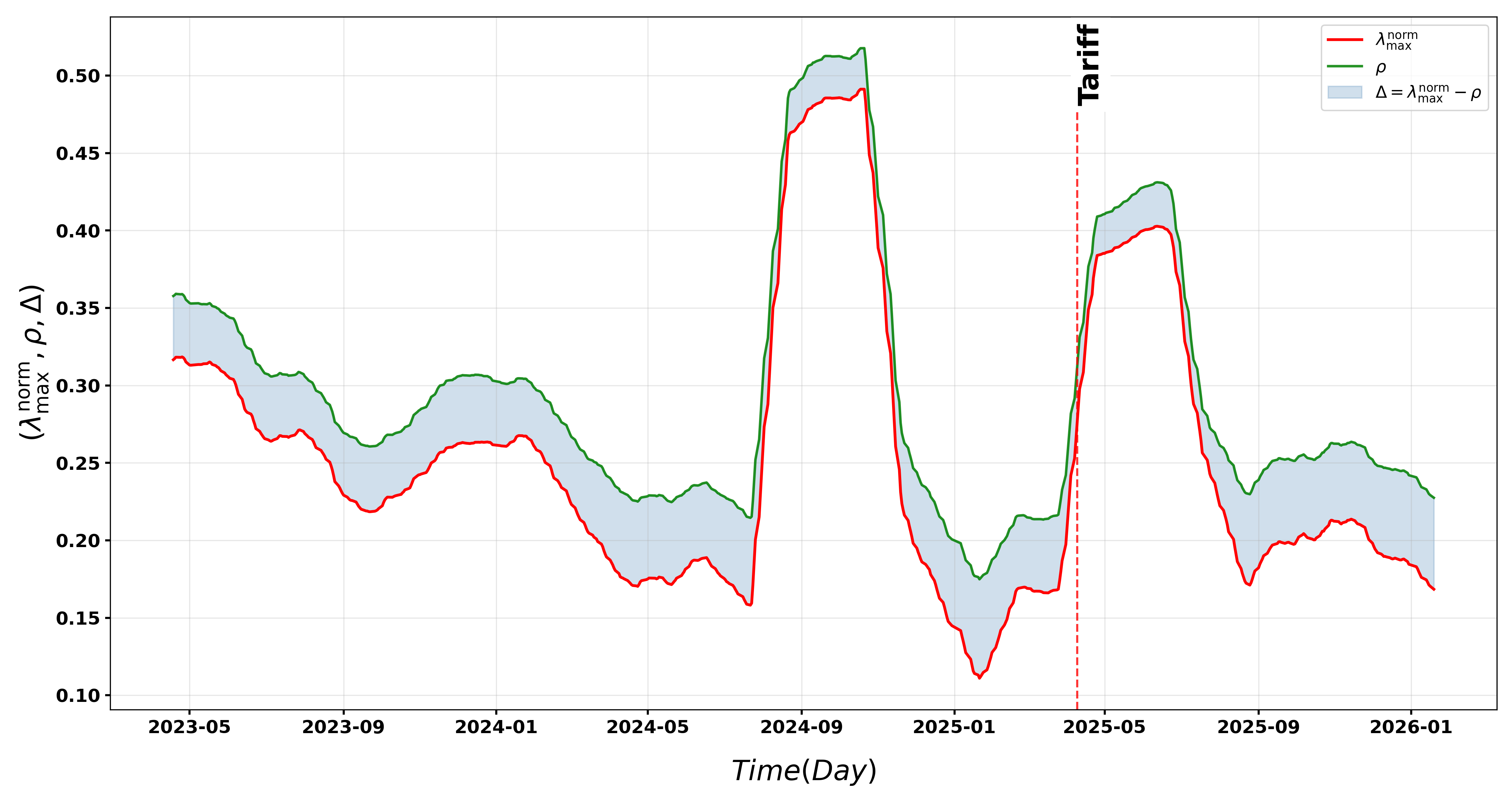}
    }

    \vspace{0.1 cm}

    \subfigure[India\label{fig:sector_Health_india}]{
        \includegraphics[width=0.45\linewidth]{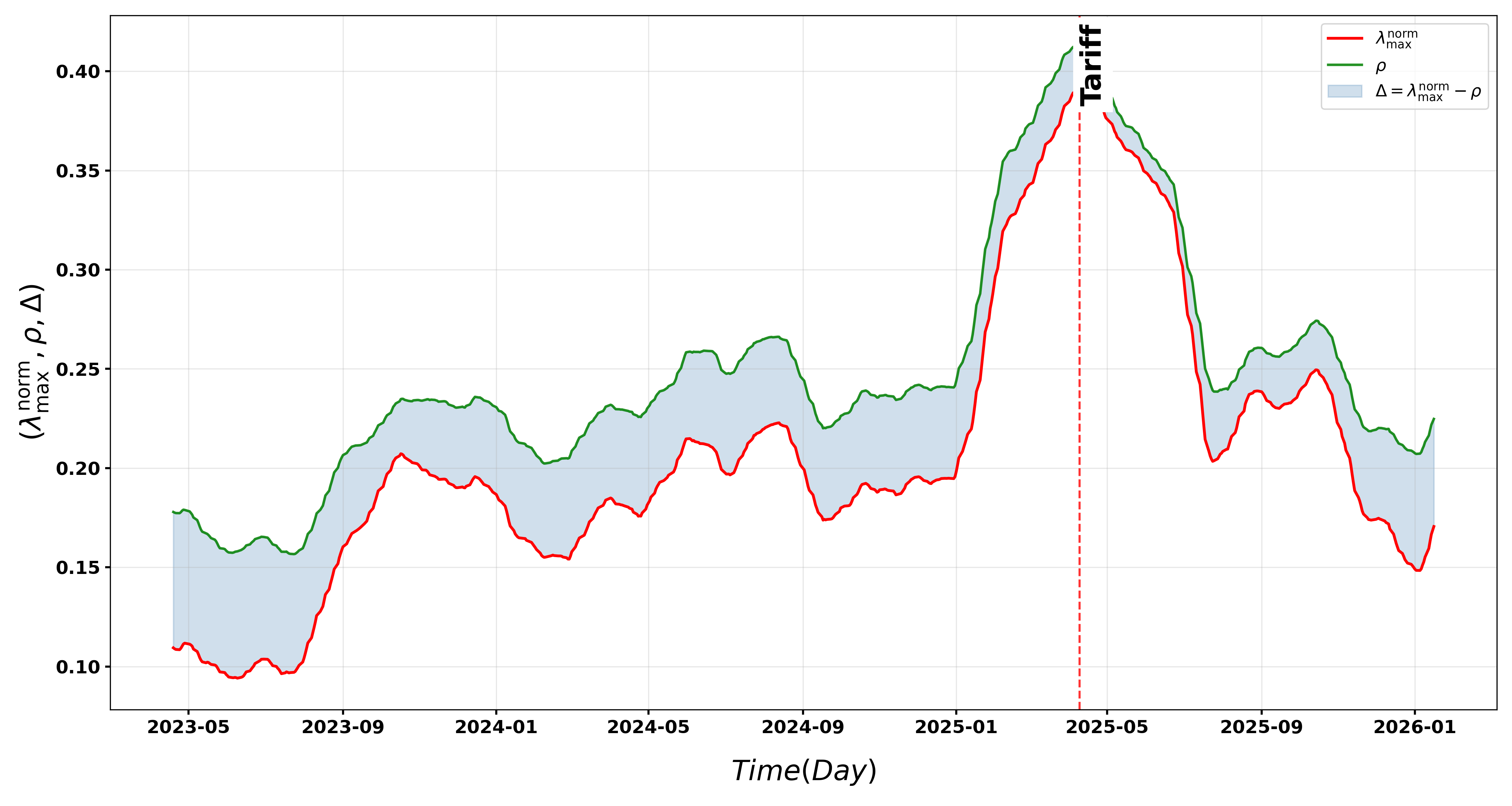}
    }
    \hspace{0.01\linewidth}
    \subfigure[ Germany\label{fig:sector_Health_germany}]{
        \includegraphics[width=0.45\linewidth]{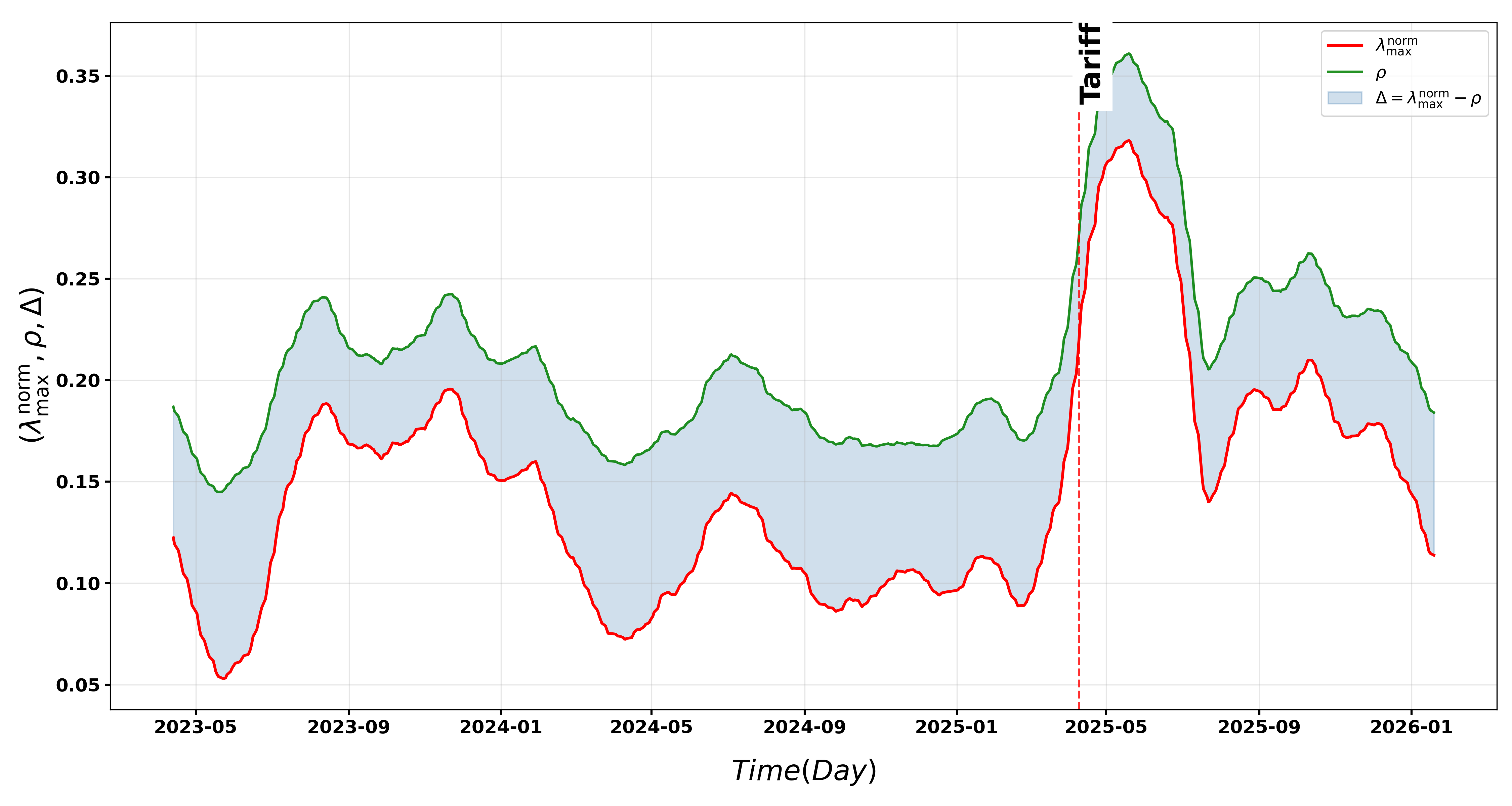}
    }

    \vspace{0.1cm}

    \subfigure[United States\label{fig:sector_Health_us}]{
        \includegraphics[width=0.80\linewidth]{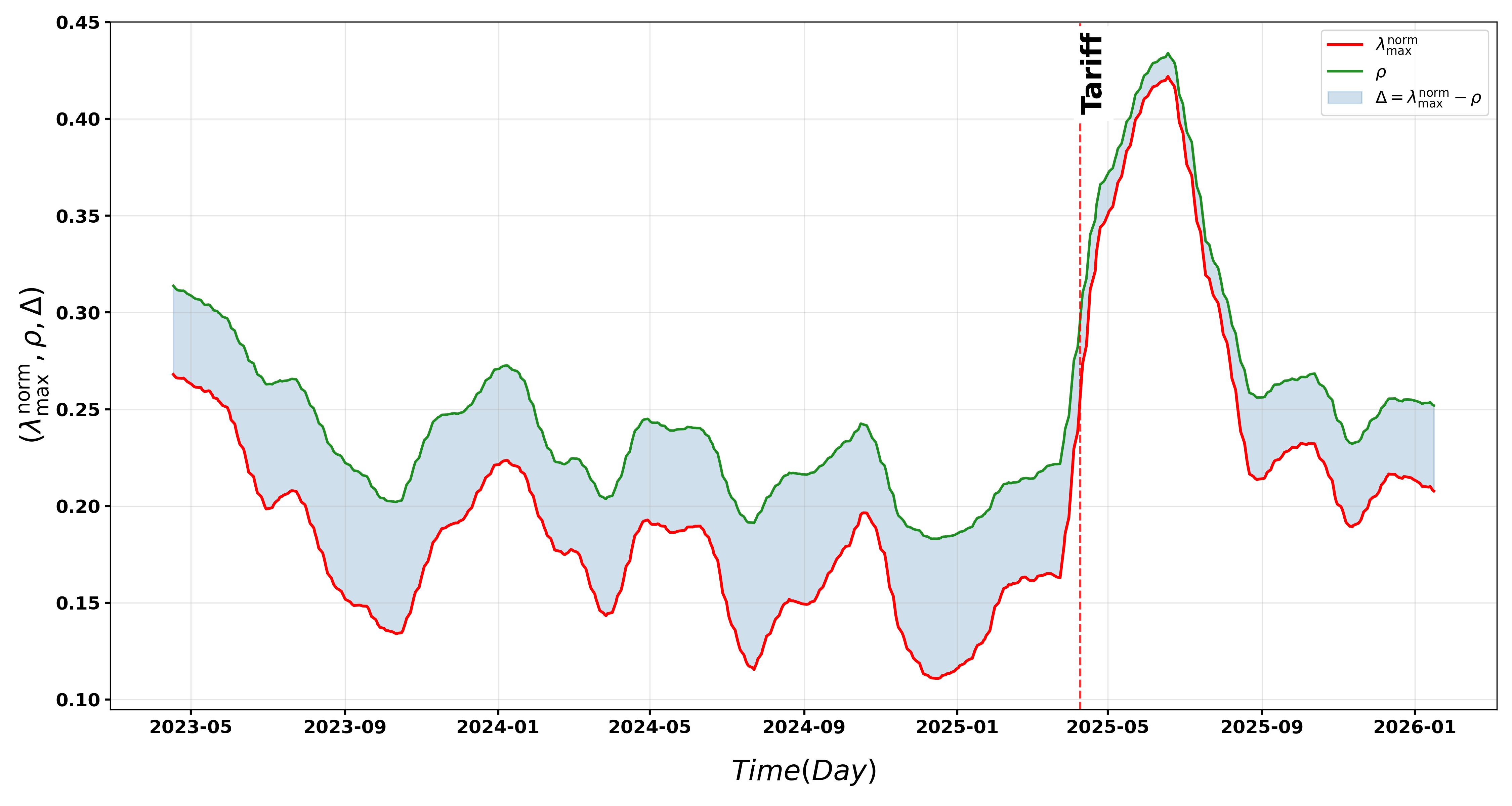}
    }

    \caption{Time evolution of RMT-based complexity metrics for G5 stock markets for the Health sector. The red line represents the normalized largest eigenvalue $\lambda_{\max}^{\text{norm}}(t)$, the green line shows the raw average correlation $\rho(t)$, and the shaded region between them denotes the complexity gap $\Delta(t)$. The vertical red dashed line marks the U.S. tariff announcement in April 2025. Across all markets, we observe a consistent three-phase pattern: a pre-event state with a persistent gap $\Delta(t) < 0$, a sharp convergence with $\Delta(t) \approx 0$ immediately following the shock, and a post-event recovery pattern characterized by gap re-widening, a secondary convergence, and finally sustained structural recovery.}
    \label{fig:sector_Health}
\end{figure}

\clearpage

\begin{figure}[htbp]
    \centering

    \subfigure[ China\label{fig:sector_Consumer_china}]{
        \includegraphics[width=0.45\linewidth]{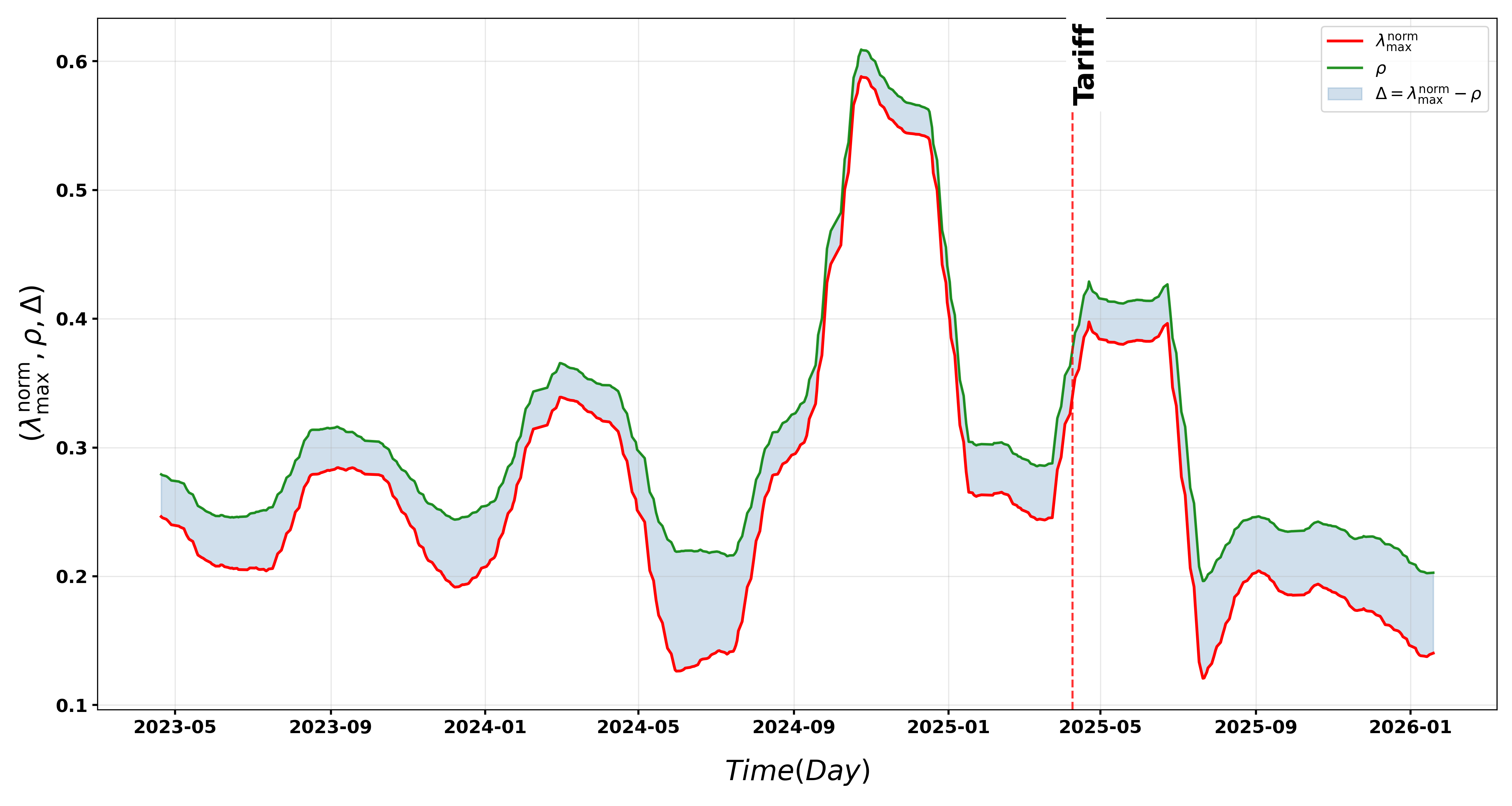}
    }
    \hspace{0.01\linewidth}
    \subfigure[ Japan\label{fig:sector_Consumer_japan}]{
        \includegraphics[width=0.45\linewidth]{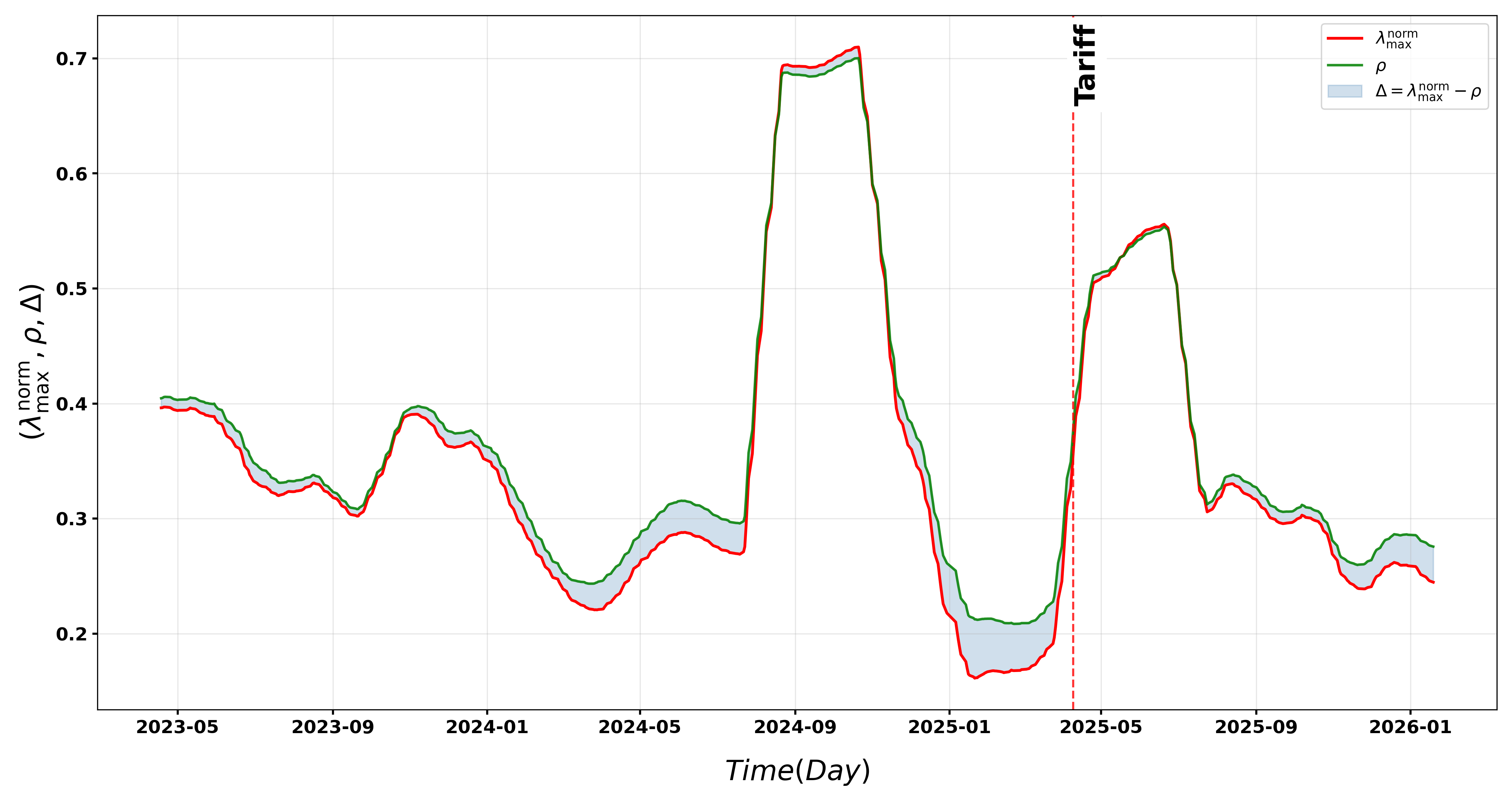}
    }

    \vspace{0.1 cm}

    \subfigure[India\label{fig:sector_Consumer_india}]{
        \includegraphics[width=0.45\linewidth]{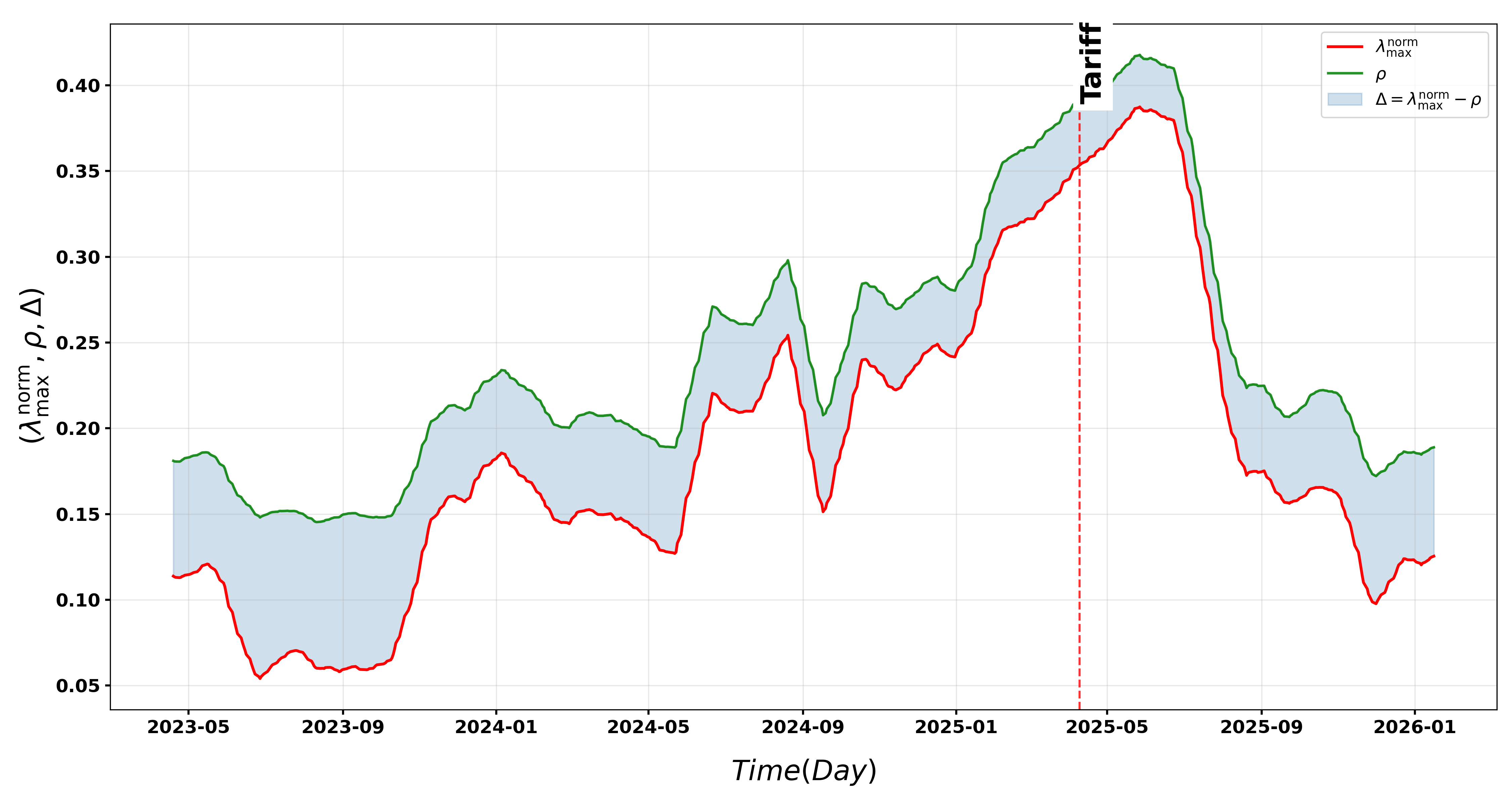}
    }
    \hspace{0.01\linewidth}
    \subfigure[ Germany\label{fig:sector_Consumer_germany}]{
        \includegraphics[width=0.45\linewidth]{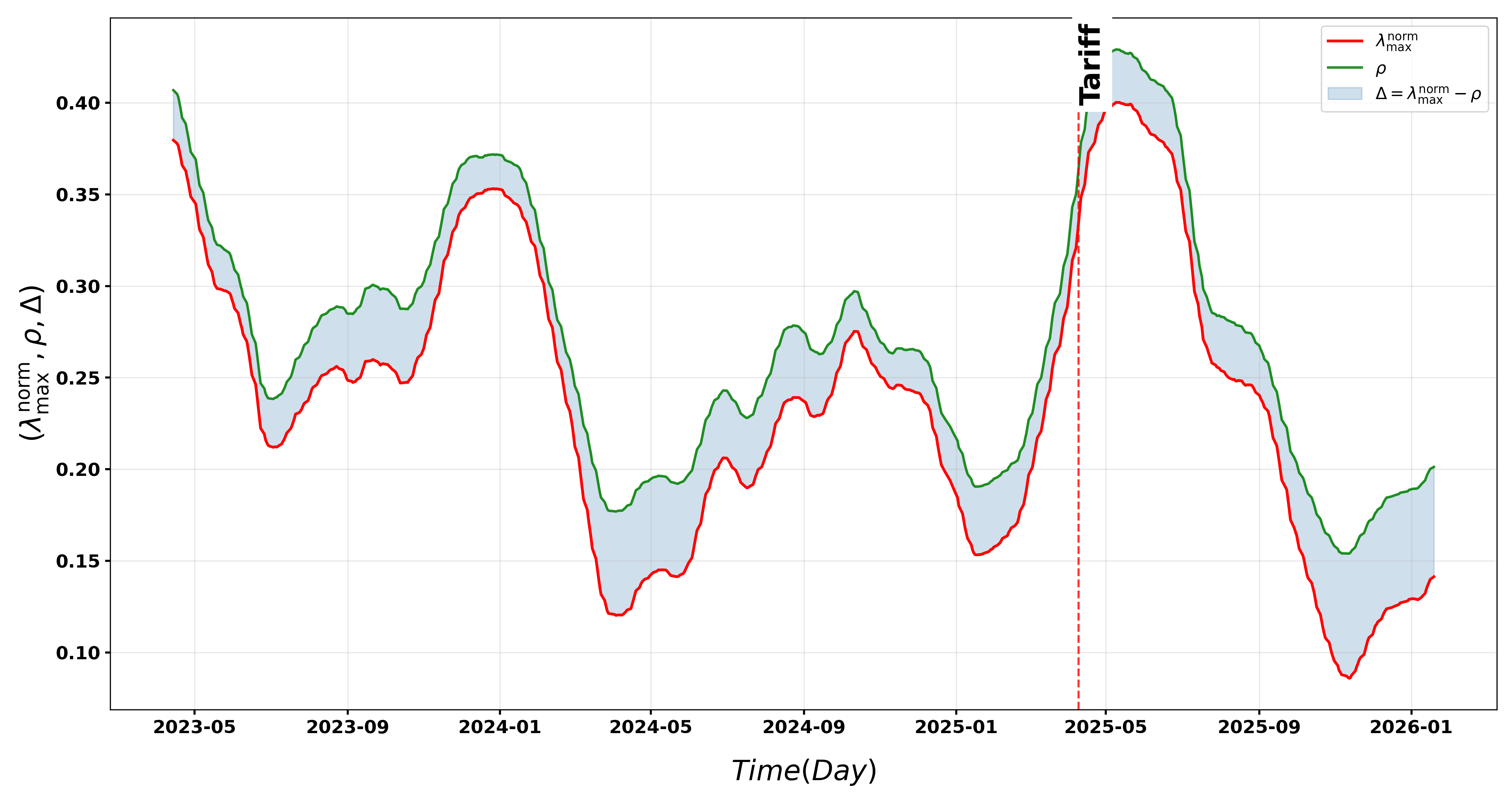}
    }

    \vspace{0.1cm}

    \subfigure[United States\label{fig:sector_Consumer_us}]{
        \includegraphics[width=0.80\linewidth]{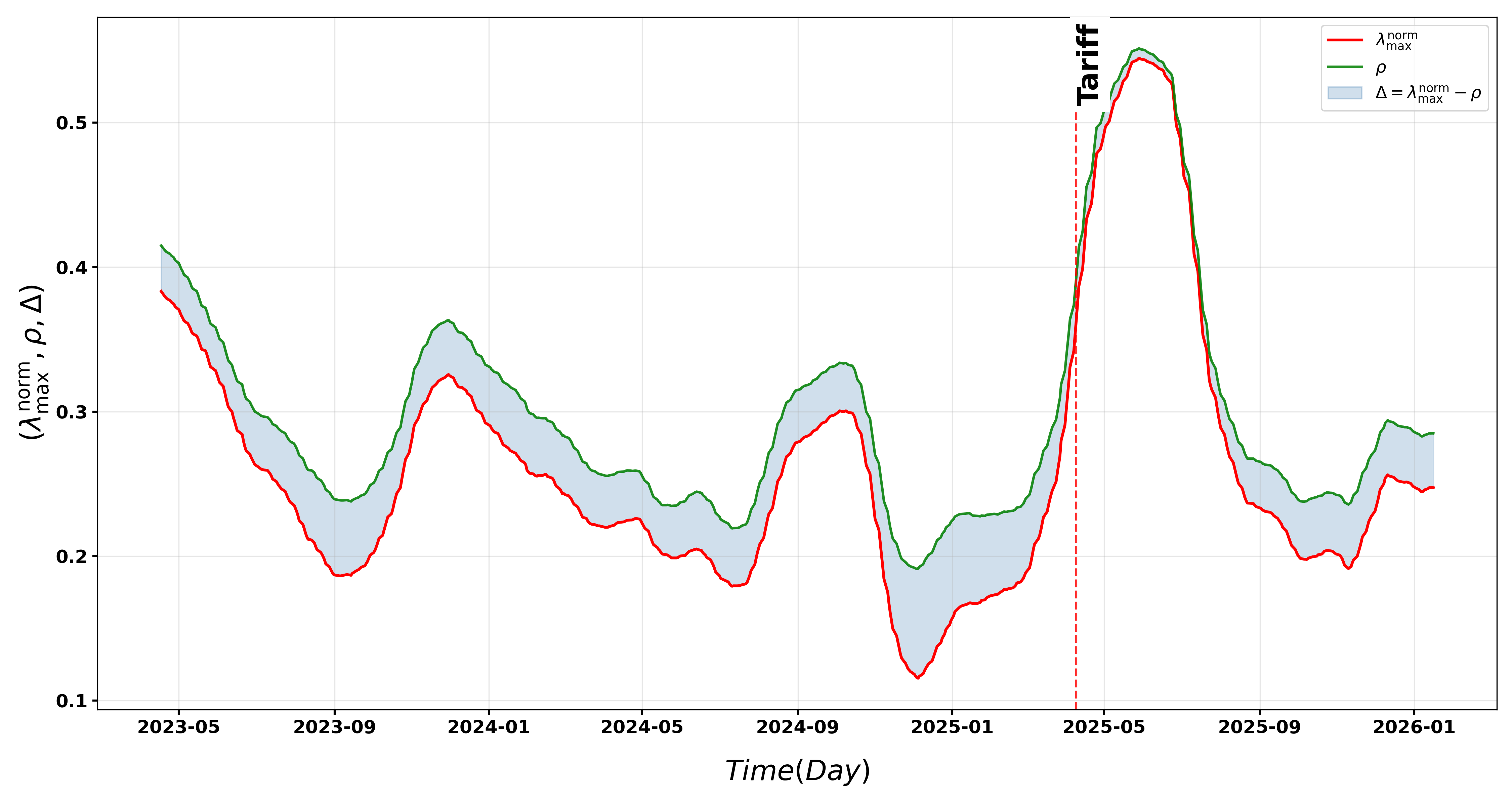}
    }

    \caption{Time evolution of RMT-based complexity metrics for G5 stock markets for the Consumer sector. The red line represents the normalized largest eigenvalue $\lambda_{\max}^{\text{norm}}(t)$, the green line shows the raw average correlation $\rho(t)$, and the shaded region between them denotes the complexity gap $\Delta(t)$. The vertical red dashed line marks the U.S. tariff announcement in April 2025. Across all markets, we observe a consistent three-phase response: a pre-event state with a persistent gap $\Delta(t) < 0$, a sharp convergence with $\Delta(t) \approx 0$ immediately following the shock, and a post-event recovery pattern characterized by gap re-widening, a secondary convergence, and finally sustained structural recovery.}
    \label{fig:sector_Consumer}
\end{figure}

\clearpage

\section{List of Stock Tickers}
\label{List of Stock Tickers}

\setlength{\tabcolsep}{10pt}
\setlength{\LTcapwidth}{\textwidth}

\begin{longtable}{cllllll}

\caption{Stock tickers across five sectors and five countries analyzed in this study. The sample comprises 125 listed equities spanning the Finance, Industrials, Consumer Discretionary, Information Technology, and Healthcare sectors drawn from the United States, India, Japan, China, and Germany, sourced from Yahoo Finance, and covering major benchmark indices in each market.}\\

\hline
\textbf{S.No.} & \textbf{Sector} & \textbf{US} & \textbf{India} & \textbf{Japan} & \textbf{China} & \textbf{Germany} \\
\hline
\endfirsthead

\multicolumn{7}{c}{\tablename\ \thetable{} \textit{}} \\
\hline
\textbf{S.No.} & \textbf{Sector} & \textbf{US} & \textbf{India} & \textbf{Japan} & \textbf{China} & \textbf{Germany} \\
\hline
\endhead

\hline
\multicolumn{7}{r}{\textit{}} \\
\endfoot

\hline
\endlastfoot

% ================= FINANCE =================
1  & Finance & JPM   & HDFCBANK.NS   & 8306.T & 601398.SS & ALV.DE \\
2  & Finance & V     & ICICIBANK.NS  & 8316.T & 601288.SS & DBK.DE \\
3  & Finance & MA    & SBIN.NS       & 8411.T & 601939.SS & MUV2.DE \\
4  & Finance & BAC   & LICI.NS       & 8766.T & 601988.SS & DB1.DE \\
5  & Finance & WFC   & AXISBANK.NS   & 8750.T & 601318.SS & HNR1.DE \\
6  & Finance & BRK-B & KOTAKBANK.NS  & 8604.T & 600036.SS & CBK.DE \\
7  & Finance & GS    & BAJFINANCE.NS & 7182.T & 601628.SS & TLX.DE \\
8  & Finance & MS    & BAJAJFINSV.NS & 6178.T & 601166.SS & DWS.DE \\
9  & Finance & AXP   & INDUSINDBK.NS & 8725.T & 600030.SS & GLJ.DE \\
10 & Finance & C     & JIOFIN.NS     & 8630.T & 601658.SS & HYQ.DE \\
11 & Finance & BLK   & PNB.NS        & 8308.T & 601328.SS & WUW.DE \\
12 & Finance & SPGI  & BANKBARODA.NS & 8591.T & 600000.SS & PBB.DE \\
13 & Finance & SCHW  & CHOLAFIN.NS   & 8309.T & 601601.SS & MLP.DE \\
14 & Finance & PGR   & SHRIRAMFIN.NS & 8473.T & 601688.SS & ARL.DE \\
15 & Finance & CB    & HDFCLIFE.NS   & 7186.T & 600919.SS & DBAN.DE \\
16 & Finance & CME   & SBILIFE.NS    & 8331.T & 000001.SZ & FTK.DE \\
17 & Finance & ICE   & MUTHOOTFIN.NS & 8354.T & 600016.SS & BNP.DE \\
18 & Finance & MMC   & CANBK.NS      & 5831.T & 601998.SS & PAT.DE \\
19 & Finance & MCO   & PFC.NS        & 8795.T & 600958.SS & G24.DE \\
20 & Finance & COF   & RECLTD.NS     & 8593.T & 600837.SS & LEG.DE \\  % replaced GFT.DE (IT) with LEG.DE
21 & Finance & AON   & UNIONBANK.NS  & 8304.T & 000166.SZ & OLB.DE \\
22 & Finance & MET   & IRFC.NS       & 8303.T & 000776.SZ & PCZ.DE \\
23 & Finance & PRU   & IDFCFIRSTB.NS & 8253.T & 601211.SS & LUS.DE \\ % replaced duplicate 8303.T with 8253.T
24 & Finance & TRV   & YESBANK.NS    & 8410.T & 601881.SS & XTP.DE \\
25 & Finance & IBKR  & ICICIPRULI.NS & 5838.T & 002142.SZ & VNA.DE \\ % replaced duplicate FTK.DE with VNA.DE

% ================= INDUSTRIALS =================
26 & Industrials & GE   & LT.NS           & 6861.T & 300750.SZ & SIE.DE \\
27 & Industrials & CAT  & ADANIPORTS.NS   & 8058.T & 601800.SS & DHL.DE \\ % replaced duplicate 002594.SZ (BYD) with 601800.SS
28 & Industrials & RTX  & HAL.NS          & 8001.T & 600031.SS & AIR.DE \\
29 & Industrials & GEV  & ADANIENT.NS     & 8031.T & 601668.SS & SND.DE \\
30 & Industrials & BA   & EICHERMOT.NS    & 6954.T & 601919.SS & SDF.DE \\ % replaced duplicate BMW.DE with SDF.DE
31 & Industrials & HON  & INDIGO.NS       & 6273.T & 601766.SS & SZG.DE \\ % replaced duplicate VOW3.DE with SZG.DE
32 & Industrials & LMT  & TATAMOTORS.NS   & 6301.T & 601390.SS & DTG.DE \\
33 & Industrials & DE   & ABB.NS          & 9022.T & 600893.SS & RHM.DE \\
34 & Industrials & UNP  & ASHOKLEY.NS     & 9020.T & 000338.SZ & MTX.DE \\
35 & Industrials & ETN  & CUMMINSIND.NS   & 6367.T & 002475.SZ & BNR.DE \\
36 & Industrials & PH   & SIEMENS.NS      & 6594.T & 601989.SS & G1A.DE \\
37 & Industrials & NOC  & POLYCAB.NS      & 6326.T & 000157.SZ & KGX.DE \\
38 & Industrials & GD   & MAZDOCK.NS      & 7011.T & 000425.SZ & TKA.DE \\
39 & Industrials & TT   & GMRAIRPORT.NS   & 8002.T & 600089.SS & FRA.DE \\
40 & Industrials & UPS  & CGPOWER.NS      & 8053.T & 600406.SS & LHA.DE \\ % LHA.DE kept here (airlines)
41 & Industrials & WM   & BHEL.NS         & 6201.T & 300014.SZ & HLAG.DE \\
42 & Industrials & EMR  & POWERINDIA.NS   & 6506.T & 688599.SS & HEI.DE \\
43 & Industrials & HWM  & GVT\&D.NS       & 6501.T & 300274.SZ & RAA.DE \\ % RAA.DE kept here (industrial kitchen equipment)
44 & Industrials & MMM  & HAVELLS.NS      & 9202.T & 002459.SZ & JUN3.DE \\
45 & Industrials & CMI  & WAAREEENER.NS   & 9201.T & 601012.SS & DUE.DE \\
46 & Industrials & TDG  & BEL.NS          & 9101.T & 601111.SS & KRN.DE \\
47 & Industrials & ITW  & ASTRAL.NS       & 9104.T & 600029.SS & GBF.DE \\
48 & Industrials & CTAS & KEI.NS          & 9107.T & 002352.SZ & HOT.DE \\
49 & Industrials & FDX  & CONCOR.NS       & 1802.T & 688187.SS & NDA.DE \\
50 & Industrials & VRT  & BHARATFORG.NS   & 1812.T & 601006.SS & DEZ.DE \\

% ================= CONSUMER =================
51 & Consumer & AMZN & MARUTI.NS      & 7203.T & 000333.SZ & MBG.DE \\
52 & Consumer & TSLA & M\&M.NS        & 6758.T & 002594.SZ & BMW.DE \\ % BMW.DE kept here, BYD also here
53 & Consumer & HD   & TITAN.NS       & 7267.T & 000651.SZ & VOW3.DE \\
54 & Consumer & MCD  & DMART.NS       & 7974.T & 600104.SS & P911.DE \\
55 & Consumer & LOW  & BAJAJ-AUTO.NS  & 9983.T & 600690.SS & ADS.DE \\
56 & Consumer & NKE  & TVSMOTOR.NS    & 6902.T & 601888.SS & CON.DE \\
57 & Consumer & SBUX & JUBLFOOD.NS    & 4661.T & 601127.SS & PUM.DE \\
58 & Consumer & TJX  & TRENT.NS       & 5108.T & 600660.SS & DHER.DE \\
59 & Consumer & BKNG & HEROMOTOCO.NS  & 7269.T & 601633.SS & ZAL.DE \\
60 & Consumer & ABNB & ASIANPAINT.NS  & 6752.T & 000625.SZ & HFG.DE \\
61 & Consumer & CMG  & INDHOTEL.NS    & 7201.T & 600741.SS & BOSS.DE \\
62 & Consumer & F    & MOTHERSON.NS   & 4755.T & 002027.SZ & TUI1.DE \\
63 & Consumer & GM   & ZOMATO.NS      & 7202.T & 300144.SZ & EVD.DE \\
64 & Consumer & MAR  & SWIGGY.NS      & 7272.T & 603515.SS & HBM.DE \\ % replaced duplicate LHA.DE with HBM.DE
65 & Consumer & LULU & PAGEIND.NS     & 7832.T & 603486.SS & FIE.DE \\
66 & Consumer & HLT  & BERGEPAINT.NS  & 4911.T & 000921.SZ & SIX2.DE \\
67 & Consumer & ROST & KALYANKJIL.NS  & 9843.T & 002032.SZ & HBH.DE \\
68 & Consumer & AZO  & APOLLOTYRE.NS  & 7270.T & 002508.SZ & ZOO.DE \\ % replaced duplicate RAA.DE with ZOO.DE
69 & Consumer & RCL  & BALKRISIND.NS  & 7261.T & 601058.SS & CEC.DE \\
70 & Consumer & LEN  & DLF.NS         & 3099.T & 601021.SS & B4B.DE \\
71 & Consumer & TGT  & METROBRAND.NS  & 7532.T & 603885.SS & PAH3.DE \\
72 & Consumer & DHI  & NYKAA.NS       & 7453.T & 300413.SZ & HLE.DE \\
73 & Consumer & EBAY & MANYAVAR.NS    & 7731.T & 002739.SZ & TRG.DE \\
74 & Consumer & ULTA & VEDANT.NS      & 6952.T & 600177.SS & VIB3.DE \\
75 & Consumer & TPR  & SHOPERSTOP.NS  & 5802.T & 600754.SS & ST5.DE \\

% ================= IT =================
76  & IT & NVDA & TCS.NS           & 8035.T & 688981.SS & SAP.DE \\
77  & IT & MSFT & INFY.NS          & 6857.T & 603501.SS & IFX.DE \\
78  & IT & AAPL & HCLTECH.NS       & 9984.T & 601138.SS & SHL.DE \\
79  & IT & GOOGL & WIPRO.NS        & 6971.T & 002415.SZ & TMV.DE \\ % fixed GOOGL\& to GOOGL
80  & IT & AVGO & LTIM.NS          & 6981.T & 000063.SZ & NEM.DE \\
81  & IT & ORCL & TECHM.NS         & 6723.T & 002230.SZ & BC8.DE \\
82  & IT & CRM  & OFSS.NS          & 6703.T & 002371.SZ & COK.DE \\ % replaced duplicate 6971.T with 6703.T
83  & IT & AMD  & PERSISTENT.NS    & 6920.T & 002049.SZ & SOW.DE \\
84  & IT & ADBE & LTTS.NS          & 6146.T & 603986.SS & AIXA.DE \\
85  & IT & IBM  & MPHASIS.NS       & 6702.T & 300782.SZ & WAF.DE \\
86  & IT & CSCO & COFORGE.NS       & 6701.T & 000977.SZ & UTDI.DE \\
87  & IT & TXN  & KPITTECH.NS      & 6963.T & 688012.SS & 1U1.DE \\
88  & IT & INTC & TATAELXSI.NS     & 6762.T & 600584.SS & AOF.DE \\
89  & IT & NOW  & NAUKRI.NS        & 7735.T & 600745.SS & JEN.DE \\
90  & IT & AMAT & CYIENT.NS        & 9613.T & 300661.SZ & COP.DE \\
91  & IT & QCOM & ZENSARTECH.NS    & 4307.T & 603019.SS & GFT.DE \\ % GFT.DE kept here (IT)
92  & IT & LRCX & SONATSOFTW.NS    & 4704.T & 600570.SS & SANT.DE \\
93  & IT & PANW & HAPPISTMND.NS    & 4684.T & 688256.SS & NA9.DE \\
94  & IT & ANET & INTELLECT.NS     & 9719.T & 688126.SS & YSN.DE \\
95  & IT & MU   & AFFLE.NS         & 4739.T & 300124.SZ & ADN1.DE \\
96  & IT & SNOW & NEWGEN.NS        & 6645.T & 002241.SZ & SMHN.DE \\
97  & IT & CRWD & TANLA.NS         & 6965.T & 000725.SZ & PFV.DE \\
98  & IT & SNPS & LATENTVIEW.NS    & 6976.T & 000100.SZ & ELG.DE \\
99  & IT & CDNS & RATEGAIN.NS      & 4062.T & 600703.SS & BSL.DE \\
100 & IT & PLTR & MAPMYINDIA.NS    & 3436.T & 688008.SS & LPK.DE \\

% ================= HEALTHCARE =================
101 & Healthcare & LLY  & SUNPHARMA.NS   & 4502.T & 600276.SS & BAYN.DE \\
102 & Healthcare & UNH  & DIVISLAB.NS    & 4503.T & 300760.SZ & MRK.DE \\
103 & Healthcare & JNJ  & CIPLA.NS       & 4568.T & 603259.SS & SHL.DE \\
104 & Healthcare & MRK  & DRREDDY.NS     & 4543.T & 600436.SS & FRE.DE \\
105 & Healthcare & ABBV & APOLLOHOSP.NS  & 4519.T & 000538.SZ & FME.DE \\
106 & Healthcare & TMO  & TORNTPHARM.NS  & 7733.T & 300015.SZ & SRT3.DE \\
107 & Healthcare & ABT  & MANKIND.NS     & 6869.T & 300122.SZ & QIA.DE \\
108 & Healthcare & DHR  & ZYDUSLIFE.NS   & 4523.T & 000661.SZ & 22UA.DE \\
109 & Healthcare & AMGN & MAXHEALTH.NS   & 4507.T & 300142.SZ & AFX.DE \\
110 & Healthcare & PFE  & LUPIN.NS       & 4578.T & 600196.SS & BEI.DE \\
111 & Healthcare & ISRG & AUROPHARMA.NS  & 2413.T & 000963.SZ & EVT.DE \\
112 & Healthcare & SYK  & ALKEM.NS       & 4151.T & 002422.SZ & GXI.DE \\
113 & Healthcare & MDT  & ABBOTT.NS      & 4528.T & 600332.SS & COP.DE \\
114 & Healthcare & ELV  & BIOCON.NS      & 4536.T & 600998.SS & MOR.DE \\
115 & Healthcare & GILD & GLENMARK.NS    & 4508.T & 300003.SZ & EUZ.DE \\
116 & Healthcare & BSX  & SYNGENE.NS     & 4581.T & 002044.SZ & DMP.DE \\
117 & Healthcare & VRTX & FORTIS.NS      & 4530.T & 002007.SZ & SAE.DE \\
118 & Healthcare & REGN & GSK.NS         & 4527.T & 002030.SZ & DRW3.DE \\
119 & Healthcare & ZTS  & IPCALAB.NS     & 4506.T & 002603.SZ & MEDP.DE \\
120 & Healthcare & HCA  & MEDANTA.NS     & 6849.T & 688363.SS & NXU.DE \\
121 & Healthcare & BDX  & NH.NS          & 4544.T & 600763.SS & SBS.DE \\
122 & Healthcare & HUM  & LAURUSLABS.NS  & 4887.T & 300759.SZ & FYB.DE \\
123 & Healthcare & MRNA & NATCOPHARM.NS  & 4516.T & 603658.SS & SYAB.DE \\
124 & Healthcare & EW   & ASTERDM.NS     & 4587.T & 002821.SZ & PHH2.DE \\
125 & Healthcare & DXCM & GRANULES.NS    & 7747.T & 600085.SS & BIO3.DE \\ % replaced -- with 600085.SS and PBG.DE with BIO3.DE

\hline
\end{longtable}

\bibliographystyle{elsarticle-num} 
\bibliography{aipsamp}

\end{document}